\documentclass[aps,prd,amsmath,two column,amssymb,showpacs]{revtex4}
\usepackage{amssymb}
\usepackage{txfonts}
\usepackage{mathbbold}
\usepackage{amsfonts}
\usepackage{mathrsfs}
\usepackage{epsfig,bm,dcolumn}
\usepackage{graphicx}
\usepackage{color}
\usepackage{amsmath}

\begin{document}

\title{Nambu--Jona-Lasinio model description of weakly interacting Bose condensate \\ and BEC-BCS crossover in dense QCD-like theories}

\author{Lianyi He\footnote{Email address: lianyi@th.physik.uni-frankfurt.de}}
\address{Frankfurt Institute for Advanced Studies and Institute
for Theoretical Physics, J.W.Goethe University, 60438 Frankfurt am
Main, Germany}

\date{\today}

\begin{abstract}
QCD-like theories possess a positively definite fermion determinant
at finite baryon chemical potential $\mu_{\text B}$ and the lattice
simulation can be successfully performed. While the chiral
perturbation theories are sufficient to describe the Bose condensate
at low density, to describe the crossover from Bose-Einstein
condensation (BEC) to BCS superfluidity at moderate density we
should use some fermionic effective model of QCD, such as the
Nambu--Jona-Lasinio model. In this paper, using two-color
two-flavor QCD as an example, we examine how the Nambu--Jona-Lasinio model describes
the weakly interacting Bose condensate at low density and the
BEC-BCS crossover at moderate density. Near the quantum phase
transition point $\mu_{\text B}=m_\pi$ ($m_\pi$ is the mass of
pion/diquark multiplet), the Ginzburg-Landau free energy at the mean-field level
 can be reduced to the Gross-Pitaevskii free energy
describing a weakly repulsive Bose condensate with a diquark-diquark
scattering length identical to that predicted by the chiral
perturbation theories. The Goldstone mode recovers the Bogoliubov
excitation in weakly interacting Bose condensates. The results of
in-medium chiral and diquark condensates predicted by chiral
perturbation theories are analytically recovered. The BEC-BCS
crossover and meson Mott transition at moderate baryon chemical
potential as well as the beyond-mean-field corrections are studied.
Part of our results can also be applied to real QCD at finite baryon
or isospin chemical potential.
\end{abstract}


\maketitle

\section {Introduction}
\label{s1}

A good understanding of quantum chromodynamics (QCD) at finite temperature
and baryon density is crucial for us to understand a wide range of
physical phenomena. For instance, to understand the evolution of the
Universe in the first few seconds, one needs the knowledge of QCD
phase transition at temperature $T\sim200$MeV and very small baryon
density. On the other hand, understanding the physics of neutron
stars requires the knowledge of QCD at high baryon density and very
low temperature \cite{itoh}. Lattice simulation of QCD at finite
temperature has been successfully performed in the past few decades;
however, no successful lattice simulation at high baryon density has
been done due to the sign problem \cite{Lr01,Lr02}: The fermion
determinant is not positively definite in presence of a nonzero
baryon chemical potential $\mu_{\text B}$.

We thus look for some special cases which have a positively definite
fermion determinant. One case is QCD at finite isospin chemical
potential $\mu_{\text I}$ \cite{ISO01,ISO02}, where the ground state
changes from a pion condensate to a BCS superfluid with
quark-antiquark condensation with increasing isospin density.
Another case is the QCD-like theories
\cite{QL01,QL02,QL03,QL04,QL05,QL06,QL07} where quarks are in a real
or pseudoreal representation of the gauge group, including
two-color QCD with quarks in the fundamental representation and QCD
with quarks in the adjoint representation. While these cases do not
correspond to the real world, they can be simulated on the lattice
and may give us some information of real QCD at finite baryon
density. For all these special cases, chiral perturbation theories
predict a continuous quantum phase transition from the vacuum to the
matter phase at baryon or isospin chemical potential equal to the
pion mass, in contrast to real QCD where the phase transition takes
place at $\mu_{\text B}$ approximately equal to the nucleon mass.
The resulting matter near the quantum phase transition is a dilute
Bose condensate of diquarks or pions with weakly repulsive
interactions \cite{Bose01}. The equations of state and elementary
excitations in such matter have been investigated many years ago by
Bogoliubov \cite{Bose02} and Lee, Huang, and Yang \cite{Bose03}.
Bose-Einstein condensation (BEC) phenomenon is believed to widely
exist in dense matter, such as pions and kaons can condense in
neutron star matter if the electron chemical potential exceeds the
effective mass for pions and kaons \cite{PiC01,PiC02,PiC03,PiC04}.
However, the condensation of pions and kaons in neutron star matter
is rather complicated due to the meson-nucleon interactions in dense
nuclear medium. On the other hand, at asymptotically high density,
perturbative QCD calculations show that the ground state is a weakly
coupled BCS superfluid with the condensation of overlapping Cooper
pairs \cite{ISO01,ISO02,pQ01,pQ02,pQ03,Yama}. It is interesting that
the dense BCS superfluid and the dilute Bose condensate have the
same symmetry breaking pattern and thus are continued with one
another. In condensed matter physics, this phenomenon was first
discussed by Eagles \cite{Eagles} and Leggett \cite{Leggett} and is
now called BEC-BCS crossover. It has been successfully realized
using ultracold fermionic atoms in the past few years
\cite{BCSBECexp}.

While the lattice simulations of two-color QCD at finite baryon
chemical potential \cite{L2C01,L2C02,L2C03,L2C04,L2C05,L2C06} and
QCD at finite isospin chemical potential
\cite{Liso01,Liso02,Liso03,Liso04} have been successfully performed,
we still ask for some effective models to link the physics of Bose
condensate and the BCS superfluidity. The chiral perturbation
theories \cite{ISO01,ISO02,QL01,QL02,QL03,QL04,QL05,QL06,CHiso} as
well as the linear sigma models \cite{Sigma}, which describe only
the physics of Bose condensate, do not meet our purpose. The
Nambu--Jona-Lasinio (NJL) model \cite{NJL} with quarks as elementary
blocks, which describes well the mechanism of chiral symmetry
breaking and low energy phenomenology of the QCD vacuum, is
generally believed to work at low and moderate temperatures and
densities \cite{NJLreview01,NJLreview02,NJLreview03}. Recently, this
model has been used to describe the superfluid transition at finite
chemical potentials
\cite{QL07,isoNJL01,isoNJL02,isoNJL03,isoNJL04,isoNJL05,isoNJL06,isoNJL07,isoNJL08,isoNJL09,2CNJL01,2CNJL02,2CNJL03,2CNJL04}
for the special cases we are interested in this paper. One finds
that the critical chemical potential for the superfluid transition
predicted by the NJL model is indeed equal to the pion mass
\cite{isoNJL06,2CNJL01}, and the chiral and diquark condensates
obtained from the mean-field calculation agree with the results from
lattice simulations and chiral perturbation theories near the
quantum phase transition \cite{isoNJL06,2CNJL01}. The NJL model also
predicts a BEC-BCS crossover when the chemical potential increases
\cite{isoNJL08,2CNJL02,2CNJL03,2CNJL04}. A natural problem arises:
how can the fermionic NJL model describe the weakly interacting Bose
condensate near the quantum phase transition? In fact, we do not
know how the repulsive interactions among diquarks or mesons enter
in the pure mean-field calculations \cite{2CNJL01,2CNJL02,2CNJL03}.
In this paper, we will focus on this problem and show that the
repulsive interaction is indeed properly included even in the mean-field
 calculations. This phenomenon is in fact analogous to the BCS
description of the molecular condensation in strongly interacting
Fermi gases studied by Leggett many years ago \cite{Leggett}.
Fermionic models have been used to describe the BEC-BCS crossover in
cold Fermi gases by the cold atom community. Recently, it has been
shown that we can recover the equation of state of the dilute Bose
condensate with correct boson-boson scattering length in the strong
coupling limit, including the Lee-Huang-Yang correction by
considering the beyond-mean-field corrections
\cite{HU,Diener,Leyron}. In Appendix \ref{app1}, we give a summary
of the many-body theoretical approach in cold atoms, which is useful
for us to understand the theoretical approach and the results of
this paper.

In this paper, using two-color two-flavor QCD as an example and
following the theoretical approach of the BEC-BCS crossover in cold
Fermi gases \cite{HU,Diener}, we examine how the NJL model describes
the weakly interacting Bose condensate and the BEC-BCS crossover.
Near the quantum phase transition point $\mu_{\text B}=m_\pi$, we
perform a Ginzburg-Landau expansion of the effective potential at
the mean-field level, and show that the Ginzburg-Landau free energy is
essentially the Gross-Pitaevskii free energy describing weakly
interacting Bose condensates via a proper redefinition of the
condensate wave function. As a by-product, we obtain a
diquark-diquark scattering length $a_{\text{dd}}=m_\pi/(16\pi
f_\pi^2)$ ($f_\pi$ is the pion decay constant) characterizing the
repulsive interaction between the diquarks, which recovers the
tree-level result predicted by chiral Lagrangian
\cite{QL01,QL02,QL03,QL04,QL05,QL06}. We also show analytically that
the Goldstone mode takes the same dispersion as the Bogoliubov
excitation in weakly interacting Bose condensates, which gives a
diquark-diquark scattering length identical to that in the
Gross-Pitaevskii free energy. The mixing between the sigma meson and
diquarks plays an important role in recovering the Bogoliubov
excitation. The results of in-medium chiral and diquark condensates
predicted by chiral perturbation theory are analytically recovered.
At high density, we find the superfluid matter undergoes a BEC-BCS
crossover at $\mu_{\text B}\simeq
(m_\sigma/m_\pi)^{1/3}m_\pi\simeq(1.6-2)m_\pi$ with $m_\sigma$ being
the mass of the sigma meson. At $\mu_{\text B}\simeq 3m_\pi$, we
find that the chiral symmetry is approximated restored and the
spectra of pions and sigma meson become nearly degenerated. Well
above the chemical potential of chiral symmetry restoration, the
degenerate pions and sigma meson undergo a Mott transition, where
they become unstable resonances. Because of the spontaneous breaking of
baryon number symmetry, mesons can decay into quark pairs in the
superfluid medium at nonzero momentum.

The beyond-mean-field corrections are studied. The thermodynamic
potential including the Gaussian fluctuations is derived. It is
shown that the vacuum state $|\mu_{\text B}|<m_\pi$ is
thermodynamically consistent in the Gaussian approximation, i.e.,
all thermodynamic quantities keep vanishing in the regime
$|\mu_{\text B}|<m_\pi$ even though the beyond-mean-field
corrections are included. Near the quantum phase transition point,
we expand the fluctuation contribution to the thermodynamic
potential in powers of the superfluid order parameter. To leading
order, the beyond-mean-field correction is quartic and its effect is
to renormalize the diquark-diquark scattering length. The correction
to the mean-field result is shown to be proportional to
$m_\pi^2/f_\pi^2$. Thus, our theoretical approach provides a new way
to calculate the diquark-diquark or meson-meson scattering lengths
in the NJL model beyond the mean-field approximation. We also find that
we can obtain a correct transition temperature of Bose condensation
in the dilute limit, including the beyond-mean-field corrections.

The paper is organized as follows: In Sec. \ref{s2}, we derive
the general effective action of the two-color NJL model at finite
temperature and density, and determine the model parameters via the
vacuum phenomenology. In Sec. \ref{s3}, we investigate the
properties of dilute Bose condensate near the quantum phase
transition at the mean-field level. In Sec. \ref{s5}, the properties
of matter at high density are discussed. Beyond-mean-field
corrections are studied in Sec. \ref{s4}. We summarize in Sec.
\ref{s6}. Natural units are used throughout.

\section {NJL model of two-color QCD}
\label{s2}
Without loss of generality, we study in this paper two-color QCD
(the number of colors $N_c=2$) at finite baryon chemical potential
$\mu_{\text B}$. For vanishing current quark mass $m_0$, two-color
QCD possesses an enlarged flavor symmetry SU$(2N_f)$ [$N_f$ is the
number of flavors], the so-called Pauli-Gursey symmetry which
connects quarks and antiquarks
\cite{QL01,QL02,QL03,QL04,QL05,QL06}. For $N_f=2$, the flavor
symmetry SU$(4)$ is spontaneously broken down to Sp$(4)$ driven by a
nonzero quark condensate $\langle\bar{q}q\rangle$ and there arise
five Goldstone bosons: three pions and two scalar diquarks. For
nonvanishing current quark mass, the flavor symmetry is explicitly
broken, resulting in five pseudo-Goldstone bosons with a small
degenerate mass $m_\pi$. At finite baryon chemical potential
$\mu_{\text B}$, the flavor symmetry SU$(2N_f)$ is explicitly broken
down to SU$_{\text L}(N_f)\otimes$SU$_{\text
R}(N_f)\otimes$U$_{\text B}(1)$. Further, a nonzero diquark
condensate $\langle qq\rangle$ can form at large enough chemical
potentials and breaks spontaneously the U$_{\text B}(1)$ symmetry.
In two-color QCD, the scalar diquarks are in fact the lightest
``baryons," and we expect a baryon superfluid phase with $\langle
qq\rangle\neq0$ for $|\mu_{\text B}|>m_\pi$.

To construct a NJL model for two-color two-flavor QCD with the above
flavor symmetry, we consider a contact current-current interaction
$G_{\text c}\sum_{{\text a}=1}^3(\bar{q}\gamma_\mu t_{\text
a}q)(\bar{q}\gamma^\mu t_{\text a}q)$ where $t_{\text a}$ (${\text
a}=1,2,3$) are the generators of color SU$_{\text c}(2)$ and
$G_{\text c}$ is a phenomenological coupling constant. After the
Fierz transformation we can obtain an effective NJL Lagrangian
density with scalar mesons and color singlet scalar diquarks
\cite{2CNJL01},
\begin{widetext}
\begin{eqnarray}\label{NJL}
{\cal L}_{\text{NJL}}=\bar{q}(i\gamma^\mu\partial_\mu-m_0)q
+G\left[(\bar{q}q)^{2}+(\bar{q}i\gamma_{5}\mbox{\boldmath{$\tau$}}q)^{2}+(\bar{q}i\gamma_5\tau_2t_2q_c)(\bar{q}_ci\gamma_5\tau_2t_2q)\right],
\end{eqnarray}
\end{widetext}
where $q_c={\cal C}\bar{q}^{\text{T}}$ and $\bar{q}_c=q^{\text
T}{\cal C}$ are the charge conjugate spinors with ${\cal
C}=i\gamma_0\gamma_2$ and $\tau_{\text i}$ (${\text i}=1,2,3$) are
the Pauli matrices in the flavor space. The four-fermion coupling
constants for the scalar mesons and diquarks are the same,
$G=3G_{\text c}/4$ \cite{2CNJL01}, which ensures the enlarged flavor
symmetry SU$(2N_f)$ of two-color QCD in the chiral limit $m_0=0$.
One can show explicitly that there are five Goldstone bosons (three
pions and two diquarks) driven by a nonzero quark condensate
$\langle\bar{q}q\rangle$.  With explicit chiral symmetry broken
$m_0\neq0$, pions and diquarks are also degenerate, and their mass
$m_\pi$ can be determined via the standard method for the NJL model
\cite{NJLreview01,NJLreview02,NJLreview03}.

\subsection {Effective action at finite temperature and density }
\label{s2-1}

The partition function of the two-color NJL model (\ref{NJL}) at
finite temperature $T$ and baryon chemical potential $\mu_{\text B}$
is
\begin{eqnarray}
Z_{\text{NJL}}=\int[d\bar{q}][dq]\exp\left[\int dx\left({\cal
L}_{\text{NJL}}+\frac{\mu_{\text
B}}{2}\bar{q}\gamma_{0}q\right)\right],
\end{eqnarray}
where we adopt the finite temperature formalism with $\tau=it$,
$x=(\tau,{\bf r})$, and $\int dx=\int_{0}^{1/T}d\tau\int d^{3}{\bf
r}$. The partition function can be bosonized after introducing the
auxiliary boson fields
\begin{eqnarray}
\sigma(x)=-2G\bar{q}(x)q(x),\ \
\mbox{\boldmath{$\pi$}}(x)=-2G\bar{q}(x)i\gamma_5\mbox{\boldmath{$\tau$}}q(x)
\end{eqnarray}
for mesons and
\begin{eqnarray}
\phi(x)=-2G\bar{q}_c(x)i\gamma_5\tau_2t_2q(x)
\end{eqnarray}
for diquarks.  With the help of the Nambu-Gor'kov representation
$\bar{\Psi} = \left(\begin{array}{cc} \bar{q} &
\bar{q}_c\end{array}\right)$, the partition function can be written
as
\begin{eqnarray}
Z_{\text{NJL}}=\int[d\bar{\Psi}][d\Psi][d\sigma][d\mbox{\boldmath{$\pi$}}][d\phi^\dagger][d\phi]\exp\left(-{\cal
A_{\text{eff}}}\right),
\end{eqnarray}
where the action ${\cal A}_{\text{eff}}$ is given by
\begin{widetext}
\begin{eqnarray}
{\cal A_{\text{eff}}}=\int
dx\frac{\sigma^2(x)+\mbox{\boldmath{$\pi$}}^2(x)+|\phi(x)|^2}{4G}-\int
dx\int dx^\prime\bar{\Psi}(x){\bf G}^{-1}(x,x^\prime)\Psi(x^\prime)
\end{eqnarray}
with the inverse quark propagator defined as
\begin{eqnarray}
{\bf
G}^{-1}(x,x^\prime)=\left(\begin{array}{cc}\gamma^0(-\partial_{\tau}+\frac{\mu_{\text
B}}{2})+i\mbox{\boldmath{$\gamma$}}\cdot\mbox{\boldmath{$\nabla$}}
-\mathcal {M}(x)&-i\gamma_5\phi(x)\tau_2t_2\\
-i\gamma_5\phi^\dagger(x)\tau_2t_2 &
\gamma^0(-\partial_{\tau}-\frac{\mu_{\text
B}}{2})+i\mbox{\boldmath{$\gamma$}}\cdot\mbox{\boldmath{$\nabla$}}-\mathcal
{M}^{\text T}(x)\end{array}\right)\delta(x-x^\prime).
\end{eqnarray}
Here $\mathcal {M}(x) =m_0+\sigma(x)+
i\gamma_5\mbox{\boldmath{$\tau$}}\cdot\mbox{\boldmath{$\pi$}}(x)$.
After integrating out the quarks, we can reduce the partition
function to
$Z_{\text{NJL}}=\int[d\sigma][d\mbox{\boldmath{$\pi$}}][d\phi^\dagger][d\phi]\exp\Big\{-{\cal
S}_{\text{eff}}[\sigma,\mbox{\boldmath{$\pi$}},\phi^\dagger,\phi]\Big\}$,
where the bosonized effective action ${\cal S}_{\text{eff}}$ is
given by
\begin{eqnarray}
{\cal
S}_{\text{eff}}[\sigma,\mbox{\boldmath{$\pi$}},\phi^\dagger,\phi]=\int
dx\frac{\sigma^2(x)+\mbox{\boldmath{$\pi$}}^2(x)+|\phi(x)|^2}{4G}-\frac{1}{2}\text{Tr}\ln{\bf
G}^{-1}(x,x^\prime).
\end{eqnarray}
\end{widetext}
Here the trace $\text{Tr}$ is taken over color, flavor, spin,
Nambu-Gor'kov and coordinate ($x$ and $x^\prime$) spaces. The
thermodynamic potential density of the system is given by
$\Omega(T,\mu_{\text B})=-\lim_{V\rightarrow\infty}(T/V)\ln
Z_{\text{NJL}}$.

\subsection {Evaluating the effective action}
\label{s2-2}

The effective action ${\cal S}_{\text{eff}}$ as well as the
thermodynamic potential $\Omega$ cannot be evaluated exactly in our
$3+1$ dimensional case. In this work, we firstly consider the saddle
point approximation, i.e., the mean-field approximation. Then we
investigate the fluctuations around the mean field.
\\
\emph{(I)Mean-field approximation.} In this approximation, all
bosonic auxiliary fields are replaced by their expectation values.
To this end, we write $\langle\sigma(x)\rangle=\upsilon$,
$\langle\phi(x)\rangle=\Delta$ and set
$\langle\mbox{\boldmath{$\pi$}}(x)\rangle=0$. While $\Delta$ can be
set to be real, we do not do this first in our derivations. We will
show in the following that all physical results depend only on
$|\Delta|^2$. The zeroth order or mean-field effective action reads
\begin{equation}
{\cal
S}_{\text{eff}}^{(0)}=\frac{V}{T}\left[\frac{\upsilon^2+|\Delta|^2}{4G}-\frac{1}{2}\sum_K\text{Trln}\frac{{\cal
G}^{-1}(K)}{T}\right].
\end{equation}
Here and in the following $K=(i\omega_n,{\bf k})$ with
$\omega_n=(2n+1)\pi T$ being the fermion Matsubara frequency, and
$\sum_K=T\sum_n\sum_{\bf k}$ with $\sum_{\bf k}=\int d^3{\bf
k}/(2\pi)^3$. The inverse of the Nambu-Gor'kov quark propagator
${\cal G}^{-1}(K)$ is given by
\begin{eqnarray}
\left(\begin{array}{cc} (i\omega_n+\frac{\mu_{\text B}}{2})\gamma^0-\mbox{\boldmath{$\gamma$}}\cdot{\bf k}-M & -i\gamma_5\Delta\tau_2 t_2\\
-i\gamma_5\Delta^\dagger\tau_2 t_2 & (i\omega_n-\frac{\mu_{\text
B}}{2})\gamma^0-\mbox{\boldmath{$\gamma$}}\cdot{\bf
k}-M\end{array}\right)\
\end{eqnarray}
with the effective quark Dirac mass $M=m_0+\upsilon$.  The mean-field thermodynamic potential $\Omega_0=(T/V){\cal
S}_{\text{eff}}^{(0)}$ can be evaluated as
\begin{eqnarray}
\Omega_0=\frac{\upsilon^2+|\Delta|^2}{4G}-2N_cN_f\sum_{\bf
k}\left[\mathcal {W}(E_{\bf k}^+)+\mathcal {W}(E_{\bf k}^-)\right]
\end{eqnarray}
with the definitions of the function
$\mathcal{W}(E)=E/2+T\ln{(1+e^{-E/T})}$ and the BCS-like
quasiparticle dispersions $E_{\bf k}^\pm=\sqrt{(E_{\bf k}\pm
\mu_{\text B}/2)^2+|\Delta|^2}$ where $E_{\bf k}=\sqrt{{\bf
k}^2+M^2}$. The signs $\pm$ correspond to quasiquark and
quasi-antiquark excitations, respectively. The integral over the
quark momentum ${\bf k}$ is divergent at large $|{\bf k}|$, and some
regularization scheme should be adopted. In this paper, we employ a
hard three-momentum cutoff $\Lambda$.

The physical values of the variational parameters $M$ (or
$\upsilon$) and $\Delta$ should be determined by the saddle point
condition
\begin{eqnarray}\label{saddle}
\frac{\delta{\cal S}_{\text{eff}}^{(0)}[\upsilon,\Delta]}{\delta
\upsilon}=0,\ \ \ \ \ \frac{\delta{\cal
S}_{\text{eff}}^{(0)}[\upsilon,\Delta]}{\delta \Delta}=0,
\end{eqnarray}
which minimizes the mean-field effective action ${\cal
S}_{\text{eff}}^{(0)}$. One can show that the saddle point condition
is equivalent to the following Green function relations
\begin{eqnarray}
\langle\bar{q}q\rangle &=&  \sum_K\text{Tr}{\cal G}_{11}(K)\ ,\nonumber\\
\langle \bar{q}_ci\gamma_5\tau_2t_2q\rangle&=&
\sum_K\text{Tr}\left[{\cal G}_{12}(K)i\gamma_5\tau_2 t_2\right]\ ,
\end{eqnarray}
where the matrix elements of ${\cal G}$ are explicitly given by
\begin{eqnarray}
{\cal G}_{11}(K) &=& {i\omega_n+\xi_{\bf k}^-\over
(i\omega_n)^2-(E_{\bf k}^-)^2}\Lambda_{\bf k}^+\gamma_0+ {i\omega_n-\xi_{\bf k}^+\over (i\omega_n)^2-(E_{\bf k}^+)^2}\Lambda_{\bf k}^-\gamma_0\ ,\nonumber\\
{\cal G}_{22}(K) &=& {i\omega_n-\xi_{\bf k}^-\over
(i\omega_n)^2-(E_{\bf k}^-)^2}\Lambda_{\bf k}^-\gamma_0+ {i\omega_n+\xi_{\bf k}^+\over (i\omega_n)^2-(E_{\bf k}^+)^2}\Lambda_{\bf k}^+\gamma_0\ ,\nonumber\\
{\cal G}_{12}(K) &=& {-i\Delta\tau_2 t_2\over
(i\omega_n)^2-(E_{\bf k}^-)^2}\Lambda_{\bf k}^+\gamma_5+ {-i\Delta\tau_2 t_2\over (i\omega_n)^2-(E_{\bf k}^+)^2}\Lambda_{\bf k}^-\gamma_5\ ,\nonumber\\
{\cal G}_{21}(K) &=& {-i\Delta^\dagger\tau_2 t_2\over
(i\omega_n)^2-(E_{\bf k}^-)^2}\Lambda_{\bf k}^-\gamma_5+
{-i\Delta^\dagger\tau_2 t_2\over (i\omega_n)^2-(E_{\bf
k}^+)^2}\Lambda_{\bf k}^+\gamma_5
\end{eqnarray}
with the help of the massive energy projectors \cite{massive}
\begin{equation}
\Lambda_{\bf k}^{\pm}= {1\over
2}\left[1\pm{\gamma_0\left(\mbox{\boldmath{$\gamma$}}\cdot{\bf
k}+M\right)\over E_{\bf k}}\right]\ .
\end{equation}
Here we have defined the notation $\xi_{\bf k}^\pm=E_{\bf
k}\pm\mu_{\text B}/2$ for convenience.
\\
\emph{(II)Derivative expansion.} Next, we consider the fluctuations
around the mean field, corresponding to the bosonic collective
excitations. Making the field shifts for the auxiliary fields,
\begin{eqnarray}
&&\sigma(x)\rightarrow \upsilon+\sigma(x),\ \
\mbox{\boldmath{$\pi$}}(x)\rightarrow
0+\mbox{\boldmath{$\pi$}}(x),\nonumber\\
&&\phi(x)\rightarrow \Delta+\phi(x), \ \ \phi^\dagger(x)\rightarrow
\Delta^\dagger+\phi^\dagger(x),
\end{eqnarray}
we can express the total effective action as
\begin{eqnarray}
{\cal S}_{\text{eff}}&=&{\cal S}_{\text{eff}}^{(0)}+\int
dx\left(\frac{\sigma^2+\mbox{\boldmath{$\pi$}}^2+|\phi|^2}{4G}+\frac{\upsilon\sigma+\Delta\phi^\dagger+\Delta^\dagger\phi}{2G}\right)\nonumber\\
&-&\frac{1}{2}\text{Tr}\ln{\left[\mathbbold{1}+\int dx_1{\cal
G}(x,x_1)\Sigma(x_1,x^\prime)\right]}.
\end{eqnarray}
Here ${\cal G}(x,x^\prime)$ is the Fourier transformation of ${\cal
G}(i\omega_n,{\bf k})$, and $\Sigma(x,x^\prime)$ is defined as
\begin{eqnarray}
\Sigma(x,x^\prime)&=&\left(\begin{array}{cc}-\sigma(x)-
i\gamma_5\mbox{\boldmath{$\tau$}}\cdot\mbox{\boldmath{$\pi$}}(x)&-i\gamma_5\phi(x)\tau_2t_2\\-i\gamma_5\phi^\dagger(x)\tau_2t_2&
-\sigma(x)- i\gamma_5\mbox{\boldmath{$\tau$}}^{\text
T}\cdot\mbox{\boldmath{$\pi$}}(x)\end{array}\right)\nonumber\\
&\times&\delta(x-x^\prime).
\end{eqnarray}
With the help of the derivative expansion
\begin{eqnarray}
\text{Tr}\ln{\left[\mathbbold{1}+{\cal
G}\Sigma\right]}=\sum_{n=1}^\infty\frac{(-1)^{n+1}}{n}\text{Tr}[{\cal
G}\Sigma]^n,
\end{eqnarray}
we can calculate the effective action in powers of the fluctuations
$\sigma(x),\mbox{\boldmath{$\pi$}}(x),\phi(x),\phi^\dagger(x)$.

The first order effective action ${\cal S}_{\text{eff}}^{(1)}$ which
includes linear terms of the fluctuations should vanish exactly,
since the expectation value of the fluctuations should be exactly
zero. In fact, ${\cal S}_{\text{eff}}^{(1)}$ can be evaluated as
\begin{eqnarray}
{\cal S}_{\text{eff}}^{(1)} &=&\int
dx\Bigg\{\left[\frac{\upsilon}{2G}+\frac{1}{2}\text{Tr}\left({\cal
G}_{11}+{\cal G}_{22}\right)\right]\sigma(x)\nonumber\\
&+&\frac{1}{2}\text{Tr}\left[i\gamma_5\left({\cal
G}_{11}\mbox{\boldmath{$\tau$}}+{\cal
G}_{22}\mbox{\boldmath{$\tau$}}^{\text
T}\right)\right]\cdot\mbox{\boldmath{$\pi$}}(x)\nonumber\\
&+&\left[\frac{\Delta}{2G}+\frac{1}{2}\text{Tr}\left(i\gamma_5\tau_2
t_2{\cal G}_{12}\right)\right]\phi^\dagger(x)\nonumber\\
&+&\left[\frac{\Delta^\dagger}{2G}+\frac{1}{2}\text{Tr}\left(i\gamma_5\tau_2
t_2{\cal G}_{21}\right)\right]\phi(x)\Bigg\}.
\end{eqnarray}
We observe that the coefficients of $\mbox{\boldmath{$\pi$}}(x)$ is
automatically zero after taking the trace in Dirac spin space. The
coefficients of $\phi(x),\phi^\dagger(x)$ and $\sigma(x)$ vanish
once the quark propagator takes the mean-field form and $M,\Delta$
take the physical values satisfying the saddle point condition. Thus,
in the present approach, the saddle point condition plays a crucial
role in having a vanishing linear term in the expansion.

The quadratic term ${\cal S}_{\text{eff}}^{(2)}$ corresponds to the
Gaussian fluctuations. It reads
\begin{widetext}
\begin{eqnarray}
{\cal S}_{\text{eff}}^{(2)}=\int
dx\frac{\sigma^2(x)+\mbox{\boldmath{$\pi$}}^2(x)+|\phi(x)|^2}{4G}+\frac{1}{4}\text{Tr}\left[\int
dx_1dx_2dx_3{\cal G}(x,x_1)\Sigma(x_1,x_2) {\cal
G}(x_2,x_3)\Sigma(x_3,x^\prime)\right].
\end{eqnarray}
\end{widetext}
For the convenience of our investigation in the following, we will
use the form of ${\cal S}_{\text{eff}}^{(2)}$ in the momentum space.
After the Fourier transformation, it can be written as
\begin{eqnarray}
{\cal
S}_{\text{eff}}^{(2)}&=&\frac{1}{2}\sum_Q\Bigg\{\frac{|\sigma(Q)|^2+|\mbox{\boldmath{$\pi$}}(Q)|^2+|\phi(Q)|^2}{2G}\nonumber\\
&+&\frac{1}{2}\sum_K\text{Tr}\left[{\cal G}(K)\Sigma(-Q){\cal
G}(K+Q)\Sigma(Q)\right]\Bigg\}.
\end{eqnarray}
where $Q=(i\nu_m,{\bf q})$ with $\nu_m=2m\pi T$ being the boson
Matsubara frequency and $\sum_Q=T\sum_m\sum_{\bf q}$. Here $A(Q)$ is
the Fourier transformation of the field $A(x)$, and $\Sigma(Q)$ is
defined as \cite{note3}
\begin{eqnarray}
\Sigma(Q)=\left(\begin{array}{cc}-\sigma(Q)-
i\gamma_5\mbox{\boldmath{$\tau$}}\cdot\mbox{\boldmath{$\pi$}}(Q)&-i\gamma_5\phi(Q)\tau_2t_2\\-i\gamma_5\phi^\dagger(-Q)\tau_2t_2&
-\sigma(Q)- i\gamma_5\mbox{\boldmath{$\tau$}}^{\text
T}\cdot\mbox{\boldmath{$\pi$}}(Q)\end{array}\right).
\end{eqnarray}
\\
\emph{(III)Gaussian fluctuations.} After taking the trace in
Nambu-Gor'kov space, we find that ${\cal S}_{\text{eff}}^{(2)}$ can
be written in the following bilinear form
\begin{eqnarray}
{\cal
S}_{\text{eff}}^{(2)}&=&\frac{1}{2}\sum_Q\left(\begin{array}{ccc}
\phi^\dagger(Q) & \phi(-Q)&
\sigma^\dagger(Q)\end{array}\right){\bf M}(Q)\left(\begin{array}{cc} \phi(Q)\\
\phi^\dagger(-Q)\\ \sigma(Q)
\end{array}\right)\nonumber\\
&+&\frac{1}{2}\sum_Q\left(\begin{array}{ccc} \pi_1^\dagger(Q) &
\pi_2^\dagger(Q)&
\pi_3^\dagger(Q)\end{array}\right){\bf N}(Q)\left(\begin{array}{cc} \pi_1(Q)\\
\pi_2(Q)\\ \pi_3(Q) \end{array}\right).
\end{eqnarray}

The matrix ${\bf M}$ takes the following nondiagonal form
\begin{equation}
{\bf M}(Q)=\left(\begin{array}{ccc} \frac{1}{4G}+\Pi_{11}(Q)&\Pi_{12}(Q)&\Pi_{13}(Q)\\
\Pi_{21}(Q)&\frac{1}{4G}+\Pi_{22}(Q)&\Pi_{23}(Q)\\
\Pi_{31}(Q)&\Pi_{32}(Q)&\frac{1}{2G}+\Pi_{33}(Q)\end{array}\right)\
.
\end{equation}
The polarization functions $\Pi_{\text{ij}}(Q)$ (${\text i}, {\text
j}=1,2,3$) are one-loop susceptibilities composed of the matrix
elements the Nambu-Gor'kov quark propagator, and can be expressed as
\begin{widetext}
\begin{eqnarray}
&&\Pi_{11}(Q)=\frac{1}{2}\sum_K\text{Tr}\left[{\cal
G}_{22}(K)\Gamma{\cal G}_{11}(P)\Gamma\right],\ \ \ \
\Pi_{22}(Q)=\frac{1}{2}\sum_K\text{Tr}\left[{\cal
G}_{11}(K)\Gamma{\cal G}_{22}(P)\Gamma\right],\nonumber\\
&&\Pi_{12}(Q)=\frac{1}{2}\sum_K\text{Tr}\left[{\cal
G}_{12}(K)\Gamma{\cal G}_{12}(P)\Gamma\right],\ \ \ \
\Pi_{21}(Q)=\frac{1}{2}\sum_K\text{Tr}\left[{\cal
G}_{21}(K)\Gamma{\cal G}_{21}(P)\Gamma\right],\nonumber\\
&&\Pi_{33}(Q)=\frac{1}{2}\sum_K\text{Tr}\left[{\cal G}_{11}(K){\cal
G}_{11}(P)+{\cal G}_{22}(K){\cal G}_{22}(P)+{\cal G}_{12}(K){\cal
G}_{21}(P)+{\cal G}_{21}(K){\cal
G}_{12}(P)\right],\nonumber\\
&&\Pi_{13}(Q)=\frac{1}{2}\sum_K\text{Tr}\left[{\cal
G}_{12}(K)\Gamma{\cal G}_{11}(P)+{\cal G}_{22}(K)\Gamma{\cal
G}_{12}(P)\right],\ \ \
\Pi_{31}(Q)=\frac{1}{2}\sum_K\text{Tr}\left[{\cal G}_{21}(K){\cal
G}_{11}(P)\Gamma+{\cal G}_{22}(K){\cal
G}_{21}(P)\Gamma\right],\nonumber\\
&&\Pi_{23}(Q)=\frac{1}{2}\sum_K\text{Tr}\left[{\cal
G}_{11}(K)\Gamma{\cal G}_{21}(P)+{\cal G}_{21}(K)\Gamma{\cal
G}_{22}(P)\right],\ \ \
\Pi_{32}(Q)=\frac{1}{2}\sum_K\text{Tr}\left[{\cal G}_{11}(K){\cal
G}_{12}(P)\Gamma+{\cal G}_{12}(K){\cal G}_{22}(P)\Gamma\right],
\end{eqnarray}
\end{widetext}
where $P=K+Q$, $\Gamma=i\gamma_5\tau_2t_2$ and the trace is taken
over color, flavor and spin spaces. Using the fact that ${\cal
G}_{22}(K,\mu_{\text B})={\cal G}_{11}(K,-\mu_{\text B})$ and ${\cal
G}_{21}(K,\mu_{\text B})={\cal G}_{12}^\dagger(K,-\mu_{\text B})$,
we can easily show that
\begin{eqnarray}
&&\Pi_{22}(Q)=\Pi_{11}(-Q),\ \ \ \ \Pi_{12}(Q)=\Pi_{21}^\dagger(Q),\nonumber\\
&&\Pi_{13}(Q)=\Pi_{31}^\dagger(Q)=\Pi_{23}^\dagger(-Q)=\Pi_{32}(-Q).
\end{eqnarray}
Therefore, only five of the polarization functions are independent.
At $T=0$, their explicit form is shown in Appendix \ref{app2}. For
general case, we can show that $\Pi_{12}\propto\Delta^2$ and
$\Pi_{13}\propto M\Delta$. Thus, in the normal phase where
$\Delta=0$, the matrix ${\bf M}$ recovers the diagonal form. The
off-diagonal elements $\Pi_{13}$ and $\Pi_{23}$ represents the
mixing between the sigma meson and diquarks. At large chemical
potentials where the chiral symmetry is approximately restored,
$M\rightarrow m_0$, this mixing can be safely neglected.

On the other hand, the matrix ${\bf N}$ of the pion sector is
diagonal and proportional to the identity matrix, i.e.,
\begin{equation}
{\bf N}_{\text{ij}}(Q)=
\delta_{\text{ij}}\left[\frac{1}{2G}+\Pi_{\pi}(Q)\right], \ \ \
\text{i,j}=1,2,3.
\end{equation}
This means pions are eigen mesonic excitations even in the
superfluid phase. The polarization function $\Pi_\pi(Q)$ is given by
\begin{eqnarray}\label{pionpo}
\Pi_{\pi}(Q)&=&\frac{1}{2}\sum_K\text{Tr}\bigg[{\cal
G}_{11}(K)i\gamma_5{\cal G}_{11}(P)i\gamma_5+{\cal
G}_{22}(K)i\gamma_5{\cal G}_{22}(P)i\gamma_5\nonumber\\
&-&{\cal G}_{12}(K)i\gamma_5{\cal G}_{21}(P)i\gamma_5-{\cal
G}_{21}(K)i\gamma_5{\cal G}_{12}(P)i\gamma_5\bigg].
\end{eqnarray}
Its explicit form at $T=0$ is shown in Appendix \ref{app2}. We find
that $\Pi_{\pi}(Q)$ and $\Pi_{33}(Q)$ is different only to a term
proportional to $M^2$. Thus, at high density where
$\langle\bar{q}q\rangle\rightarrow 0$, the spectra of pions and
sigma meson become nearly degenerate which represents the
approximate restoration of chiral symmetry.
\\
\emph{(IV)Goldstone's theorem.} The U$_{\text B}(1)$ baryon number
symmetry is spontaneously broken by the nonzero diquark condensate
$\langle qq\rangle$ in the superfluid phase, resulting in one
Goldstone boson. In our model, this is ensured by the condition
$\det{\bf M}(Q=0)=0$. From the explicit form of the polarization
functions shown in Appendix \ref{app2}, we find that this condition
holds if and only if the saddle point condition (\ref{saddle}) for
$\upsilon$ and $\Delta$ is satisfied. We thus emphasize that in our
theoretical framework, the condensates $\upsilon$ and $\Delta$
should be determined by the saddle point condition, and the
beyond-mean-field corrections are possible only through the
thermodynamics, i.e., the equations of state.

\subsection {Vacuum and model parameter fixing}
\label{s2-3}
For a better understanding our derivation in the following, it is
useful to review the vacuum state at $T=\mu_{\text B}=0$. In the
vacuum, it is evident that $\Delta=0$ and the mean-field effective
potential $\Omega_{\text{vac}}$ can be evaluated as
\begin{eqnarray}
\Omega_{\text{vac}}(M)=\frac{(M-m_0)^2}{4G}-2N_cN_f\sum_{\bf
k}E_{\bf k}.
\end{eqnarray}
The physical value of $M$, denoted by $M_*$, satisfies the saddle
point condition $\partial\Omega_{\text{vac}}/\partial M=0$ and
minimizes $\Omega_{\text{vac}}$.

The meson and diquark excitations can be obtained from ${\cal
S}_{\text{eff}}^{(2)}$, which in the vacuum can be expressed as
\begin{eqnarray}
&&{\cal S}_{\text{eff}}^{(2)}=
-\frac{1}{2}\int\frac{d^4Q}{(2\pi)^4}\bigg[\sigma(-Q){\cal
D}_\sigma^{*-1}(Q)\sigma(Q)\nonumber\\
&&\ \ \ \ \ \ \ \ \ \ \ +\sum_{i=1}^3\pi_i(-Q){\cal
D}_{\pi}^{*-1}(Q)\pi_i(Q)\nonumber\\
&&\ \ \ \ \ \ \ \ \ \ \ +\sum_{i=1}^2\phi_i(-Q){\cal
D}_{\phi}^{*-1}(Q)\phi_i(Q)\bigg],
\end{eqnarray}
where $\phi_1,\phi_2$ are the real and imaginary parts of $\phi$,
respectively. The inverse propagators in vacuum can be expressed in
a symmetrical form \cite{NJLreview02}
\begin{eqnarray}
&&{\cal D}^{*-1}_{l}(Q)=\frac{1}{2G}+\Pi_{l}^*(Q),\ \ \ \
l=\sigma,\pi,\phi\nonumber\\
&&\Pi_{l}^*(Q)=2iN_cN_f(Q^2-\epsilon_l^2)I(Q^2)\nonumber\\
&&\ \ \ \ \ \ \ \ \ \ \
-4iN_cN_f\int\frac{d^4K}{(2\pi)^4}\frac{1}{K^2-M_*^2},
\end{eqnarray}
where $\epsilon_\sigma=2M_*$, $\epsilon_{\pi}=\epsilon_\phi=0$, and
the function $I(Q^2)$ is defined as
\begin{eqnarray}
I(Q^2)=\int\frac{d^4K}{(2\pi)^4}\frac{1}{(K_+^2-M_*^{2})(K_-^2-M_*^{2})},
\end{eqnarray}
with $K_\pm=K\pm Q/2$. Keeping in mind that $M_*$ satisfies the
saddle point condition, we find that the pions and diquarks are
Nambu-Goldstone bosons in the chiral limit, corresponding to the
symmetry breaking pattern SU$(4)\rightarrow$Sp$(4)$. Using the gap
equation of $M_*$, we find that the masses of mesons and diquarks
can be determined by the equation
\begin{eqnarray}\label{mesonmass}
m_{l}^2=-\frac{m_0}{M_*}\frac{1}{4iGN_cN_fI(m_{l}^2)}+\epsilon_{l}^2.
\end{eqnarray}
Since the $Q^2$ dependence of the function $I(Q^2)$ is very weak, we
find $m_\pi^2\sim m_0$ and $m_\sigma^2\simeq 4M_*^2+m_\pi^2$.

Since pions and diquarks are deep bound states, their propagators
can be well approximated by ${\cal D}_\pi^*(Q)\simeq -g_{\pi
qq}^2/(Q^2-m_\pi^2)$ with $g_{\pi qq}^{-2}\simeq -2iN_cN_f I(0)$.
The pion decay constant $f_\pi$ can be determined by the matrix
element of the axial current,
\begin{eqnarray}
&&iQ_\mu f_\pi\delta_{\text{ij}}\nonumber\\
&=&-\frac{1}{2}\text{Tr}\int\frac{d^4K}{(2\pi)^4}\left[\gamma_\mu
\gamma_5\tau_{\text i}{\cal G}(K_+)g_{\pi
qq}\gamma_5\tau_{\text j}{\cal G}(K_-)\right]\nonumber\\
&=&2N_cN_fg_{\pi qq}M_*Q_\mu I(Q^2)\delta_{\text{ij}}.
\end{eqnarray}
Here ${\cal G}(K)=(\gamma^\mu K_\mu-M_*)^{-1}$. Thus, the pion decay
constant can be expressed as
\begin{eqnarray}\label{piondecay}
f_\pi^2\approx-2iN_cN_fM_*^{2}I(0).
\end{eqnarray}
Finally, together with (\ref{mesonmass}) and (\ref{piondecay}), we
recover the well-known Gell-Mann--Oakes--Renner relation $m_\pi^2
f_\pi^2=-m_0\langle\bar{q}q\rangle_0$.

\begin{table}[b!]
\begin{center}
\begin{tabular}{|c| c c c | c c c |}
\hline
&&&&&&\\[-3mm]
 Set & $\Lambda$ [MeV] & $G\Lambda^2$ & $m_0$ [MeV]  & $\langle\bar{u}u\rangle_0^{\frac{1}{3}}$ [MeV]  &
 $M_*$ [MeV]
 & $m_\pi$ [MeV]
\\[1mm]
\hline
&&&&&&\\[-3mm]
1 & 657.9 & 3.105 & 4.90 & -217.4 & 300 & 133.6 \\
2 & 583.6 & 3.676 & 5.53 & -209.1 & 400 & 134.0 \\
3 & 565.8 & 4.238 & 5.43 & -210.6 & 500 & 134.2 \\
4 & 565.4 & 4.776 & 5.11 & -215.1 & 600 & 134.4 \\
\hline
\end{tabular}
\end{center}
\caption{\small Model parameters (3-momentum cutoff $\Lambda$,
coupling constant $G$, and current quark mass $m_0$) and related
quantities (quark condensate $\langle\bar{u}u\rangle_0$, constituent
quark mass $M_*$ and pion mass $m_\pi$ ) for the two-flavor
two-color NJL model (\ref{NJL}). The pion decay constant is fixed to
be $f_\pi = 75$~MeV.} \label{fit}
\end{table}

There are three parameters in our model, the current quark mass
$m_0$, the coupling constant $G$ and the cutoff $\Lambda$. In
principle they should be determined from the known values of the
pion mass $m_\pi$, the pion decay constant $f_\pi$ and the quark
condensate $\langle\bar{q}q\rangle_0$. Since two-color QCD does not
correspond to our real world, we get the above values from the 
empirical values $f_\pi\simeq93$MeV,
$\langle\bar{u}{u}\rangle_0\simeq(250$MeV$)^3$ in the $N_c=3$ case,
according to the relation $f_\pi^2, \langle\bar{q}{q}\rangle_0\sim
N_c$. To obtain the model parameters, we fix the values of the pion
decay constant $f_\pi$ and slightly vary the values of the chiral
condensate $\langle\bar{q}q\rangle_0$ and the pion mass $m_\pi$.
Thus, we can obtain different sets of model parameters corresponding
to different values of effective quark mass $M_*$ and hence
different values of the sigma meson mass $m_\sigma$. Four sets of
model parameters are shown in Table.~\ref{fit}. As we will show in
the following, the physics near the quantum phase transition point
$\mu_{\text B}=m_\pi$ is not sensitive to different model parameter
sets, since the low energy dynamics is dominated by the
pseudo-Goldstone bosons (i.e., the diquarks). However, at high
density, the physics becomes sensitive to different model parameter
sets corresponding to different sigma meson masses. The predictions
by the chiral perturbation theories should be recovered in the limit
$m_\sigma/m_\pi\rightarrow\infty$.


\section {Dilute Bose Condensate: Mean Field Theory}
\label{s3}
Now we begin to study the properties of two-color matter at finite
baryon density. Without loss of generality, we set $\mu_{\text
B}>0$. In this section, we study the two-color baryonic matter in
the dilute limit, which forms near the quantum phase transition
point $\mu_{\text B}=m_\pi$. Since the diquark condensate is
vanishingly small near the quantum phase transition point, we can
make a Ginzburg-Landau expansion for the effective action. As we
will see below, this corresponds to the mean-field theory of weakly
interacting dilute Bose condensates.

\subsection {Ginzburg-Landau free energy near the quantum phase
transition} \label{s3-1}
Since the diquark condensate $\Delta$ is vanishingly small near the
quantum phase transition, we can derive the Ginzburg-Landau free
energy functional $V_{\text{GL}}[\Delta(x)]$ at $T=0$ for the order
parameter field $\Delta(x)=\langle\phi(x)\rangle$ in the static and
long-wavelength limit. The general form of
$V_{\text{GL}}[\Delta(x)]$ can be written as
\begin{eqnarray}
&&V_{\text{GL}}[\Delta(x)]=\int
dx\Bigg[\Delta^\dagger(x)\left(-\delta\frac{\partial^2}{\partial\tau^2}+
\kappa\frac{\partial}{\partial\tau}-\gamma\mbox{\boldmath{$\nabla$}}^2\right)\Delta(x)\nonumber\\
&&\ \ \ \ \ \ \ \ \ \ \ \ \ \ \ \ \ \ \
+\alpha|\Delta(x)|^2+\frac{1}{2}\beta|\Delta(x)|^4\Bigg],
\end{eqnarray}
where the coefficients $\alpha,\beta,\gamma,\delta,\kappa$ should be
low energy constants which \emph{depend only on the vacuum
properties}. The calculation is somewhat similar to the derivation
of Ginzburg-Landau free energy of a superconductor from the
microscopic BCS theory\cite{nao}, but for our case there is a
difference in that we have another variational parameter, i.e., the
effective quark mass $M$ which should be a function of $|\Delta|^2$
determined by the saddle point condition.
\\
\emph{(I) The potential terms.} In the static and long-wavelength
limit, the coefficients $\alpha,\beta$ of the potential terms can be
obtained from the effective action ${\cal S}_{\text{eff}}$ in the
mean-field approximation. At $T=0$, the mean-field effective action
reads ${\cal S}_{\text{eff}}^{(0)}=\int dx\Omega_0$, where the mean-field
 thermodynamic potential is given by
\begin{eqnarray}
\Omega_0(|\Delta|^2,M)=\frac{(M-m_0)^2+|\Delta|^2}{4G}-N_cN_f\sum_{\bf
k}(E_{\bf k}^++E_{\bf k}^-).
\end{eqnarray}
The Ginzburg-Landau coefficients $\alpha,\beta$ can be obtained via
a Taylor expansion of $\Omega_0$ in terms of $|\Delta|^2$,
\begin{equation}
\Omega_0=\Omega_{\text{vac}}(M_*)+\alpha|\Delta|^2+\frac{1}{2}\beta|\Delta|^4+O(|\Delta|^6),
\end{equation}
where $\Omega_{\text{vac}}(M_*)$ is the vacuum contribution which
should be subtracted. One should keep in mind that \emph{the
effective quark mass $M$ is not a fixed parameter, but a function of
$|\Delta|^2$ via its saddle point condition or gap equation
$\partial\Omega_0/\partial M=0$}.

For convenience, we define $y\equiv|\Delta|^2$. The Ginzburg-Landau
coefficient $\alpha$ is defined as
\begin{eqnarray}
\alpha&=&\frac{d\Omega_0(y,M)}{dy}\Bigg|_{y=0}\nonumber\\
&=&\frac{\partial\Omega_0(y,M)}{\partial y}\Bigg|_{y=0}
+\frac{\partial\Omega_0(y,M)}{\partial M}\frac{dM}{d
y}\Bigg|_{y=0}\nonumber\\
&=&\frac{\partial\Omega_0(y,M)}{\partial y}\Bigg|_{y=0}
\end{eqnarray}
where the indirect derivative term vanishes due to the saddle point
condition for $M$. After some simple algebra, we get
\begin{equation}
\alpha=\frac{1}{4G}-N_cN_f\sum_{\bf k}\frac{E_{\bf k}^*}{E_{\bf
k}^{*2}-\mu_{\text B}^2/4},
\end{equation}
where $E_{\bf k}^*=\sqrt{{\bf k}^2+M_*^2}$. We can make the above
expression more meaningful using the pion mass equation in the same
three-momentum regularization scheme\cite{NJLreview02,NJLreview03},
\begin{equation}
\frac{1}{4G}-N_cN_f\sum_{\bf k}\frac{E_{\bf k}^*}{E_{\bf
k}^{*2}-m_\pi^2/4}=0.
\end{equation}
We therefore obtain a $G$-independent result
\begin{equation}
\alpha=\frac{1}{4}N_cN_f(m_\pi^2-\mu_{\text B}^2)\sum_{\bf
k}\frac{E_{\bf k}^*}{(E_{\bf k}^{*2}-m_\pi^2/4)(E_{\bf
k}^{*2}-\mu_{\text B}^2/4)}.
\end{equation}
From the fact that $m_\pi\ll 2M_*$ and $\beta>0$ (see below), we see
clearly that a second order quantum phase transition takes place at
exactly $\mu_{\text B}=m_\pi$. Thus, the Ginzburg-Landau free energy
is meaningful only near the quantum phase transition point, i.e.,
$|\mu_{\text B}-m_\pi|\ll m_\pi$, and $\alpha$ can be further
simplified as
\begin{equation}\label{alpha}
\alpha\simeq(m_\pi^2-\mu_{\text B}^2){\cal J},
\end{equation}
where the factor ${\cal J}$ is defined as
\begin{eqnarray}
{\cal J}=\frac{1}{4}N_cN_f\sum_{\bf k}\frac{E_{\bf k}^*}{(E_{\bf
k}^{*2}-m_\pi^2/4)^2}.
\end{eqnarray}

The coefficient $\beta$ of the quartic term can be evaluated via the
definition
\begin{eqnarray}
\beta&=&\frac{d^2\Omega_0(y,M)}{dy^2}\Bigg|_{y=0}\nonumber\\
&=&\frac{\partial^2\Omega_0(y,M)}{\partial y^2}\Bigg|_{y=0}
+\frac{\partial^2\Omega_0(y,M)}{\partial M\partial y}\frac{dM}{d
y}\Bigg|_{y=0}.
\end{eqnarray}
Notice that the last indirect derivative term does not vanish here
and \emph{will be important for us to obtain a correct
diquark-diquark scattering length}. The derivative $dM/dy$ can be
analytically derived from the gap equation for $M$. From the fact that
$\partial \Omega_0/\partial M=0$, we obtain
\begin{equation}
\frac{\partial}{\partial y}\left(\frac{\partial
\Omega_0(y,M)}{\partial M}\right)+ \frac{\partial}{\partial
M}\left(\frac{\partial \Omega_0(y,M)}{\partial
M}\right)\frac{dM}{dy}=0.
\end{equation}
Thus, we find
\begin{equation}
\frac{dM}{dy}= -\frac{\partial^2\Omega_0(y,M)}{\partial M\partial
y}\left(\frac{\partial^2\Omega_0(y,M)}{\partial M^2}\right)^{-1}.
\end{equation}
Then the practical expression for $\beta$ can be written as
\begin{eqnarray}
\beta=\beta_1+\beta_2,
\end{eqnarray}
where $\beta_1$ is the direct derivative term
\begin{equation}
\beta_1=\frac{\partial^2\Omega_0(y,M)}{\partial y^2}\Bigg|_{y=0},
\end{equation}
and $\beta_2$ is the indirect term
\begin{equation}
\beta_2=-\left(\frac{\partial^2\Omega_0(y,M)}{\partial M\partial
y}\right)^2\left(\frac{\partial^2\Omega_0(y,M)}{\partial
M^2}\right)^{-1}\Bigg|_{y=0}.
\end{equation}

Near the quantum phase transition, all chemical potential dependence
can be absorbed into the coefficient $\alpha$, and we can set
$\mu_{\text B}=m_\pi$ in $\beta$. After a simple algebra, the
explicit form of $\beta_1$ and $\beta_2$ can be evaluated as
\begin{eqnarray}
\beta_1=\frac{1}{4}N_cN_f\sum_{e=\pm}\sum_{\bf k}\frac{1}{(E_{\bf
k}^*-em_\pi/2)^3}
\end{eqnarray}
and
\begin{eqnarray}\label{beta2}
\beta_2&=&-\left\{\frac{1}{2}N_cN_f\sum_{e=\pm}\sum_{\bf
k}\frac{M_*}{E_{\bf k}^*}\frac{1}{(E_{\bf
k}^*-em_\pi/2)^2}\right\}^2\nonumber\\
&\times&\left(\frac{m_0}{2GM_*}+2N_cN_f\sum_{\bf
k}\frac{M_*^2}{E_{\bf k}^{*3}}\right)^{-1}.
\end{eqnarray}
The $G$-dependent term $m_0/(2GM_*)$ in (\ref{beta2}) can be
approximated as $m_\pi^2 f_\pi^2/M_*^2$ using the relation
$m_\pi^2f_\pi^2=-m_0\langle\bar{q}{q}\rangle$.
\\
\emph{(II) The kinetic terms.} The kinetic terms in the
Ginzburg-Landau free energy can be derived from the inverse of the
diquark propagator \cite{nao}. In the general case with $\Delta\neq0$,
the diquarks are mixed with the sigma meson. However, approaching
the quantum phase transition point, $\Delta\rightarrow 0$, the
problem is simplified. After the analytical continuation
$i\nu_m\rightarrow \omega+i0^+$, the inverse of the diquark
propagator in the limit $\mu_{\text B}\rightarrow m_\pi$ can be
evaluated as
\begin{eqnarray}\label{dipro}
{\cal D}_{\text d}^{-1}(\omega,{\bf q})=\frac{1}{4G}+\Pi_{\text
d}(\omega,{\bf q}),
\end{eqnarray}
where the polarization function $\Pi_{\text d}(\omega,{\bf q})$ is
given by
\begin{eqnarray}
&&\Pi_{\text d}(\omega,{\bf q})=N_cN_f\sum_{\bf k}\frac{E^*_{{\bf
k}}+E^*_{{\bf k}+{\bf
q}}}{(\omega+\mu_{\text B})^2-(E^*_{{\bf k}}+E^*_{{\bf k}+{\bf q}})^2}\nonumber\\
&&\ \ \ \ \ \ \ \ \ \ \ \ \ \ \ \times\left(1+\frac{{\bf
k}\cdot({\bf k}+{\bf q})+M_*^2}{E^*_{{\bf k}}E^*_{{\bf k}+{\bf
q}}}\right).
\end{eqnarray}

In the static and long-wavelength limit ($\omega,|{\bf
q}|\rightarrow 0$), the coefficients $\kappa,\delta,\gamma$ can be
determined by the Taylor expansion ${\cal D}_{\text
d}^{-1}(\omega,{\bf q})={\cal D}_{\text d}^{-1}(0,{\bf
0})-\delta\omega^2-\kappa\omega+\gamma{\bf q}^2$. Notice that
$\alpha$ is identical to ${\cal D}_{\text d}^{-1}(0,{\bf 0})$ which
is in fact the Thouless criterion for the superfluid transition. On
the other hand, keeping in mind that ${\cal D}_{\text
d}^{-1}(\omega,{\bf q})$ can be related to the pion propagator in
the vacuum, i.e., ${\cal D}_{\text d}^{-1}(\omega,{\bf
q})=(1/2){\cal D}_\pi^{*-1}(\omega+\mu_{\text B},{\bf q})$, in the
static and long-wavelength limit and for $\mu_{\text B}\rightarrow
m_\pi\ll 2M_*$ we can well approximate it as\cite{NJLreview02}
\begin{eqnarray}\label{diproa}
{\cal D}_{\text d}^{-1}(\omega,{\bf q})\simeq-{\cal
J}\left[(\omega+\mu_{\text B})^2-{\bf q}^2-m_\pi^2\right],
\end{eqnarray}
where ${\cal J}$ is the same factor defined in (\ref{alpha}), and one
can show that ${\cal J}\simeq g_{\pi qq}^{-2}/2$. We thus find that
$\delta\simeq\gamma\simeq{\cal J}$ which ensures the Lorentz
invariance of the vacuum, and $\kappa\simeq2\mu_{\text B}{\cal J}$.

\subsection {From Ginzburg-Landau to Gross-Pitaevskii free energy}
\label{s3-2}

\begin{table}[b!]
\begin{center}
\begin{tabular}{|c|c|c|c|c|}
  \hline
  Set & 1 & 2 & 3 & 4 \\
  \hline
  $a_{\text{dd}}$ according to (\ref{add}) [$m_\pi^{-1}$]& 0.0631 & 0.0635 & 0.0637 & 0.0639 \\
  \hline
  $a_{\text{dd}}$ according to (\ref{adda}) [$m_\pi^{-1}$]& 0.0624 & 0.0628 & 0.0630 & 0.0633 \\
  \hline
\end{tabular}
\end{center}
\caption{\small The values of diquark-diquark scattering length
$a_{\text{dd}}$ (in units of $m_\pi^{-1}$) for different model
parameter sets.} \label{scatt}
\end{table}

We now show how the Ginzburg-Landau free energy can be reduced to
the theory describing weakly repulsive Bose condensates, i.e., the
Gross-Pitaevskii free energy \cite{GP01,GP02}.

\emph{(I) Nonrelativistic version.} First, since the Bose
condensed matter is indeed dilute, let us consider the
nonrelativistic version, where $\omega\ll m_\pi$ and the kinetic
term $\propto \partial^2/\partial\tau^2$ is neglected. To this end,
we define the nonrelativistic chemical potential $\mu_{\text d}$
for diquarks, $\mu_{\text d}=\mu_{\text B}-m_\pi$, and further
simplify the coefficient $\alpha$ as
\begin{eqnarray}
\alpha&\simeq&-\mu_{\text d}(2m_\pi{\cal J}).
\end{eqnarray}
Then the Ginzburg-Landau free energy can be reduced to the
Gross-Pitaevskii free energy of a dilute repulsive Bose gas, if we
define a new condensate wave function $\psiup(x)$ as
\begin{eqnarray}
\psiup(x)=\sqrt{2m_\pi{\cal J}}\Delta(x).
\end{eqnarray}
The resulting Gross-Pitaevskii free energy is given by
\begin{eqnarray}\label{fgp}
&&V_{\text{GP}}[\psiup(x)]=\int
dx\bigg[\psiup^\dagger(x)\left(\frac{\partial}{\partial\tau}-\frac{\mbox{\boldmath{$\nabla$}}^2}{2m_\pi}\right)\psiup(x)\nonumber\\
&&\ \ \ \ \ \ \ \ \ \ \ \ \ \ \ \ \ \ \ \ \ \ \ \  -\mu_{\text
d}|\psiup(x)|^2+\frac{1}{2}g_0|\psiup(x)|^4\bigg],
\end{eqnarray}
where $g_0=4\pi a_{\text{dd}}/m_\pi$. The repulsive diquark-diquark
interaction is characterized by a positive scattering length
$a_{\text{dd}}$ defined as
\begin{eqnarray}\label{add}
a_{\text{dd}}=\frac{\beta}{16\pi m_\pi}{\cal J}^{-2}.
\end{eqnarray}
Keep in mind that the scattering length obtained here is \emph{at
the mean-field level}. We will discuss the possible
beyond-mean-field corrections in Sec. \ref{s5}. Thus, for a dilute
medium with density $n$ satisfying $na_{\text{dd}}^3\ll1$, the
system is indeed a weakly interacting Bose condensate
\cite{Bose01,Bose02,Bose03}.

\emph{(II) Diquark-diquark scattering length.} Even though we have
shown that the Ginzburg-Landau free energy is indeed a
Gross-Pitaevskii version near the quantum phase transition, a key
problem is whether the obtained diquark-diquark scattering length
$a_{\text{dd}}$ is quantitatively correct. A numerical calculation
for (\ref{add}) is straightforward. The obtained values of
$a_{\text{dd}}$ for the four model parameter sets are shown in Table
\ref{scatt}. We can also give an analytical expression based on the
formula of the pion decay constant in the three-momentum cutoff
scheme,
\begin{eqnarray}
f_\pi^2=N_cM_*^2\sum_{\bf k}\frac{1}{E_{\bf k}^{*3}}.
\end{eqnarray}
According to the fact that $m_\pi\ll 2M_*$, $\beta$ and ${\cal J}$
can be well approximated as
\begin{eqnarray}
&&\beta\simeq\frac{f_\pi^2}{M_*^2}-\frac{(2f_\pi^2/M_*)^2}{m_\pi^2f_\pi^2/M_*^2+4f_\pi^2}\simeq\frac{f_\pi^2m_\pi^2}{4M_*^4},\nonumber\\
&&{\cal J}\simeq\frac{f_\pi^2}{2M_*^2}.
\end{eqnarray}
Thus, the diquark-diquark scattering length $a_{\text{dd}}$ in the
limit $m_\pi/(2M_*)\rightarrow 0$ is related only to the pion mass
and decay constant,
\begin{eqnarray}\label{adda}
a_{\text{dd}}=\frac{m_\pi}{16\pi f_\pi^2}.
\end{eqnarray}
The values of $a_{\text{dd}}$ for the four model parameter sets
according to the above expression are also listed in Table
\ref{scatt}. The errors are always about $1\%$ comparing with the
exact numerical results, which means that the expression
(\ref{adda}) is a good approximation for the diquark-diquark
scattering length. The error should come from the finite value of
$m_\pi/(2M^*)$. We can obtain a correction in powers of
$m_\pi/(2M^*)$ \cite{schulze}, but it is obviously small, and its
explicit form is not shown here.

The result $a_{\text{dd}}\propto m_\pi$ is universal for the
scattering lengths of the pseudo-Goldstone bosons. Eventhough the SU$(4)$ flavor symmetry
is explicitly broken in presence of a nonzero quark mass, a descrete symmetry $\phi_1,\phi_2\leftrightarrow\pi_1,\pi_2$
holds exactly for arbitrary quark mass. This also means that the partition function of two-color QCD has a descrete
symmetry $\mu_{\text B}\leftrightarrow\mu_{\text I}$ \cite{QL03}. Because of
this descrete symmetry of two-color QCD, the analytical
expression (\ref{adda}) of $a_{\text{dd}}$ (which is in fact the
diquark-diquark scattering length in the $B=2$ channel) should be identical
to the pion-pion scattering length at tree level in the $I=2$
channel which was first obtained by Weinberg many years ago
\cite{pipi}. Therefore, the mean-field theory can describe not only
the quantum phase transition to a dilute diquark condensate but also
the effect of repulsive diquark-diquark interaction.

\begin{widetext}

\begin{figure}[!htb]
\begin{center}
\includegraphics[width=7cm]{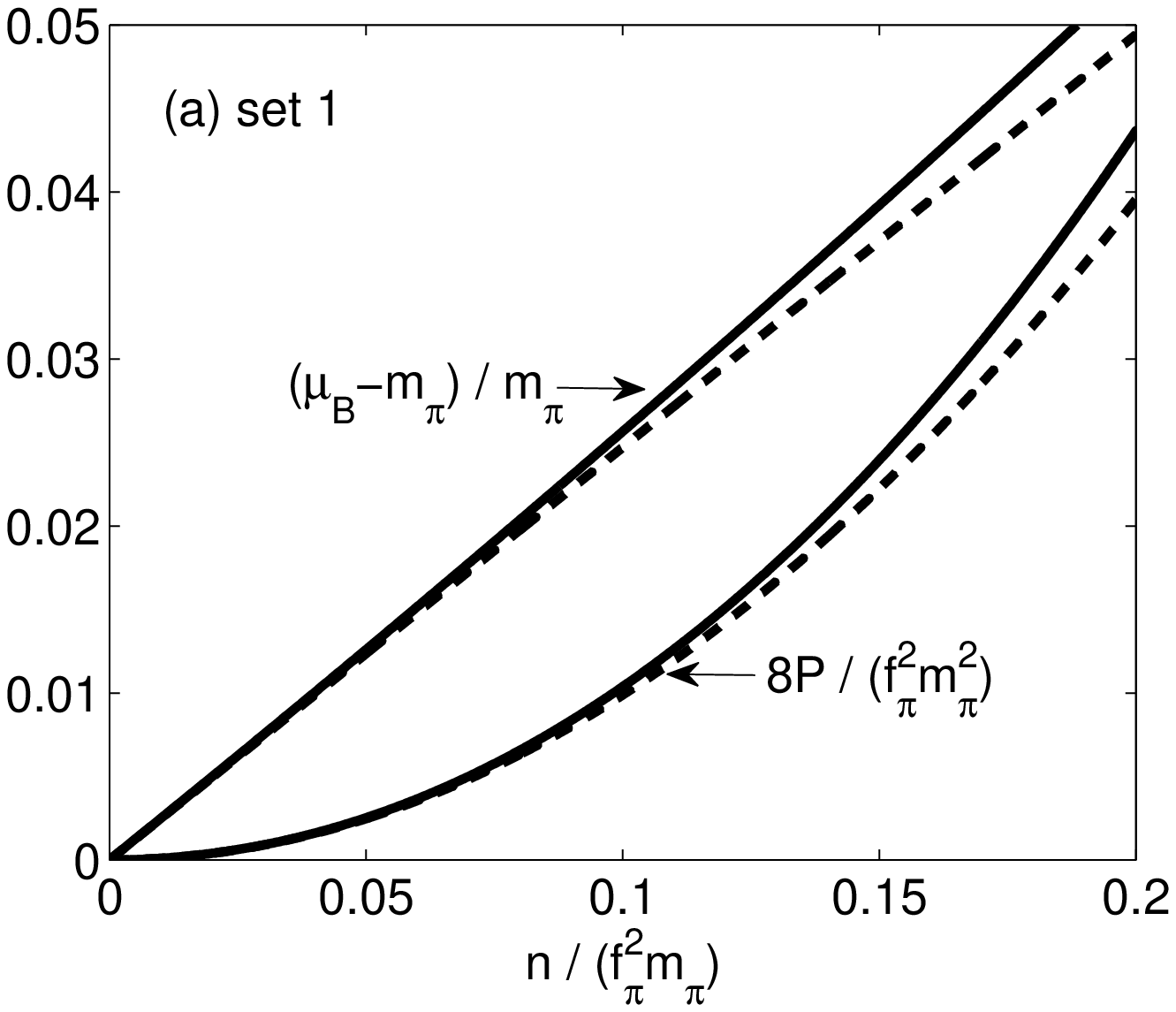}
\includegraphics[width=7cm]{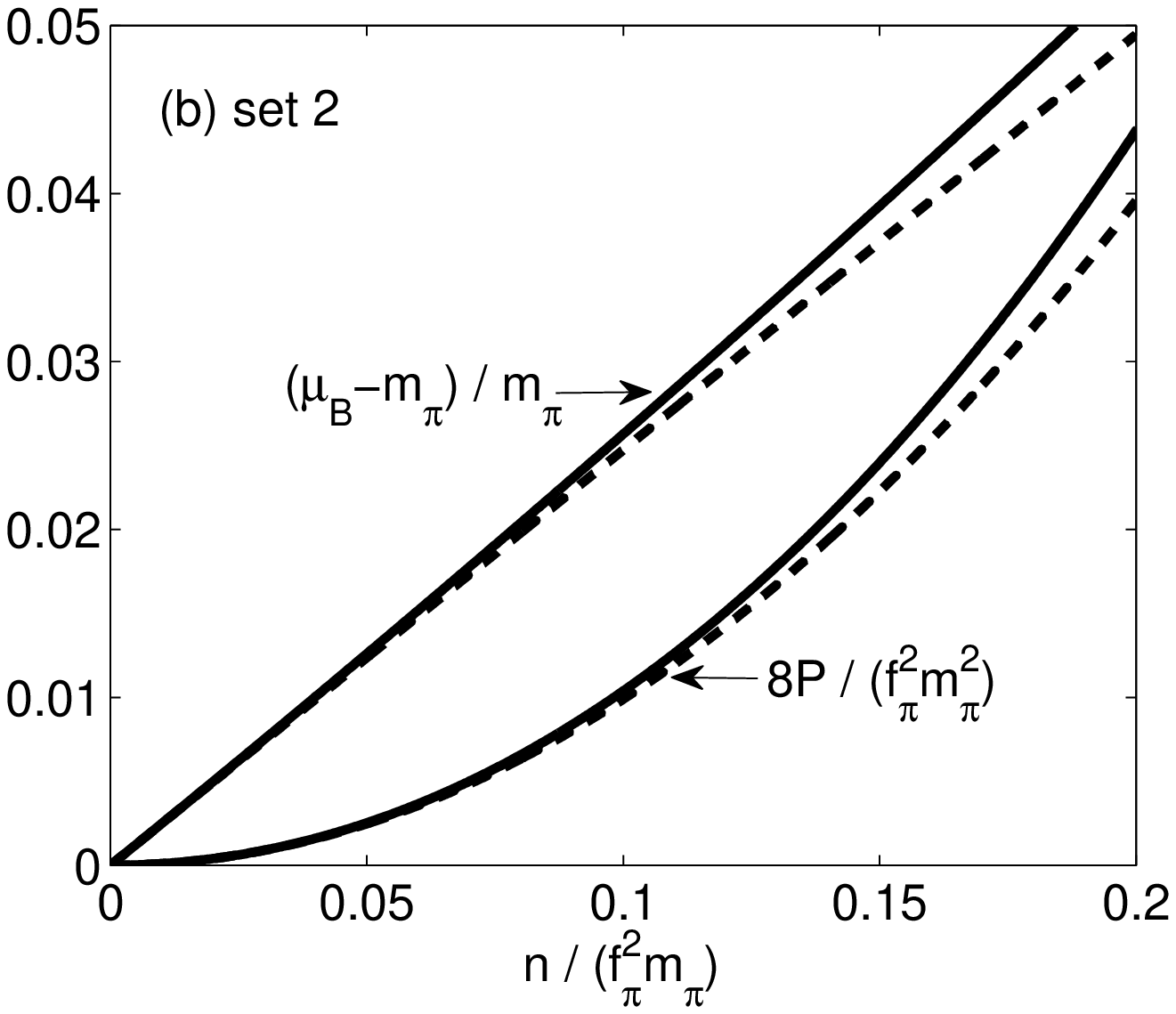}
\includegraphics[width=7cm]{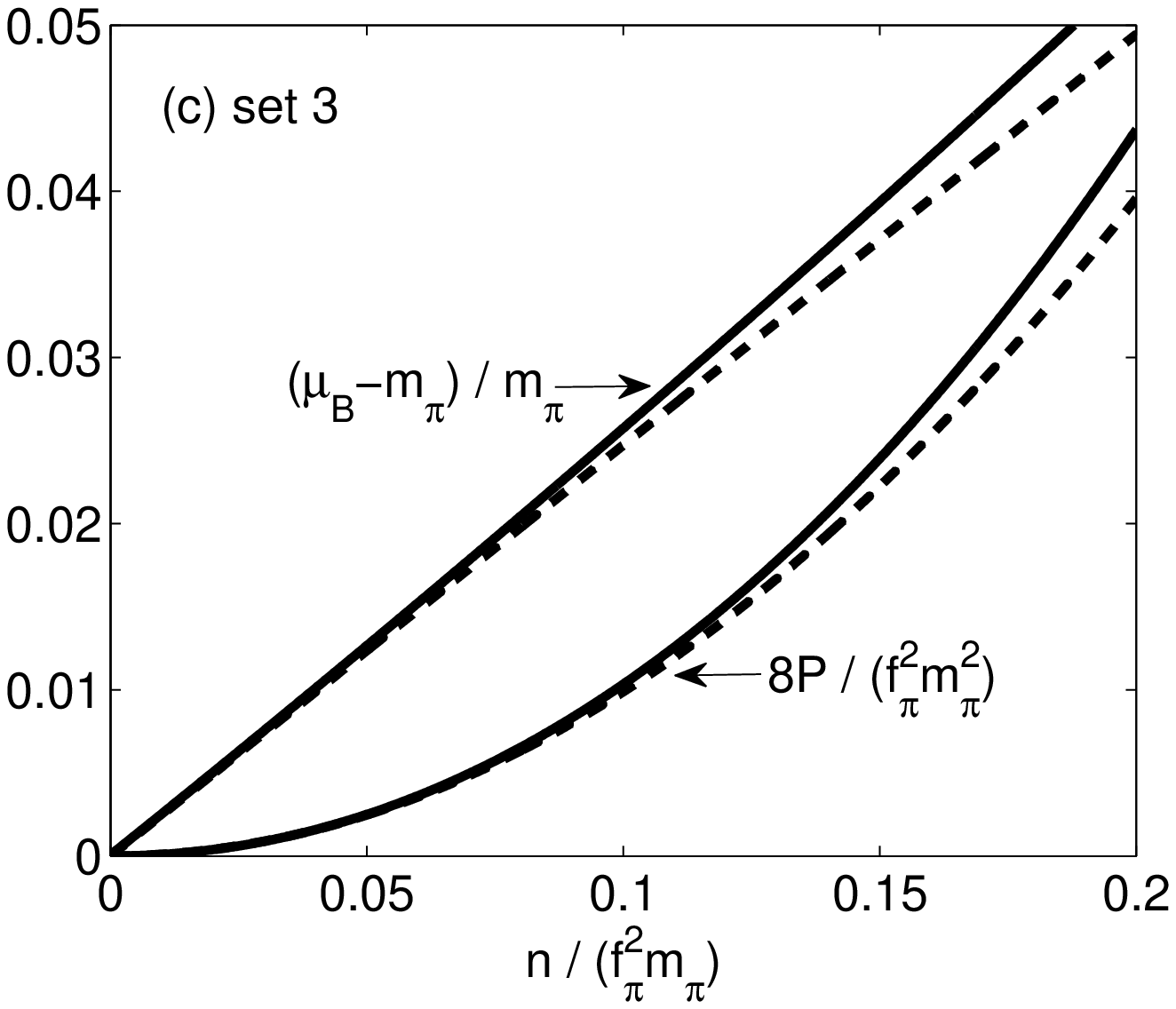}
\includegraphics[width=7cm]{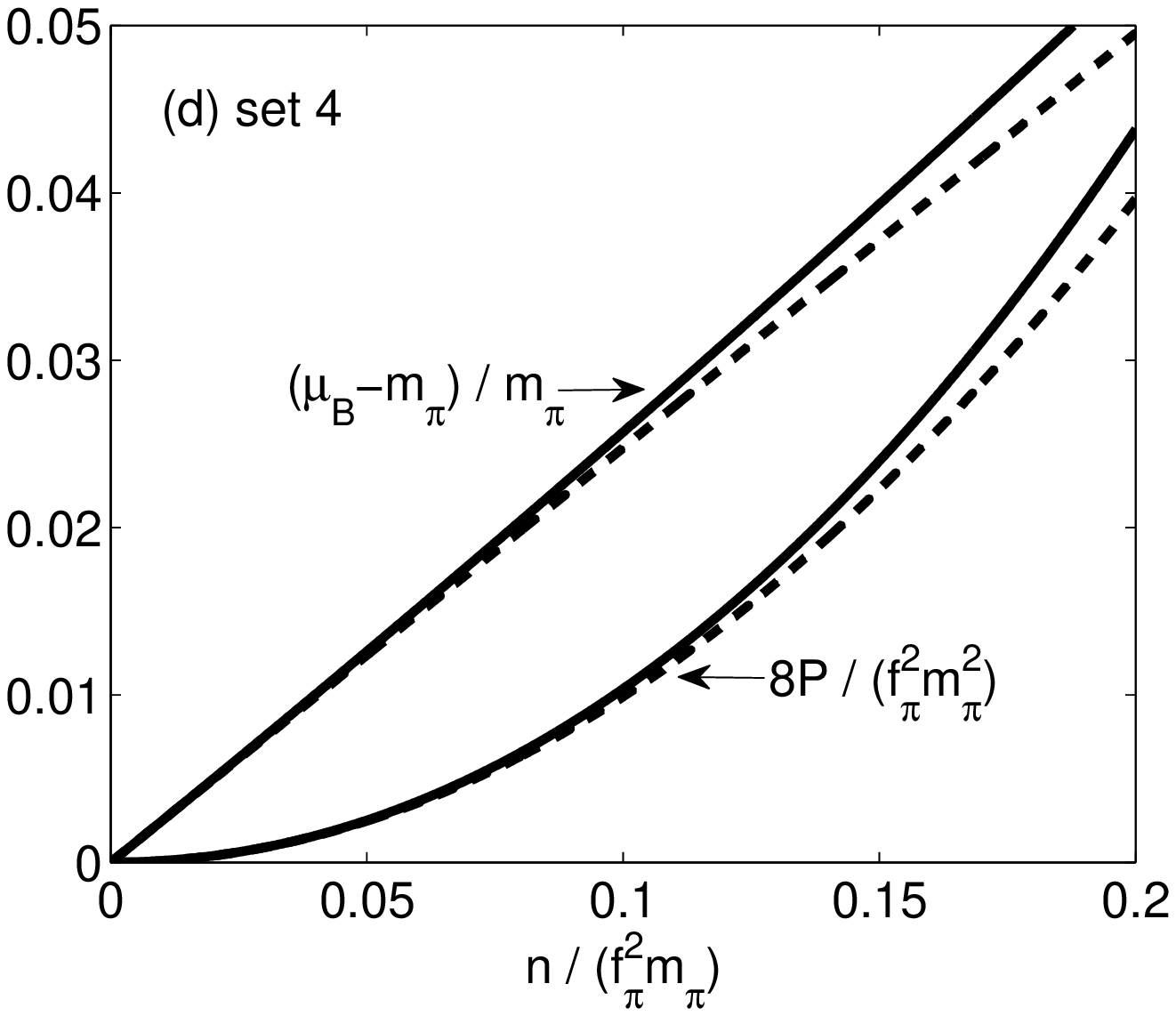}
\caption{The baryon chemical potential $\mu_{\text B}$ and the
pressure $P$ as functions of the baryon density $n$ for different
model parameter sets. The solid lines correspond to the direct mean-field
 calculation and the dashed lines are given by (\ref{eos}).
\label{fig1}}
\end{center}
\end{figure}

\end{widetext}

\emph{(III) Equations of state.} The mean-field equations of state
of the dilute diquark condensate are thus determined by the
Gross-Pitaevskii free energy (\ref{fgp}). Minimizing
$V_{\text{GP}}[\psiup(x)]$ with respect to a uniform condensate
$\psiup$, we find the physical minimum is given by
\begin{eqnarray}\label{chemi}
|\psiup_0|^2=\frac{\mu_{\text d}}{g_0},
\end{eqnarray}
and the baryon density is $n=|\psiup_0|^2$. Using the thermodynamic
relations, we therefore get the well-known results for the pressure
$P$, the energy density ${\cal E}$ and the chemical potential
$\mu_{\text B}$ in terms of the baryon density $n$,
\begin{eqnarray}\label{eos}
&&P(n)=\frac{2\pi a_{\text{dd}}}{m_\pi}n^2,\nonumber\\
&&{\cal E}(n)=m_\pi n+\frac{2\pi a_{\text{dd}}}{m_\pi}n^2,\nonumber\\
&&\mu_{\text B}(n)=m_\pi+\frac{4\pi a_{\text{dd}}}{m_\pi}n,
\end{eqnarray}
which were first obtained by Bogoliubov many years ago
\cite{Bose02}. We can examine the above results through a direct
numerical calculation with the mean-field thermodynamic potential.
The pressure is given by $P=-(\Omega_0-\Omega_{\text{vac}})$ and the
baryon density reads $n=-\partial\Omega_0/\partial\mu_{\text B}$. In
Fig.\ref{fig1} we show the numerical results for the pressure and
the chemical potential as functions of the density for the four
model parameter sets. At low enough density, the equations of state
are indeed consistent with the results (\ref{eos}) with the
scattering length given by (\ref{add}). It is evident that the
results at low density are not sensitive to different model
parameter sets, since the physics at low density should be dominated
by the pseudo-Goldstone bosons.

In fact, we can derive the equations of state (\ref{eos})
analytically from the mean-field thermodynamic potential $\Omega_0$.
For example, the baryon number density reads
\begin{eqnarray}
n&=&\frac{1}{2}N_cN_f\sum_{\bf k}\left[\left(1-\frac{\xi_{\bf
k}^-}{E_{\bf k}^-}\right)-\left(1-\frac{\xi_{\bf k}^+}{E_{\bf
k}^+}\right)\right]\nonumber\\
&=&\frac{1}{2}N_cN_f\sum_{\bf k}\left[\frac{|\Delta|^2}{E_{\bf
k}^-(E_{\bf k}^-+\xi_{\bf k}^-)}-\frac{|\Delta|^2}{E_{\bf
k}^+(E_{\bf k}^++\xi_{\bf k}^+)}\right].
\end{eqnarray}
Near the quantum phase transition point and to leading order of
$|\Delta|^2$, we obtain
\begin{eqnarray}
n&\simeq&\frac{1}{4}N_cN_f\sum_{\bf
k}\left[\frac{|\Delta|^2}{(E_{\bf
k}^*-m_\pi/2)^2}-\frac{|\Delta|^2}{(E_{\bf
k}^*+m_\pi/2)^2}\right]\nonumber\\
&=&2m_\pi{\cal J}|\Delta|^2=|\psiup_0|^2.
\end{eqnarray}

Further, since our treatment is only at the mean-field level, the
Lee-Huang-Yang corrections \cite{Bose03} which are proportional to
$(na_{\text{dd}}^3)^{1/2}$ are absent in the equations of state. As
we have shown in Appendix \ref{app1}, to obtain such corrections, it
is necessary to go beyond the mean field, and a beyond-mean-field
correction to the scattering length $a_{\text{dd}}$ is also
possible\cite{HU,Diener}.
\\
\emph{(IV) Relativistic version.} We can also consider a
relativistic version of the Gross-Pitaevskii free energy via
defining the condensate wave function
\begin{eqnarray}
\Phi(x)=\sqrt{{\cal J}}\Delta(x).
\end{eqnarray}
In this case, the Ginzburg-Landau free energy is reduced to a
relativistic version of the Gross-Pitaevskii free energy,
\begin{eqnarray}\label{RGP}
&&V_{\text{RGP}}[\Phi(x)]=\int
dx\bigg[\Phi^\dagger(x)\left(-\frac{\partial^2}{\partial\tau^2}+2\mu_{\text
B}\frac{\partial}{\partial\tau}
-\mbox{\boldmath{$\nabla$}}^2\right)\Phi(x)\nonumber\\
&&\ \ \ \ \ \ \ \ \ \ \ \ \ \ \ \ \ \ \ \ \ \ \ \ \
+(m_\pi^2-\mu_{\text
B}^2)|\Phi(x)|^2+\frac{\lambda}{2}|\Phi(x)|^4\bigg].
\end{eqnarray}
The self-interacting coupling $\lambda=\beta {\cal J}^{-2}$ is now
dimensionless and can be approximated by $\lambda\simeq
m_\pi^2/f_\pi^2$. For realistic values of $m_\pi$ and $f_\pi$, we
find $\lambda\sim O(1)$. In this sense, the Bose condensate is not
weakly interacting, except for the low density limit
$na_{\text{dd}}^3\ll1$. One should keep in mind that this result 
cannot be applied to high density, since it is valid only near the
quantum phase transition point.

%
%

\subsection {Bogoliubov excitation in a dilute diquark condensate}
\label{s3-3}
An ideal Bose-Einstein condensate is not a superfluid. In presence
of weakly repulsive interactions among the bosons, a Goldstone mode
which has a linear dispersion in the low energy limit appears, and
the condensate becomes a superfluid according to Landau's criterion
$\min_{\bf q}[\omega({\bf q})/|{\bf q}|]>0$. The Goldstone mode
which is also called the Bogoliubov mode here should have a
dispersion given by \cite{Bose01,Bose02,Bose03}
\begin{eqnarray}\label{bog}
\omega({\bf q})=\sqrt{\frac{{\bf q}^2}{2m_\pi}\left(\frac{{\bf
q}^2}{2m_\pi}+\frac{8\pi a_{\text{dd}}n}{m_\pi}\right)},\ \ \ \
|{\bf q}|\ll m_\pi.
\end{eqnarray}

Since the Gross-Pitaevskii free energy obtained above is at the
classical level, to study the bosonic collective excitations we
should consider the fluctuations around the mean field
\cite{BCSBEC0,RBCSBEC01,RBCSBEC02}. The propagator of the bosonic
collective modes is given by ${\bf M}^{-1}(Q)$ and ${\bf
N}^{-1}(Q)$. The Bogoliubov mode corresponds to the lowest
excitation obtained from the equation $\det{\bf M}(\omega,{\bf
q})=0$. With the explicit form of the matrix elements of ${\bf M}$
in the superfluid phase, we can analytically show that $\det{\bf
M}(0,{\bf 0})=0$ which ensures the Goldstone's theorem. In fact, for
$(\omega,{\bf q})=(0,{\bf 0})$, we find that $\det{\bf M}=({\bf
M}_{11}^2-|{\bf M}_{12}|^2){\bf M}_{33}+2|{\bf M}_{13}|^2(|{\bf
M}_{12}|-{\bf M}_{11})$. Using the saddle point condition for
$\Delta$, we can show that ${\bf M}_{11}(0,{\bf 0})=|{\bf
M}_{12}(0,{\bf 0})|$ and hence the Goldstone's theorem holds in the
superfluid phase. Further, we may obtain an analytical expression of
the velocity of the Bogoliubov mode via a Taylor expansion for ${\bf
M}(\omega,{\bf q})$ around $(\omega,{\bf q})=(0,{\bf 0})$ like those
done in \cite{BCSBEC0,RBCSBEC01,RBCSBEC02}. Such a calculation for
our case is more complicated due to the mixing between the sigma
meson and diquarks, and it cannot give the full dispersion
(\ref{bog}).

On the other hand, since $\Delta\rightarrow 0$ near the quantum
phase transition point, we can expand the matrix elements of ${\bf
M}$ in powers of $|\Delta|^2$. The advantage of such an expansion is
that it cannot only give the full dispersion (\ref{bog}) but also
link the meson properties in the vacuum. Formally, we can write down
the following expansions:
\begin{eqnarray}\label{bexp}
&&{\bf M}_{11}(\omega,{\bf q})={\cal D}_{\text d}^{-1}(\omega,{\bf
q})+|\Delta|^2A(\omega,{\bf q})+O(|\Delta|^4),\nonumber\\
&&{\bf M}_{22}(\omega,{\bf q})={\cal D}_{\text d}^{-1}(-\omega,{\bf
q})+|\Delta|^2A(-\omega,{\bf q})+O(|\Delta|^4),\nonumber\\
&&{\bf M}_{12}(\omega,{\bf q})={\bf M}_{21}^\dagger(\omega,{\bf
q})=\Delta^2B(\omega,{\bf q})+O(|\Delta|^4),\nonumber\\
&&{\bf M}_{13}(\omega,{\bf q})={\bf M}_{31}^\dagger(\omega,{\bf
q})=\Delta H(\omega,{\bf q})+O(|\Delta|^3),\nonumber\\
&&{\bf M}_{23}(\omega,{\bf q})={\bf M}_{32}^\dagger(\omega,{\bf
q})=\Delta^\dagger H(-\omega,{\bf q})+O(|\Delta|^3),\nonumber\\
&&{\bf M}_{33}(\omega,{\bf q})={\cal D}_\sigma^{*-1}(\omega,{\bf
q})+O(|\Delta|^2).
\end{eqnarray}
Notice that the effective quark mass $M$ is regarded as a function
of $|\Delta|^2$ as we have done in deriving the Ginzburg-Landau free
energy. Since we are interested in the dispersion in the low energy
limit, i.e., $\omega,|{\bf q}|\ll m_\pi$, we can approximate the
coefficients of the leading order terms as their values at
$(\omega,{\bf q})=(0,{\bf 0})$. That is,
\begin{eqnarray}
&&A(\omega,{\bf q})\simeq A(-\omega,{\bf q})\simeq A(0,{\bf 0})\equiv A_0,\nonumber\\
&&B(\omega,{\bf q})\simeq B(0,{\bf 0})\equiv B_0,\nonumber\\
&&H(\omega,{\bf q})\simeq H(-\omega,{\bf q})\simeq H(0,{\bf
0})\equiv H_0.
\end{eqnarray}
Further, since $m_\sigma\gg m_\pi$, we can approximate the inverse
sigma propagator ${\cal D}_\sigma^{*-1}(\omega,{\bf q})$ as its
value at $(\omega,{\bf q})=(0,{\bf 0})$. Therefore, the dispersion of
the Goldstone mode in the low energy limit can be determined by the
following equation:
\begin{widetext}
\begin{equation}\label{dispersion}
\det\left(\begin{array}{ccc} {\cal D}_{\text d}^{-1}(\omega,{\bf
q})+|\Delta|^2A_0&\Delta^2B_0&\Delta H_0\\
\Delta^{\dagger 2}B_0&{\cal D}_{\text d}^{-1}(-\omega,{\bf
q})+|\Delta|^2A_0&\Delta^\dagger H_0\\
\Delta^\dagger H_0&\Delta H_0&{\cal D}_\sigma^{*-1}(0,{\bf
0})\end{array}\right)=0.
\end{equation}
\end{widetext}

Now we can link the coefficients $A_0,B_0,H_0$ and ${\cal
D}_\sigma^{-1}(0,{\bf 0})$ to the derivatives of the mean-field
thermodynamic potential $\Omega_0$ and its Ginzburg-Landau
coefficients. Firstly, using the explicit form of ${\bf M}_{12}$, we
find that
\begin{eqnarray}
|{\bf M}_{12}(0,{\bf 0})|=|\Delta|^2\beta_1\Longrightarrow
B_0=\beta_1.
\end{eqnarray}
Secondly, using the fact that
\begin{eqnarray}
{\bf M}_{11}(0,{\bf 0})-|{\bf M}_{12}(0,{\bf
0})|=\frac{\partial\Omega_0}{\partial |\Delta|^2},
\end{eqnarray}
and together with the definition for $A(\omega,{\bf q})$,
\begin{eqnarray}
A(\omega,{\bf q})&=&\frac{d{\bf M}_{11}(y,M)}{d
y}\Bigg|_{y=0}\nonumber\\
&=&\frac{\partial{\bf M}_{11}(y,M)}{\partial
y}\Bigg|_{y=0}+\frac{\partial{\bf M}_{11}(y,M)}{\partial M}\frac{d
M}{dy}\Bigg|_{y=0},
\end{eqnarray}
we find the following exact relation:
\begin{eqnarray}
A_0=\beta+B_0=\beta+\beta_1.
\end{eqnarray}
On the other hand, we have the following relations for $H_0$ and
${\cal D}_\sigma^{*-1}(0,{\bf 0})$,
\begin{eqnarray}
H_0&=&\frac{\partial^2\Omega_0(y,M)}{\partial M\partial
y}\Bigg|_{y=0},\nonumber\\
{\cal D}_\sigma^{*-1}(0,{\bf
0})&=&\frac{\partial^2\Omega_0(y,M)}{\partial M^2}\Bigg|_{y=0}.
\end{eqnarray}
One can check the above results from the explicit forms of ${\bf
M}_{13}$ and ${\bf M}_{33}$ in Appendix \ref{app2} directly. Thus we
have
\begin{eqnarray}
-\frac{H_0^2}{{\cal D}_\sigma^{*-1}(0,{\bf 0})}=\beta_2.
\end{eqnarray}

According to the above relations, Eq. (\ref{dispersion}) can be
reduced to
\begin{eqnarray}
&&3\beta^2|\Delta|^4+2\beta|\Delta|^2[{\cal D}_{\text
d}^{-1}(\omega,{\bf
q})+{\cal D}_{\text d}^{-1}(-\omega,{\bf q})]\nonumber\\
&+&{\cal D}_{\text d}^{-1}(\omega,{\bf q}){\cal D}_{\text
d}^{-1}(-\omega,{\bf q})=0.
\end{eqnarray}
It is evident that only the coefficient $\beta$ appears in the final
equation. Further, in the nonrelativistic limit $\omega,|{\bf
q}|\ll m_\pi$ and near the quantum phase transition point, ${\cal
D}_{\text d}^{-1}(\omega,{\bf q})$ can be approximated as
\begin{eqnarray}
{\cal D}_{\text d}^{-1}(\omega,{\bf q})\simeq -2m_\pi {\cal
J}\left(\omega-\frac{{\bf q}^2}{2m_\pi}+\mu_{\text d}\right).
\end{eqnarray}
Together with the mean-field results for the chemical potential
$\mu_{\text d}=g_0|\psiup_0|^2=\beta|\Delta|^2/(2m_\pi {\cal J})$
and for the baryon density $n=|\psiup_0|^2$, we finally get the
Bogoliubov dispersion (\ref{bog}).

We should emphasize that \emph{the mixing between the sigma meson
and the diquarks, denoted by the terms $\Delta H_0$ and
$\Delta^\dagger H_0$, plays an important role in recovering the
correct Bogoliubov dispersion}. Even though we do get this
dispersion, we find the procedure is quite different to the standard
theory of weakly interacting Bose gas
\cite{Bose01,Bose02,GP01,GP02}. There, the elementary excitation is
given only by the diquark-diquark sectors, i.e.,
\begin{widetext}
\begin{eqnarray}
\det\left(\begin{array}{cc} {\bf M}_{11}(Q)&{\bf M}_{12}(Q)\\
{\bf M}_{21}(Q)&{\bf M}_{22}(Q)\end{array}\right)=0 \Longrightarrow
\det\left(\begin{array}{cc} -\omega+\frac{{\bf q}^2}{2m_\pi}-\mu_{\text d}+2g_0|\psiup_0|^2&g_0|\psiup_0|^2\\
g_0|\psiup_0|^2&\omega+\frac{{\bf q}^2}{2m_\pi}-\mu_{\text
d}+2g_0|\psiup_0|^2\end{array}\right)=0.
\end{eqnarray}
\end{widetext}
But in our case, we cannot get the correct Bogoliubov excitation if
we simply set $H_0=0$ and consider only the diquark-diquark sector.
In fact, this requires $A_0=2B_0=2\beta$ which is not true in our
case.

One can also check how the momentum dependence of $A,B,H$ and ${\cal
D}_\sigma^{*-1}$ modifies the dispersion. This needs direct
numerical solution of the equation $\det{\bf M}(\omega,{\bf q})=0$.
We have examined that for $|\mu_{\text B}-m_\pi|$ up to $0.01m_\pi$,
the numerical result agrees well with the Bogoliubov formula
(\ref{bog}). However, at higher density, a significant deviation is
observed. This is in fact a signature of BEC-BCS crossover which
will be discussed in Sec. \ref{s4}.

\subsection {In-medium chiral condensate}
\label{s3-4}
Up to now we have studied the properties of the dilute Bose
condensate induced by a small diquark condensate $\langle
qq\rangle$. The chiral condensate $\langle\bar{q}{q}\rangle$ will be
modified in the medium. In such a dilute Bose condensate, we can
study the response of the chiral condensate to the baryon density
$n$.

\begin{widetext}

\begin{figure}[!htb]
\begin{center}
\includegraphics[width=7cm]{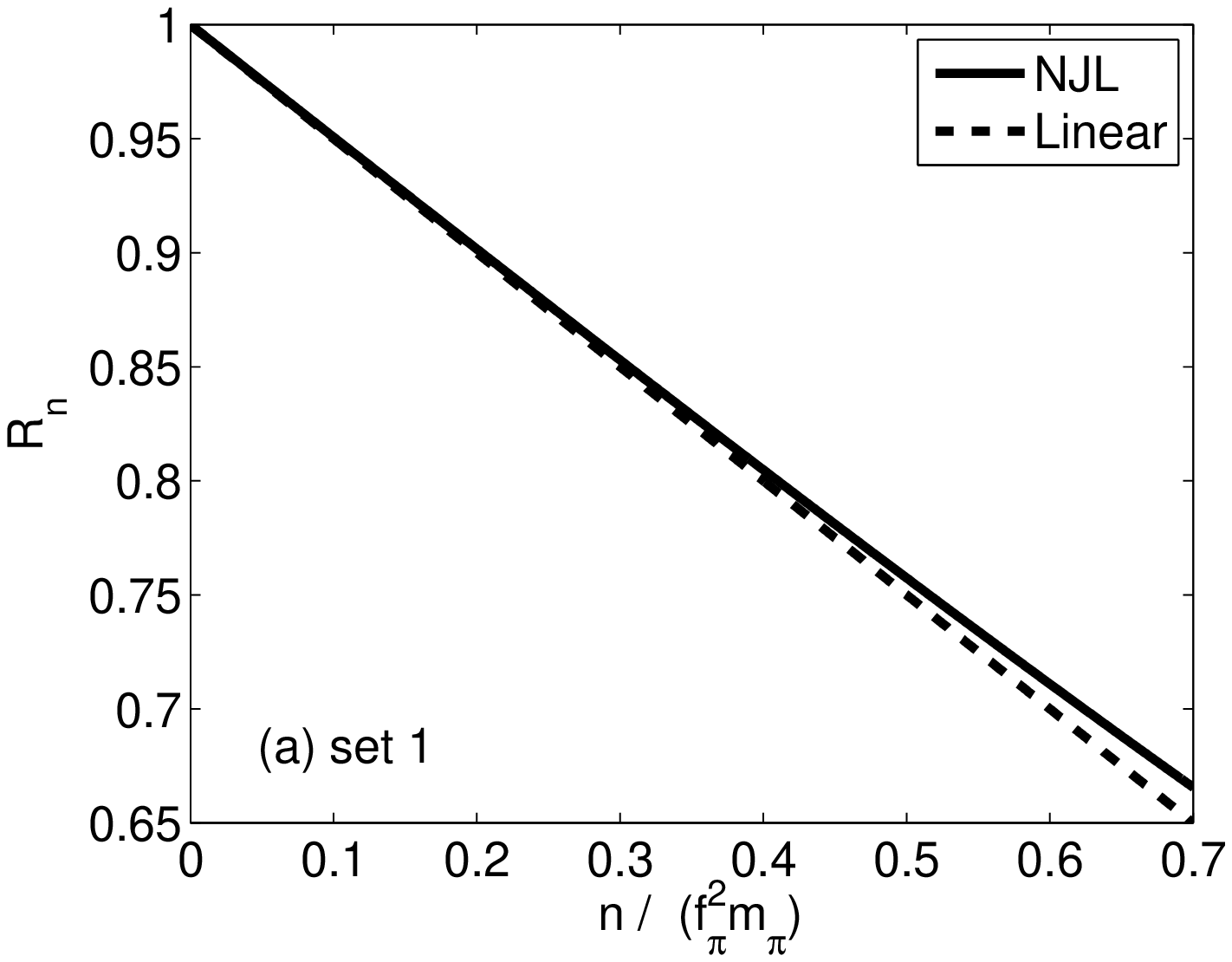}
\includegraphics[width=7cm]{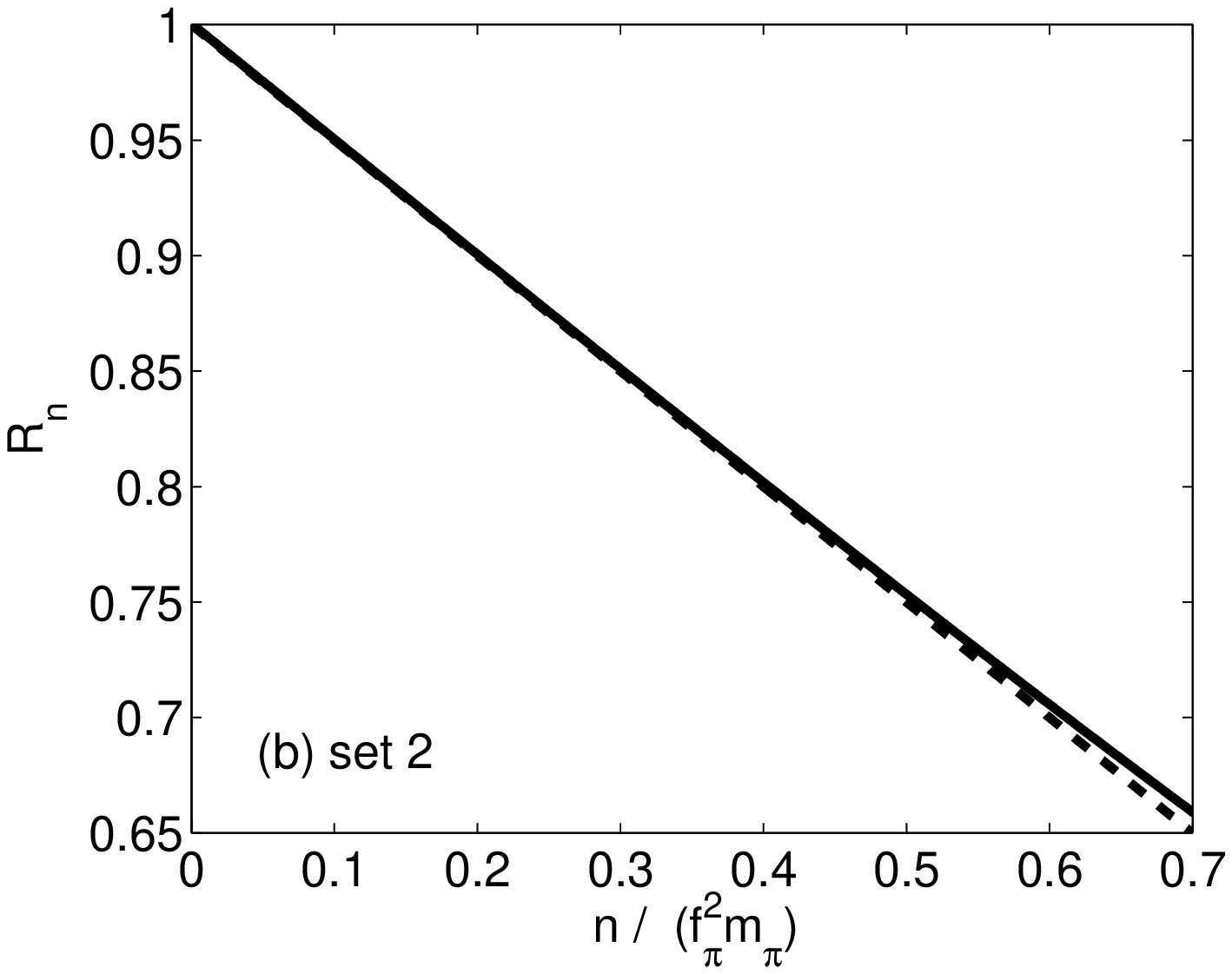}
\includegraphics[width=7cm]{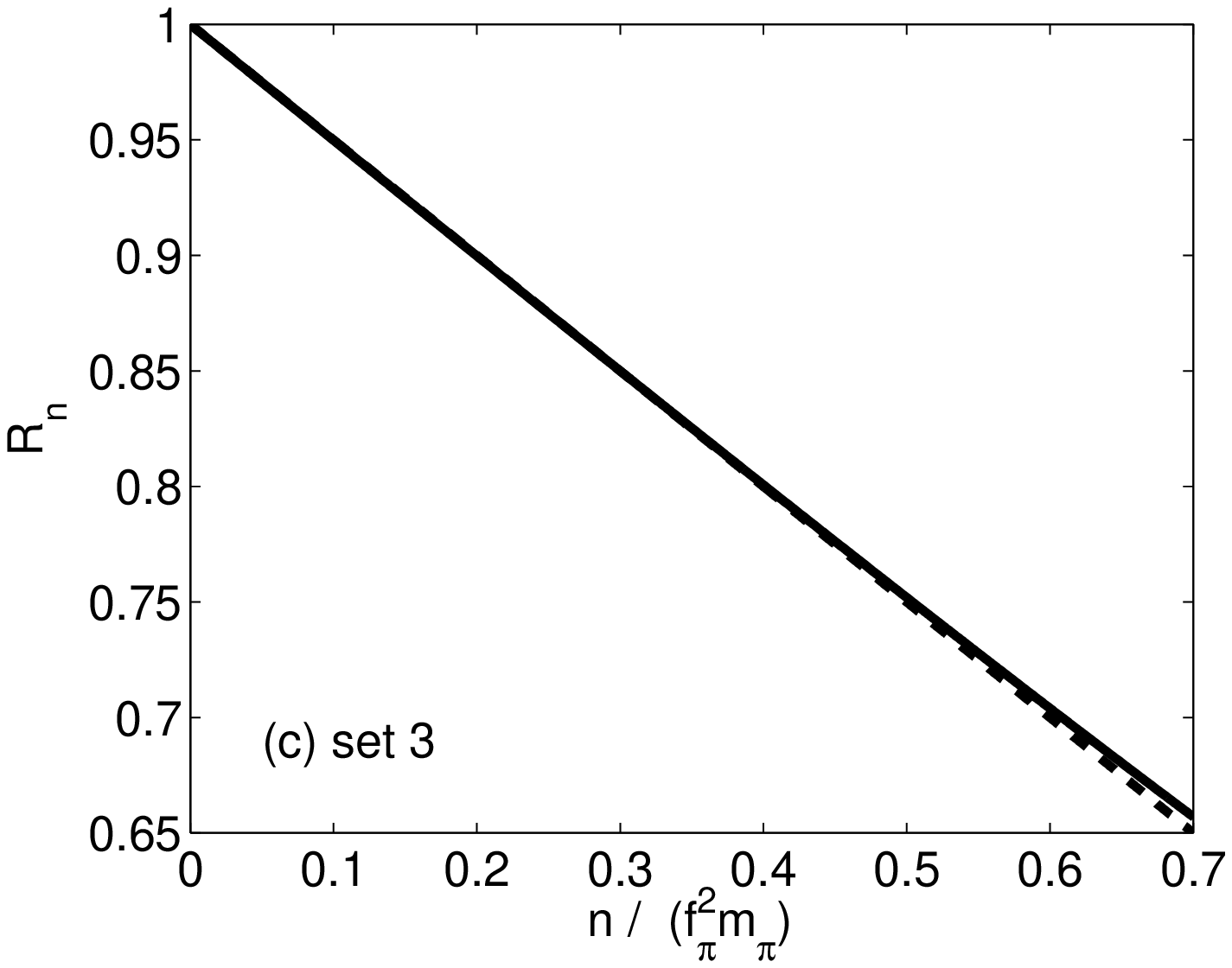}
\includegraphics[width=7cm]{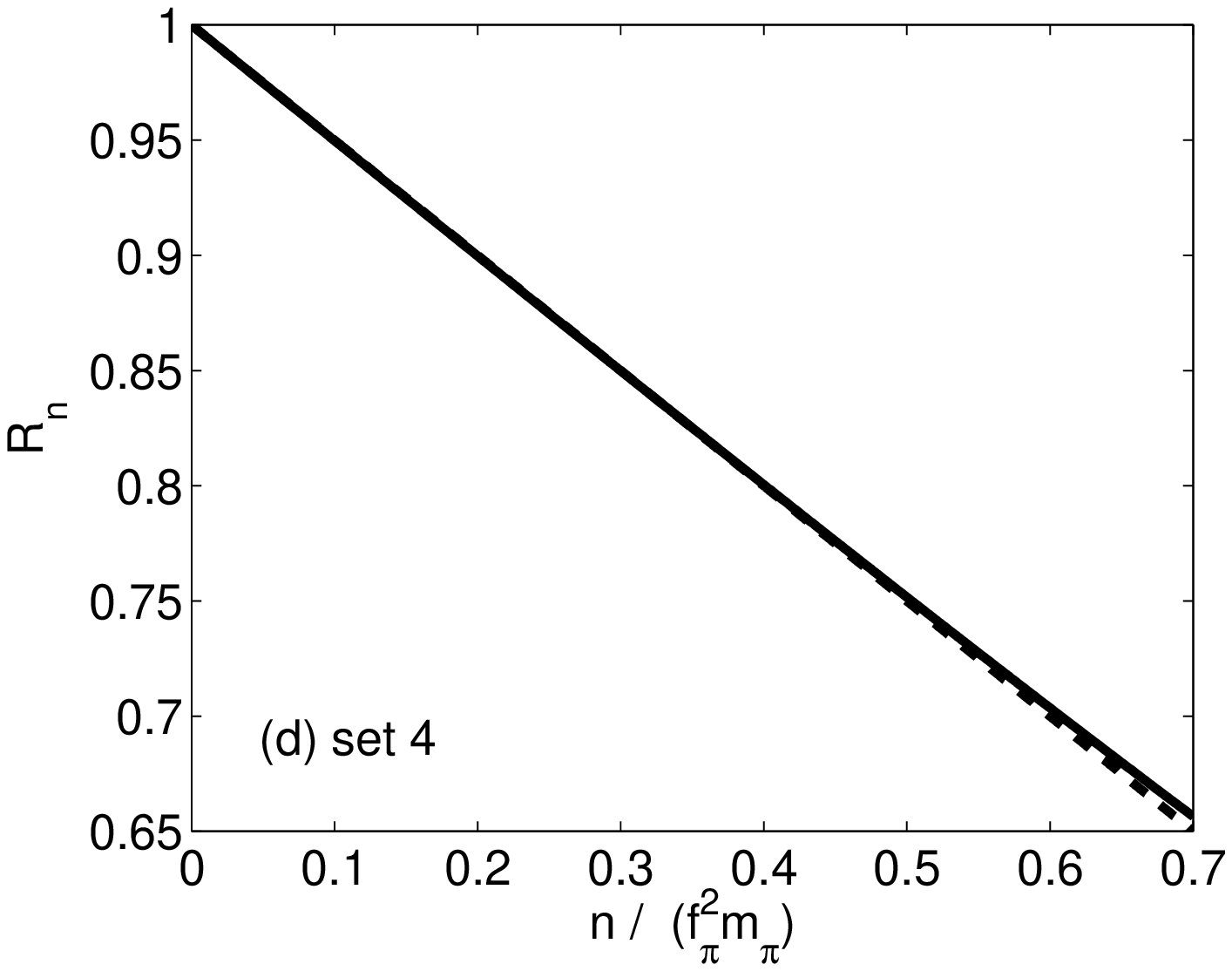}
\caption{The ratio
$R_n=\langle\bar{q}{q}\rangle_n/\langle\bar{q}{q}\rangle_0$ as a
function of $n/(f_\pi^2m_\pi)$ for different model parameter sets.
The dashed line is the linear behavior given by (\ref{linear}).
\label{fig2}}
\end{center}
\end{figure}

\end{widetext}

To this end, we expand the effective quark mass $M$ in terms of
$y=|\Delta|^2$. We have
\begin{equation}
M-M_*=\frac{dM}{dy}\bigg|_{y=0}y+O(y^2)
\end{equation}
The expansion coefficient can be approximated as
\begin{eqnarray}
\frac{dM}{dy}\bigg|_{y=0}&\simeq&
-\frac{2f_\pi^2/M_*}{m_\pi^2f_\pi^2/M_*^2+4f_\pi^2}\nonumber\\
&=&-\frac{1}{2M_*}\left[1+O\left(\frac{m_\pi^2}{4M_*^2}\right)\right].
\end{eqnarray}
Using the definition of the effective quark mass,
$M=m_0-2G\langle\bar{q}{q}\rangle$, we find that
\begin{equation}
\frac{\langle\bar{q}{q}\rangle_n}{\langle\bar{q}{q}\rangle_0}=1-\frac{|\Delta|^2}{4G\langle\bar{q}{q}\rangle_0
M_*}\simeq1-\frac{|\Delta|^2}{2M_*^2}.
\end{equation}
Since the baryon number density reads $n=|\psiup_0|^2=2m_\pi {\cal
J}|\Delta|^2$, using the fact that ${\cal J}\simeq
f_\pi^2/(2M_*^2)$, we obtain to leading order
\begin{equation}\label{linear}
\frac{\langle\bar{q}{q}\rangle_n}{\langle\bar{q}{q}\rangle_0}\simeq1-\frac{n}{2f_\pi^2
m_\pi}.
\end{equation}
This formula is in fact a two-color analogue of the density
dependence of the chiral condensate in the $N_c=3$ case, where we
have \cite{cohen,cohen2}
\begin{equation}
\frac{\langle\bar{q}{q}\rangle_n}{\langle\bar{q}{q}\rangle_0}\simeq1-\frac{\Sigma_{\pi
{\text N}}}{f_\pi^2 m_\pi^2}n
\end{equation}
with $\Sigma_{\pi \text{N}}$ being the pion-nucleon sigma term.
In Fig.\ref{fig2}, we show the numerical results via solving the
mean-field gap equations. One finds that the chiral condensate has a
perfect linear behavior at low density. For large value of $M_*$ (
and hence the sigma meson mass $m_\sigma$), the linear behavior
persists even at higher density.

In fact, the Eq. (\ref{linear}) can be obtained in a model
independent way. Applying the Hellmann-Feynman theorem to a dilute
diquark gas with energy density ${\cal E}(n)$ given by (\ref{eos}),
we can obtain (\ref{linear}) directly. According to the
Hellmann-Feynman theorem, we have
\begin{equation}
2m_0(\langle\bar{q}{q}\rangle_n-\langle\bar{q}{q}\rangle_0)=m_0\frac{d{\cal
E}}{dm_0}.
\end{equation}
The derivative $d{\cal E}/dm_0$ can be evaluated via the chain rule
$d{\cal E}/dm_0=(d{\cal E}/dm_\pi)(dm_\pi/dm_0)$. Together with the
Gell-Mann--Oakes--Renner relation
$m_\pi^2f_\pi^2=-m_0\langle\bar{q}{q}\rangle_0$ and the fact that
$da_{\text {dd}}/dm_\pi\simeq a_{\text{dd}}/m_\pi$, we can obtain to
leading order Eq. (\ref{linear}). Beyond the leading order,
we find the correction of order $O(n^2)$ vanishes. Thus, the
next-to-leading order correction should be  $O(n^{5/2})$ coming from
the Lee-Huang-Yang correction to the equation of state \cite{HFiso}.

Finally, we can show analytically that the ``chiral rotation"
behavior \cite{ISO01,ISO02,QL01,QL02,QL03,QL04,QL05,QL06} predicted
by the chiral perturbation theories is valid in the NJL model near
the quantum phase transition. In the chiral perturbation theories,
the chemical potential dependence of the chiral and diquark
condensates can be analytically expressed as
\begin{equation}
\frac{\langle\bar{q}{q}\rangle_{\mu_{\text
B}}}{\langle\bar{q}{q}\rangle_0}=\frac{m_\pi^2}{\mu_{\text B}^2},\ \
\ \ \frac{\langle q{q}\rangle_{\mu_{\text
B}}}{\langle\bar{q}{q}\rangle_0}=\sqrt{1-\frac{m_\pi^4}{\mu_{\text
B}^4}}.\label{chpt}
\end{equation}
Near the phase transition point, we can expand the above formula in
powers of $\mu_{\text d}=\mu_{\text B}-m_\pi$. To leading order, we
have
\begin{equation}
\frac{\langle\bar{q}{q}\rangle_{\mu_{\text
B}}}{\langle\bar{q}{q}\rangle_0}\simeq1-\frac{2\mu_{\text
d}}{m_\pi},\ \ \ \ \frac{\langle q{q}\rangle_{\mu_{\text
B}}}{\langle\bar{q}{q}\rangle_0}\simeq 2\sqrt{\frac{\mu_{\text
d}}{m_\pi}}.
\end{equation}
Using the mean-field result (\ref{chemi}) for the chemical potential
$\mu_{\text d}$, one can easily check that the above relations are
also valid in our NJL model.

\subsection {Chiral limit}
\label{s3-5}
In the above studies we focused on the ``physical point" where
$m_0\neq 0$. In the final part of this section, we briefly discuss
the chiral limit with $m_0=0$.

We may naively expect that the results at $m_0\neq0$ can be directly
generalized to the chiral limit via setting $m_\pi=0$. The ground
state is a noninteracting Bose condensate of massless diquarks,
since $m_\pi=0$ and $a_{\text{dd}}=0$. However, this cannot be true
since many divergences develop due to the vanishing pion mass. In
fact, the conclusion of second order phase transition is not correct
since the Ginzburg-Landau coefficient $\beta$ vanishes. Instead, the
superfluid phase transition is of strongly first order in the chiral
limit \cite{isoNJL06,2CNJL01}.

In the chiral limit, the effective action in the vacuum should
depend only on the combination
$\sigma^2+\mbox{\boldmath{$\pi$}}^2+|\phi|^2$ due to the exact
flavor symmetry SU$(4)\simeq$ SO$(6)$. The vacuum is chosen to be
associated with a nonzero chiral condensate $\langle\sigma\rangle$
without loss of generality. At zero and at finite chemical
potential, the thermodynamic potential $\Omega_0(M,|\Delta|)$ has
two minima locating at $(M,|\Delta|)=(a,0)$ and
$(M,|\Delta|)=(0,b)$. At zero chemical potential, these two minima
are degenerate due to the exact flavor symmetry. However, at nonzero
chemical potential (even arbitrarily small), the minimum $(0,b)$ has
the lowest free energy. Analytically, we can show that $b\rightarrow
M_*$ at $\mu_{\text B}=0^+$. This means the superfluid phase
transition in the chiral limit is of strongly first order, and takes
place at arbitrarily small chemical potential. Since the effective
quark mass $M$ keeps vanishing in the superfluid phase, a low
density Bose condensate does not exist in the chiral limit.

%
%

\section {Matter at High Density: BEC-BCS crossover and Mott Transition}
\label{s4}
The investigations in Sec. \ref{s3} are restricted near the
quantum phase transition point $\mu_{\text B}=m_\pi$. Generally the
state of matter at high density should not be a relativistic Bose
condensate described by (\ref{RGP}). In fact, perturbative QCD
calculations show that the matter is a weakly coupled BCS superfluid
at asymptotic density \cite{pQ01,pQ02,pQ03,Yama}. In this section,
we will discuss the evolution of the superfluid matter as the baryon
density increases from the NJL model point of view. While some
results presented in the following have been published elsewhere
\cite{2CNJL01,2CNJL02,2CNJL03,2CNJL04}, we will still show them for
the sake of completeness.

%
\begin{widetext}

\begin{figure}[!htb]
\begin{center}
\includegraphics[width=7cm]{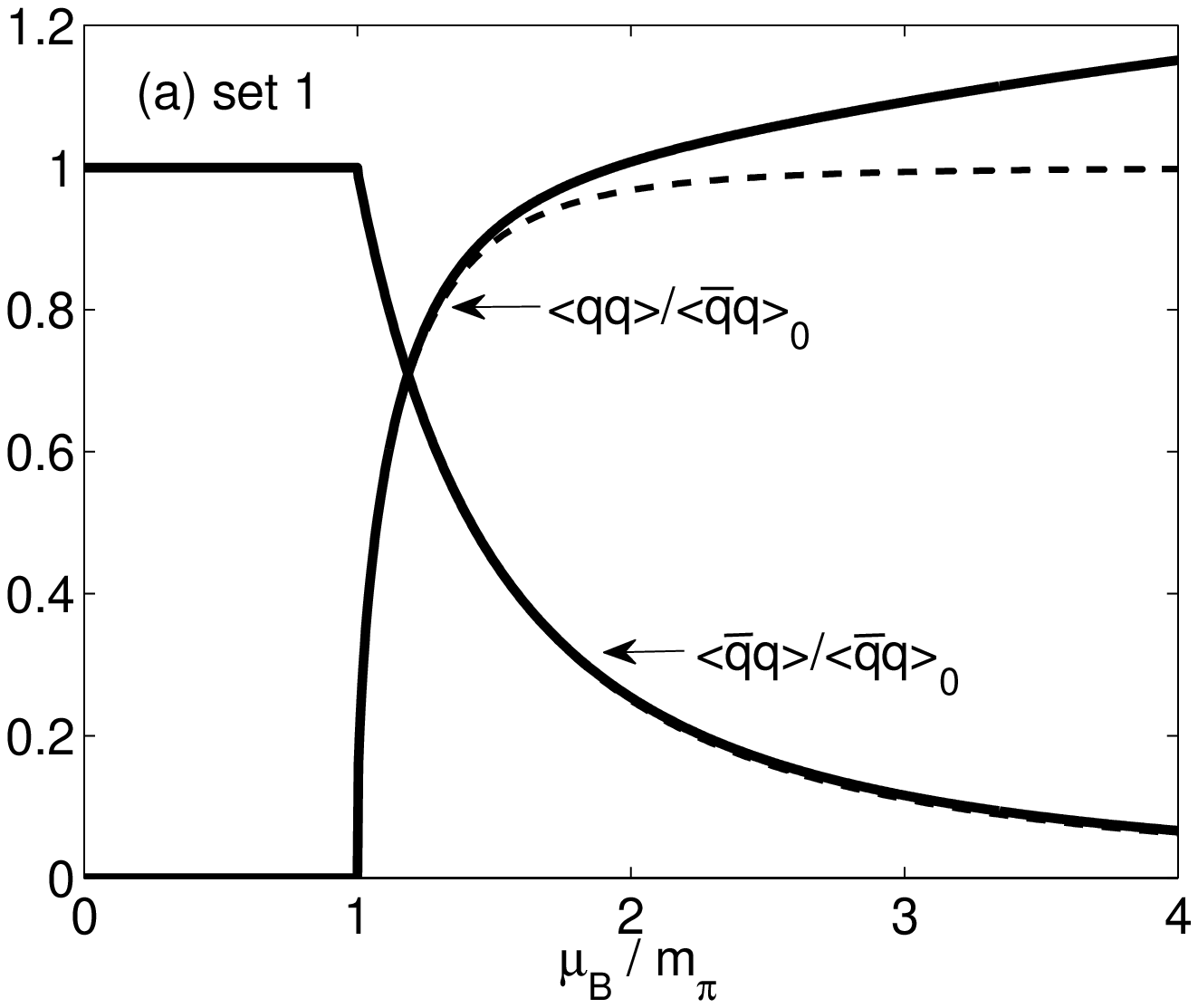}
\includegraphics[width=7cm]{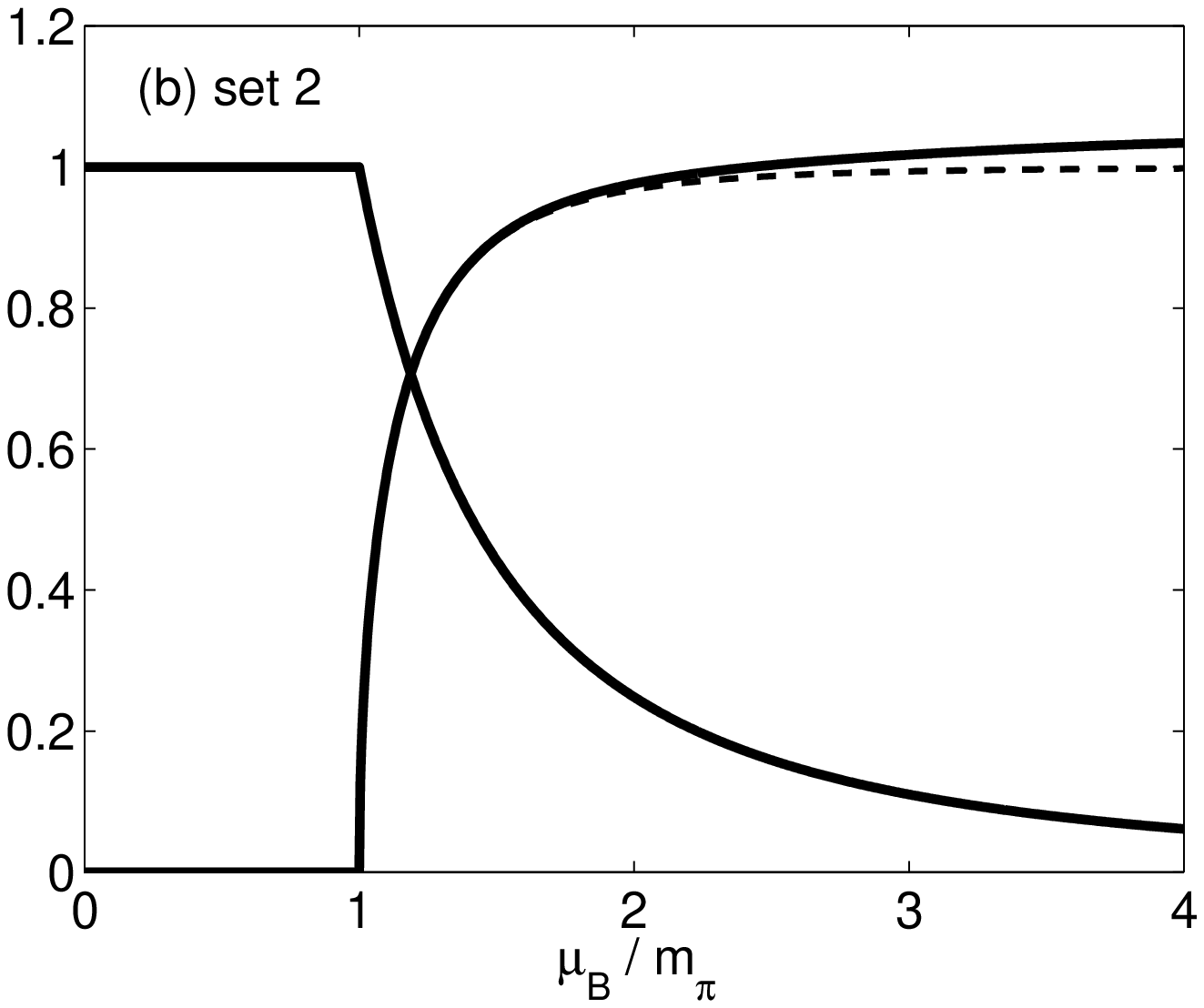}
\includegraphics[width=7cm]{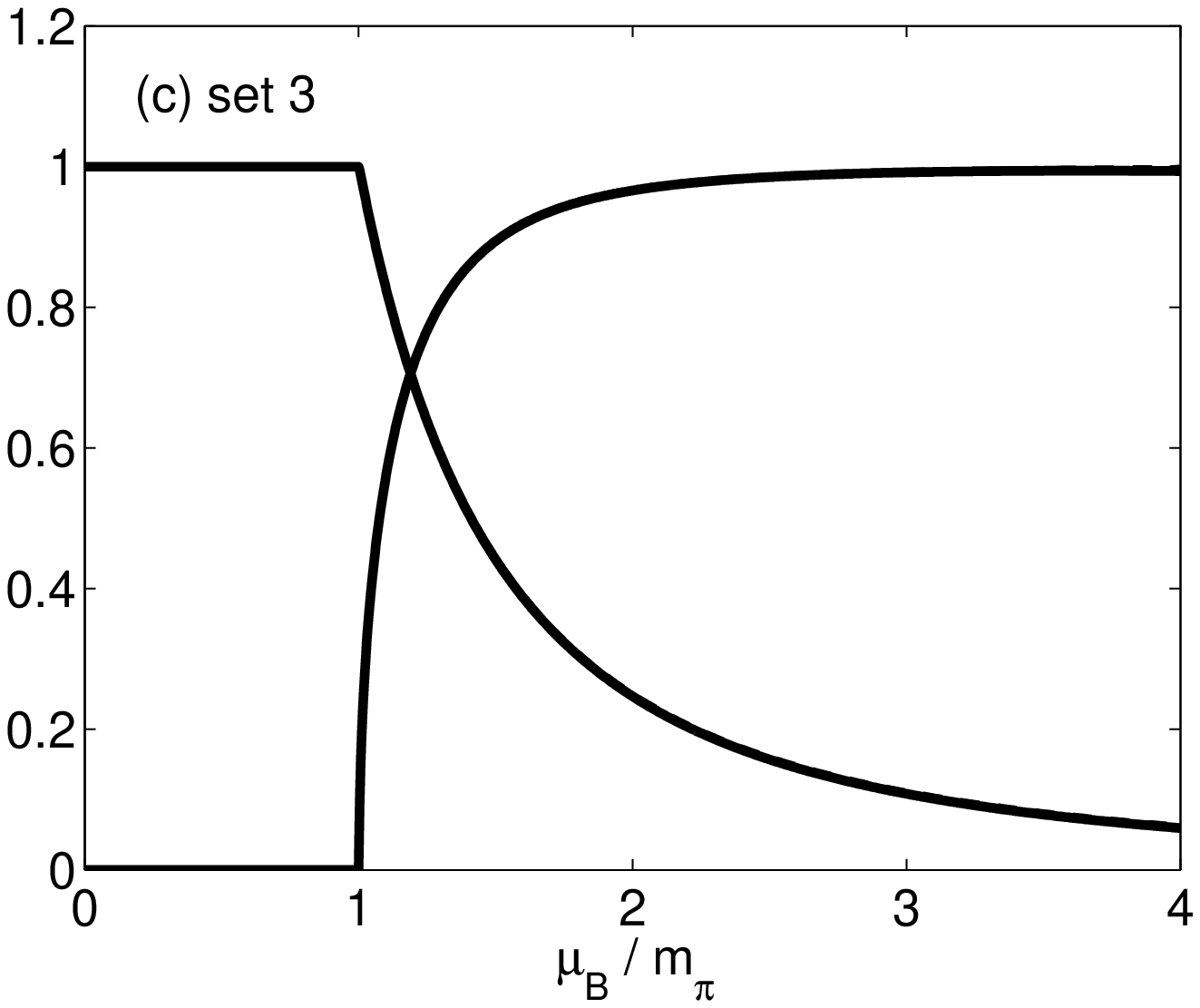}
\includegraphics[width=7cm]{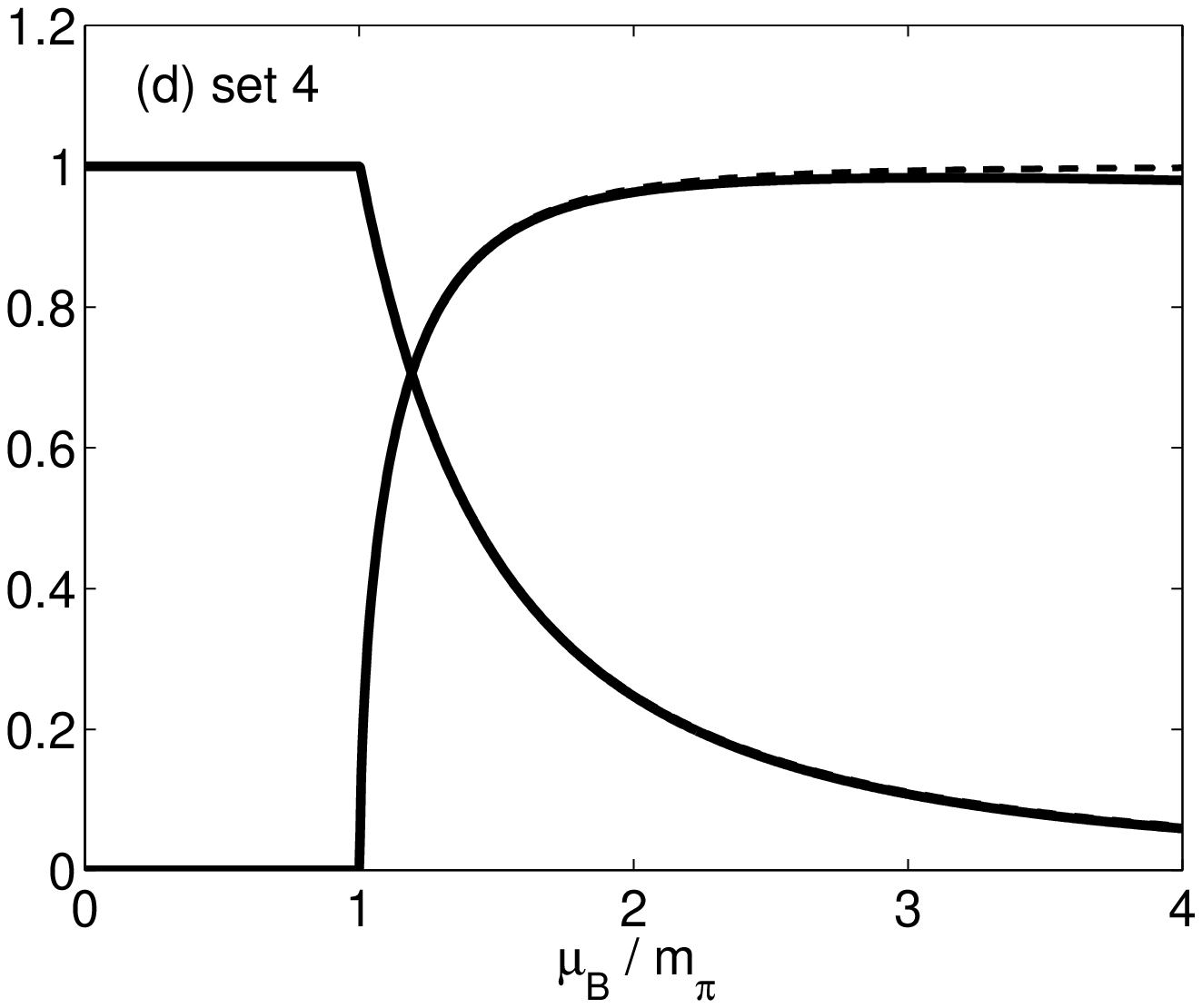}
\caption{The chiral and diquark condensates (in units of
$\langle\bar{q}{q}\rangle_0$) as functions of the baryon chemical
potential(in units of $m_\pi$) for different model parameter sets.
\label{fig3}}
\end{center}
\end{figure}

\end{widetext}
%

\subsection {Chiral and diquark condensates}
\label{s4-1}
The numerical results for the chiral condensate
$\langle\bar{q}q\rangle$ and diquark condensate $\langle qq\rangle$
are shown in Fig.\ref{fig3}. As a comparison, we also show the
analytical result (\ref{chpt}) predicted by chiral perturbation
theories. While the behavior of the chiral condensate is in good
agreement with the chiral perturbation theories, the diquark
condensate deviates significantly from the result (\ref{chpt}) for
small values of $M_*$.

This deviation can be understood from the fact that the chiral
perturbation theories correspond to the nonlinear sigma model limit
$m_\sigma\rightarrow \infty$. For finite value of $m_\sigma$, one
should consider the O$(6)$ linear sigma model \cite{2CNJL04}
\begin{equation}
{\cal L}_{\text{LSM}}=\frac{1}{2}(\partial_\mu
\mbox{\boldmath{$\varphi$}})^2-\frac{1}{2}m^2\mbox{\boldmath{$\varphi$}}^2+\frac{1}{4}\lambda\mbox{\boldmath{$\varphi$}}^4-H\sigma
\end{equation}
where $\mbox{\boldmath{$\varphi$}}=(\sigma, \mbox{\boldmath{$\pi$}},
\phi_1,\phi_2)$ and $m^2<0$. The model parameters $m^2,\lambda,H$
can be determined from the vacuum phenomenology. In this model, we
can show that the chiral and diquark condensates are given by
\cite{isoNJL06,2CNJL04}
\begin{eqnarray}
&&\frac{\langle\bar{q}{q}\rangle_{\mu_{\text
B}}}{\langle\bar{q}{q}\rangle_0}=\frac{m_\pi^2}{\mu_{\text
B}^2},\nonumber\\
&&\frac{\langle q{q}\rangle_{\mu_{\text
B}}}{\langle\bar{q}{q}\rangle_0}=\sqrt{1-\frac{m_\pi^4}{\mu_{\text
B}^4}+2\frac{\mu_{\text
B}^2-m_\pi^2}{m_\sigma^2-m_\pi^2}}.\label{lsm}
\end{eqnarray}
In the nonlinear sigma model limit $m_\sigma\rightarrow\infty$, the
above results are indeed reduced to the result (\ref{chpt})
predicted by chiral perturbation theories. However, for finite
values of $m_\sigma$, the results can be significantly different
from (\ref{chpt}) at large chemical potential.

\subsection {BEC-BCS crossover}
\label{s4-2}

While the Ginzburg-Landau free energy can be reduced to the
Gross-Pitaevskii free energy near the quantum phase transition
point, it is not the case at arbitrary $\mu_{\text B}$.

\begin{figure}[!htb]
\begin{center}
\includegraphics[width=8.5cm]{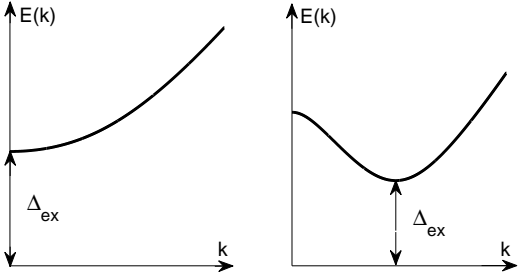}
\caption{A schematic plot of the fermionic excitation spectrum in
the BEC state (left) and the BCS state (right). \label{fig4}}
\end{center}
\end{figure}

When $\mu_{\text B}$ increases, we find that the fermionic
excitation spectra $E_{\bf k}^\pm$ undergo a characteristic change.
Near the quantum phase transition $\mu_{\text B}=m_\pi$ they are
nearly degenerate since $m_\pi\ll 2M_*$ and their minima are located
at $|{\bf k}|=0$. However, at very large $\mu_{\text B}$ the minimum
of $E_{\bf k}^-$ moves to $|{\bf k}|\simeq \mu_{\text B}/2$ since
$M\rightarrow m_0$. Meanwhile the excitation energy of the
antifermion excitations become much larger than that of the fermion
excitations and can be neglected. This characteristic change of the
fermionic excitation spectra takes place when the minimum of the
lowest band excitation $E_{\bf k}^-$ moves from $|{\bf k}|=0$ to
$|{\bf
k}|\neq0$\cite{2CNJL02,2CNJL03,dson,BCSBEC0,RBCSBEC01,RBCSBEC02,kita,NSR01,NSR02,RNSR01,RNSR02,G0G},
i.e., $\mu_{\text B}/2=M(\mu_{\text B})$\cite{note1}. A schematic
plot of this characteristic change is shown in Fig.\ref{fig4}. The
equation $\mu_{\text B}/2=M(\mu_{\text B})$ defines the so-called
crossover point $\mu_{\text B}=\mu_0$ which can be numerically
determined by the mean-field gap equations. The numerical results of
the crossover chemical potential $\mu_0$ for the four model
parameter sets are shown in Table.\ref{crossover}. For reasonable
parameter sets, the crossover chemical potential is in the range
$(1.6-2)m_\pi$.

In fact, an analytical expression for $\mu_0$ can be achieved
according to the fact that the chiral rotation behavior
$\langle\bar{q}q\rangle_{\mu_{\text
B}}/\langle\bar{q}q\rangle_0\simeq m_\pi^2/\mu_{\text B}^2$ is still
valid in the NJL model at large chemical potentials as shown in
Fig.\ref{fig3}. We obtain \cite{2CNJL02}
\begin{eqnarray}
\frac{\mu_0}{2}\simeq\frac{m_\pi^2}{\mu_0^2}M_*\ \ \Longrightarrow \
\ \mu_0\simeq (2M_* m_\pi^2)^{1/3}. \label{crossmu}
\end{eqnarray}
Using the fact that $m_\sigma\simeq 2M_*$, we find that $\mu_0$ can
be expressed as
\begin{eqnarray}
\frac{\mu_0}{m_\pi}\simeq\left(\frac{m_\sigma}{m_\pi}\right)^{1/3}.
\end{eqnarray}
Thus, in the nonlinear sigma model limit
$m_\sigma/m_\pi\rightarrow\infty$, there should be no BEC-BCS
crossover. On the other hand, this means the physical prediction
power of the chiral perturbation theories is restricted near the
quantum phase transition point.

\begin{table}[b!]
\begin{center}
\begin{tabular}{|c|c|c|c|c|}
  \hline
  Set & 1 & 2 & 3 & 4 \\
  \hline
  Crossover chemical potential $\mu_0$  [$m_\pi$]& 1.65 & 1.81 & 1.95 & 2.07 \\
  \hline
\end{tabular}
\end{center}
\caption{\small The crossover chemical potential $\mu_0$ (in units
of $m_\pi$) for different model parameter sets. } \label{crossover}
\end{table}

The fermionic excitation gap $\Delta_{\text{ex}}$ (as shown in
Fig.\ref{fig4}), defined as the minimum of the fermionic excitation
energy, i.e., $\Delta_{\text{ex}}=\min_{\bf k}\{E_{\bf k}^-,E_{\bf
k}^+\}$, can be evaluated as
\begin{equation}
\Delta_{\text{ex}}=
 \left\{ \begin{array}
{r@{\quad,\quad}l}
 \sqrt{(M-\frac{\mu_{\text B}}{2})^2+|\Delta|^2}&
 \mu_{\text B}<\mu_0 \\
 |\Delta| & \mu_{\text B}>\mu_0.
\end{array}
\right.
\end{equation}
It is evident that the fermionic excitation gap is equal to the
superfluid order parameter only in the BCS regime. This is similar
to the BEC-BCS crossover in nonrelativistic systems \cite{BCSBEC0},
and we find that the corresponding fermion chemical potential $\mu$
can be defined as $\mu=\mu_{\text B}/2-M$. The numerical results of
the fermionic excitation gap $\Delta_{\text{ex}}$ for different
model parameter sets are shown in Fig.\ref{fig5}. We find that for a
wide range of the baryon chemical potential, it is of order
$O(M_*)$. The fermionic excitation gap is equal to the pairing gap
$|\Delta|$ only at the BCS side of the crossover, and exhibits a
minimum at the quantum phase transition point.

On the other hand, the momentum distributions of quarks (denoted by
$n({\bf k})$) and antiquarks (denoted by $\bar{n}({\bf k})$) can be
evaluated using the quark Green function ${\cal G}_{11}(K)$. We
obtain
\begin{eqnarray}
&&n({\bf k})=\frac{1}{2}\left(1-\frac{\xi_{\bf k}^-}{E_{\bf
k}^-}\right),\ \ \ \ \text{for quarks},\nonumber\\
&&\bar{n}({\bf k})=\frac{1}{2}\left(1-\frac{\xi_{\bf k}^+}{E_{\bf
k}^+}\right),\ \ \ \ \text{for antiquarks}.
\end{eqnarray}
The numerical results for $n({\bf k})$ and $\bar{n}({\bf k})$ (for
model parameter set 1) are shown in Fig.\ref{fig6}. Near the quantum
phase transition point, the quark momentum distribution $n({\bf k})$
is a very smooth function in the whole momentum space. In the
opposite limit, i.e., at large chemical potentials, it approaches
unity at $|{\bf k}|=0$ and decreases rapidly around the effective
``Fermi surface" at $|{\bf k}|\simeq|\mu|$. For the antiquarks, we
find that the momentum distribution $\bar{n}({\bf k})$ exhibits a
nonmonotonous behavior: it is suppressed at both low and high
densities and is visible only at moderate chemical potentials.
However, even at very large chemical potentials, e.g., $\mu_{\text
B}=10m_\pi$, the momentum distribution $n({\bf k})$ does not
approach the standard BCS behavior, which means the dense matter is
not a weakly coupled BCS superfluid for a wide range of the baryon
chemical potential. In Fig.\ref{fig7}, we show the ratio
$|\Delta|/\mu$ up to $\mu_{\text B}\simeq 10m_\pi$. It is clear that
the ratio is not small even at large chemical potentials. At
$\mu_{\text B}=10m_\pi$, it is about $0.5$, which means the dense
matter is still a strongly coupled BCS superfluid.

The Goldstone mode also undergoes a characteristic change in the
BEC-BCS crossover. Near the quantum phase transition point, i.e., in
the dilute limit, the Goldstone mode recovers the Bogoliubov
excitation of weakly interacting Bose condensates.  In the opposite
limit, we expect the Goldstone mode approaches the
Anderson-Bogoliubov mode of a weakly coupled BCS superfluid, which
takes a dispersion $\omega({\bf q})=|{\bf q}|/\sqrt{3}$ up to the
two-particle continuum $\omega\simeq 2|\Delta|$. In fact, at large
chemical potentials, we can safely neglect the mixing between the
sigma meson and diquarks. The Goldstone boson dispersion is thus
determined by the equation
\begin{equation}
\det\left(\begin{array}{cc} {\bf M}_{11}(Q)&{\bf M}_{12}(Q)\\
{\bf M}_{21}(Q)&{\bf M}_{22}(Q)\end{array}\right)=0 .
\end{equation}
The problem is totally the same as that has been investigated in
\cite{RBCSBEC01,RBCSBEC02}. Therefore, at very large chemical
potentials where $|\Delta|/\mu$ becomes small enough, the Goldstone
mode recovers the Anderson-Bogoliubov mode of a weakly coupled BCS
superfluid.

Finally, we should emphasize that the existence of a smooth
crossover from the Bose condensate to the BCS superfluid depends on
whether there exists a deconfinement phase transition at finite
$\mu_{\text B}$ \cite{L2C05,L2C06,decon} and where it takes place.
Recent lattice calculation predicts a deconfinement crossover which
occurs at a baryon chemical potential larger than that of the
BEC-BCS crossover \cite{L2C06}.
%
\begin{widetext}

\begin{figure}[!htb]
\begin{center}
\includegraphics[width=7cm]{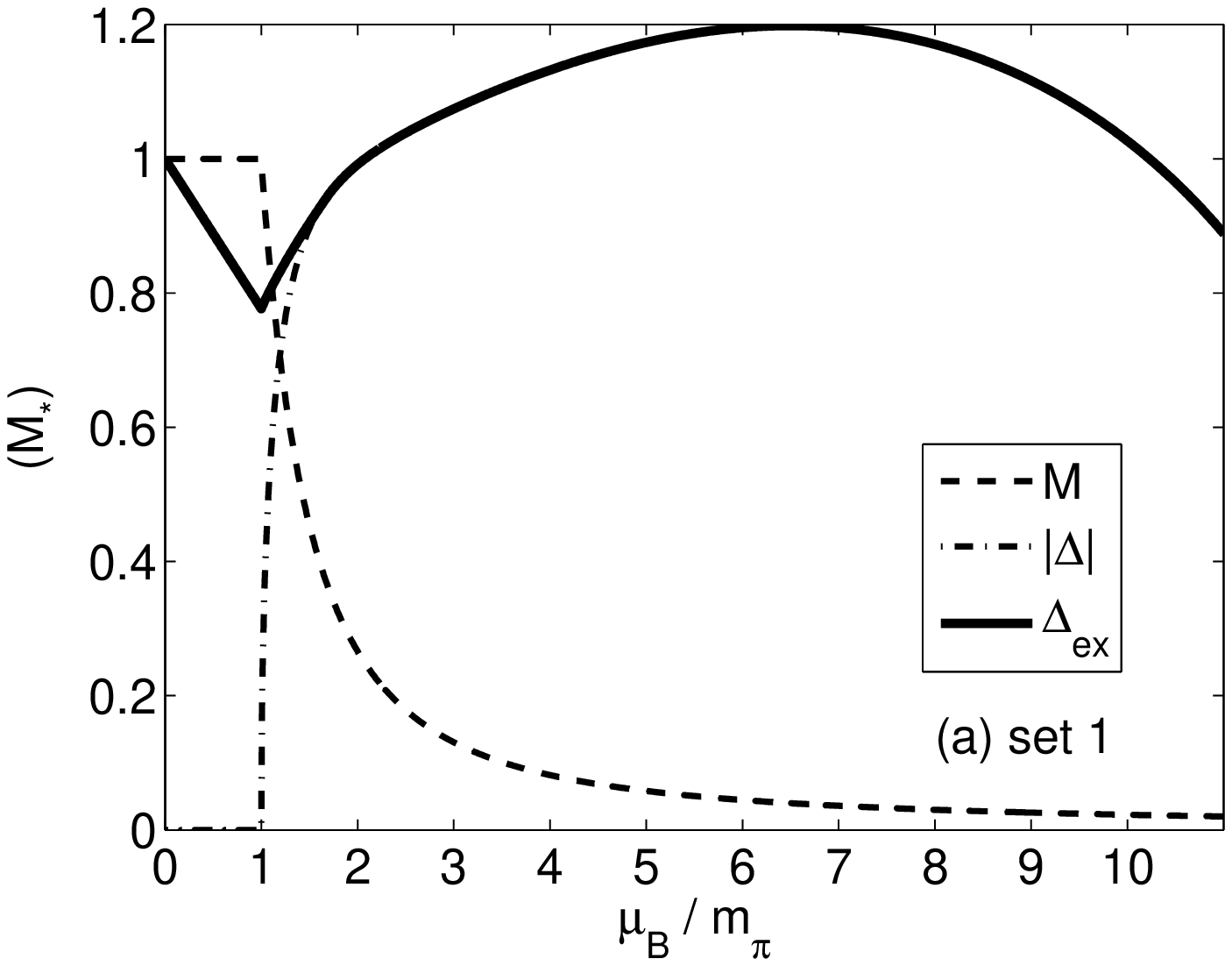}
\includegraphics[width=7cm]{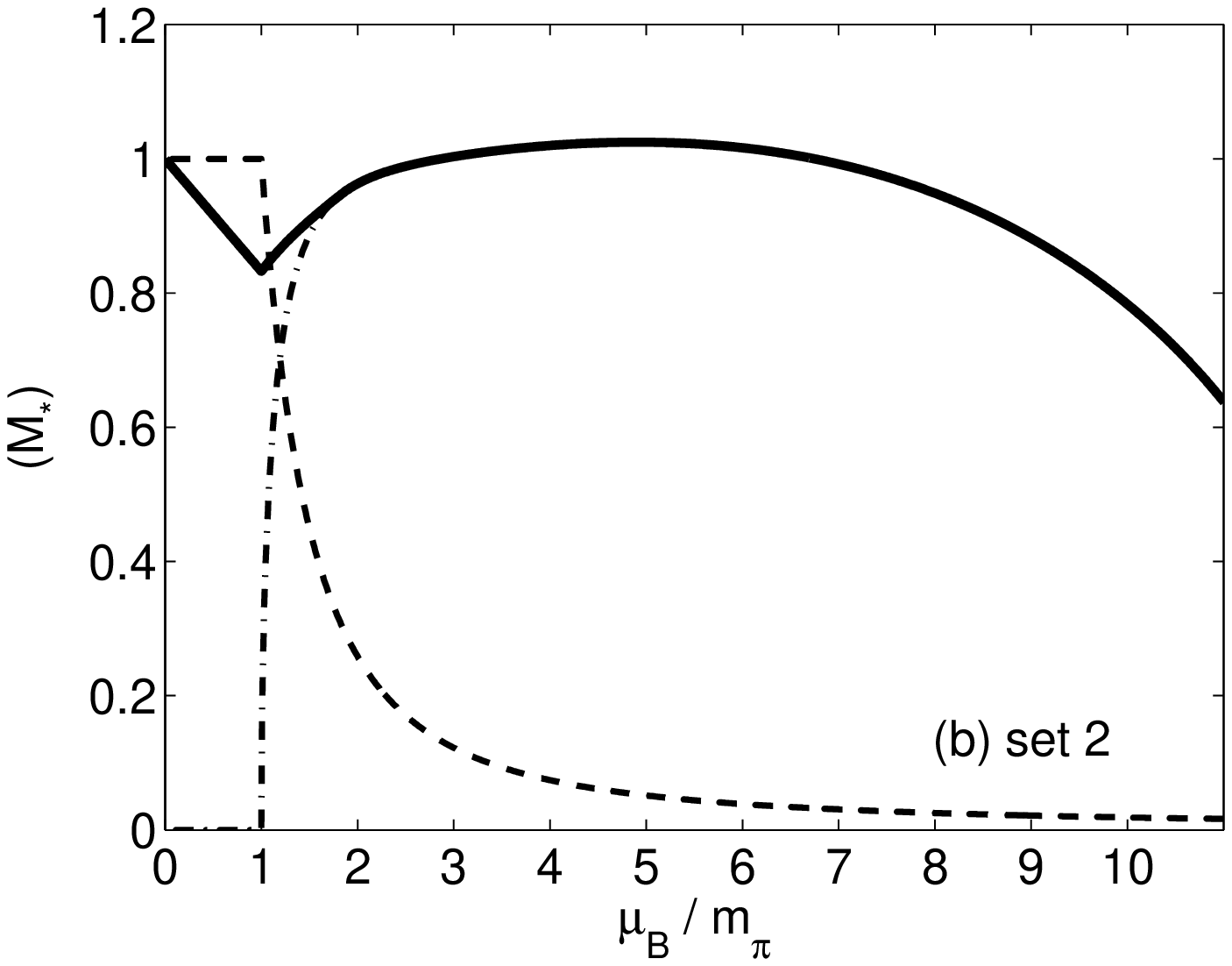}
\includegraphics[width=7cm]{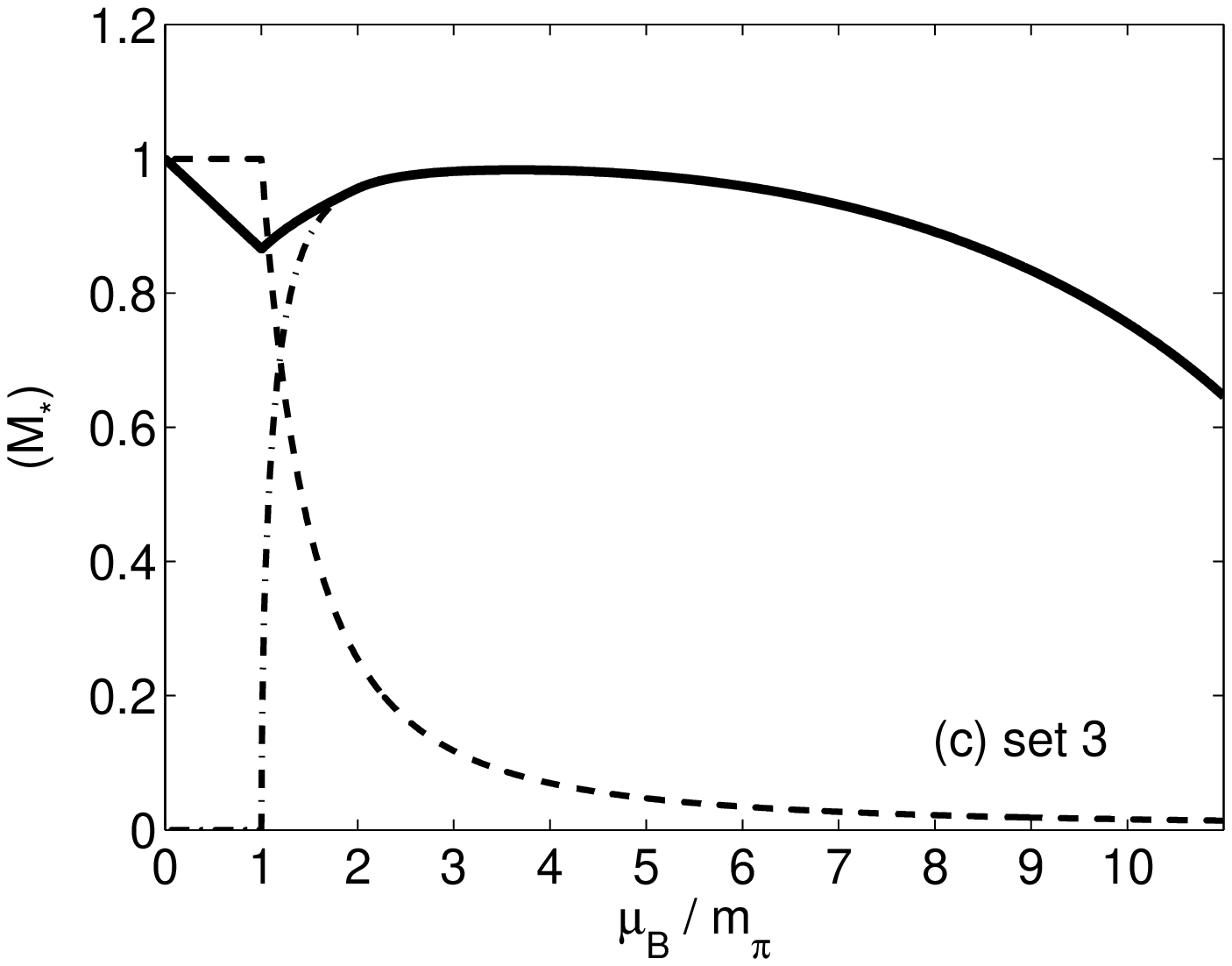}
\includegraphics[width=7cm]{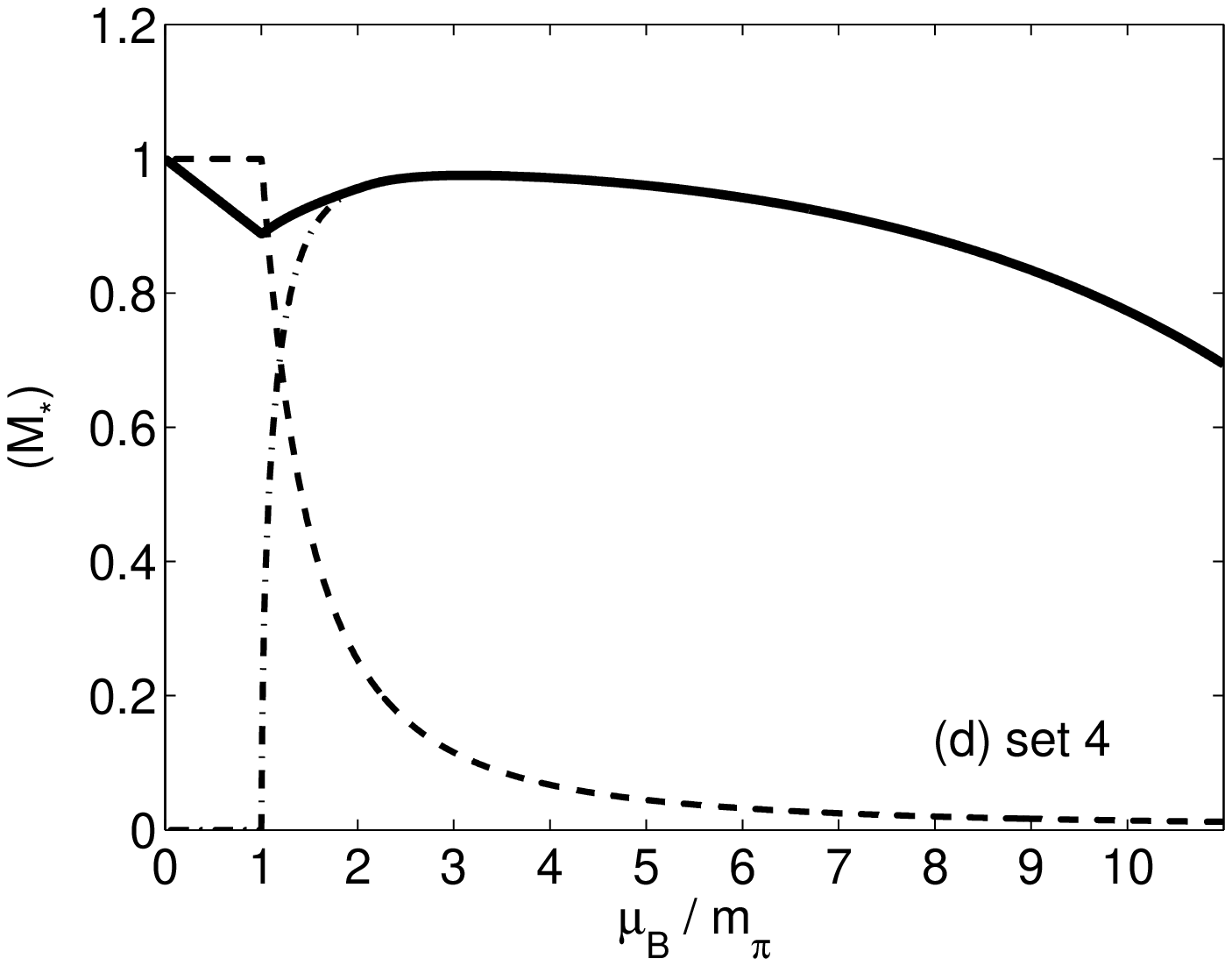}
\caption{The fermionic excitation gap $\Delta_{\text{ex}}$ (in units
of $M_*$) as a function of the baryon chemical potential (in units
of $m_\pi$) for different model parameter sets. The effective quark
mass $M$ and the pairing gap $|\Delta|$ are also shown by dashed and
dash-dotted lines, respectively. \label{fig5}}
\end{center}
\end{figure}

\begin{figure}[!htb]
\begin{center}
\includegraphics[width=6.8cm]{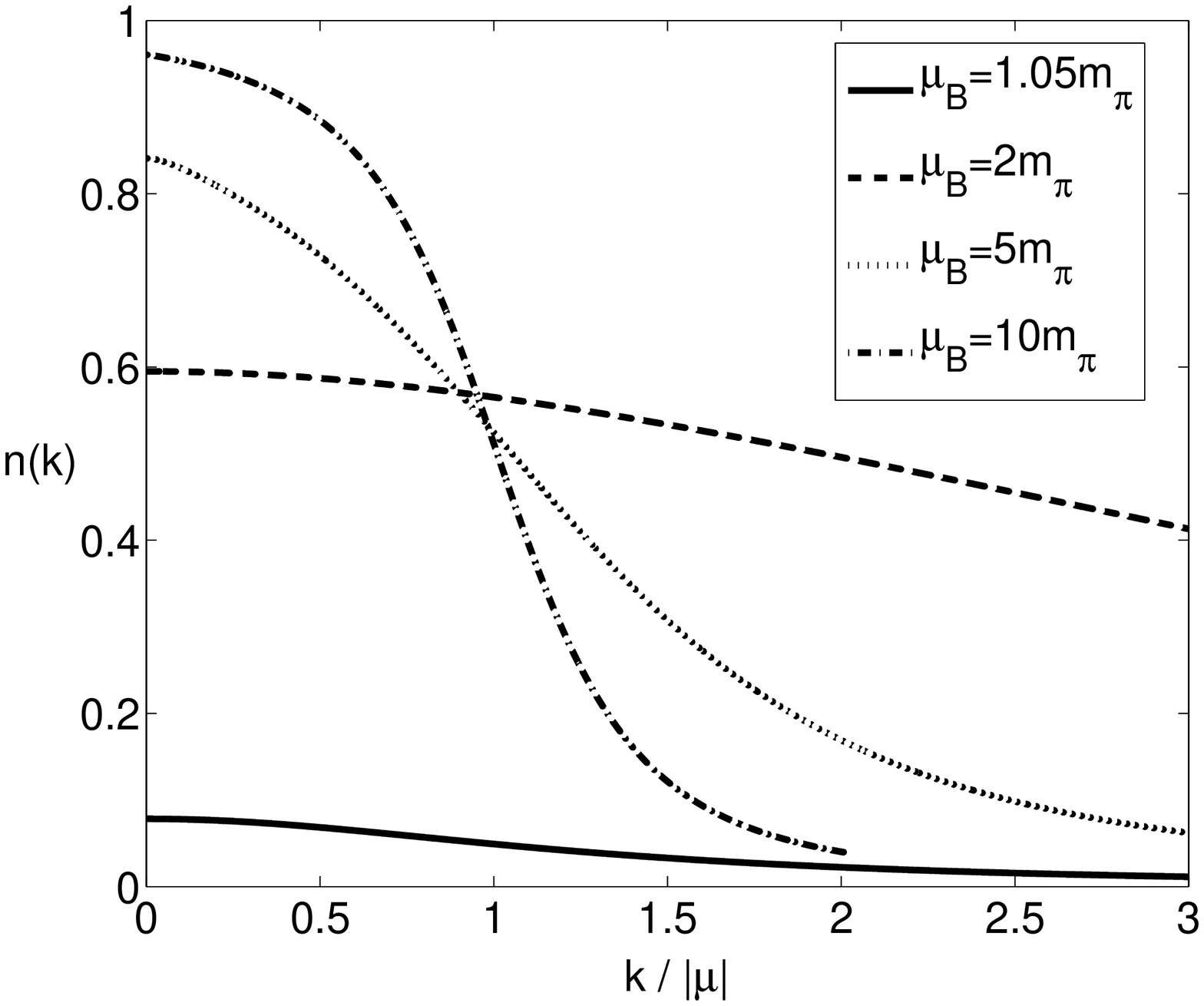}
\includegraphics[width=7.1cm]{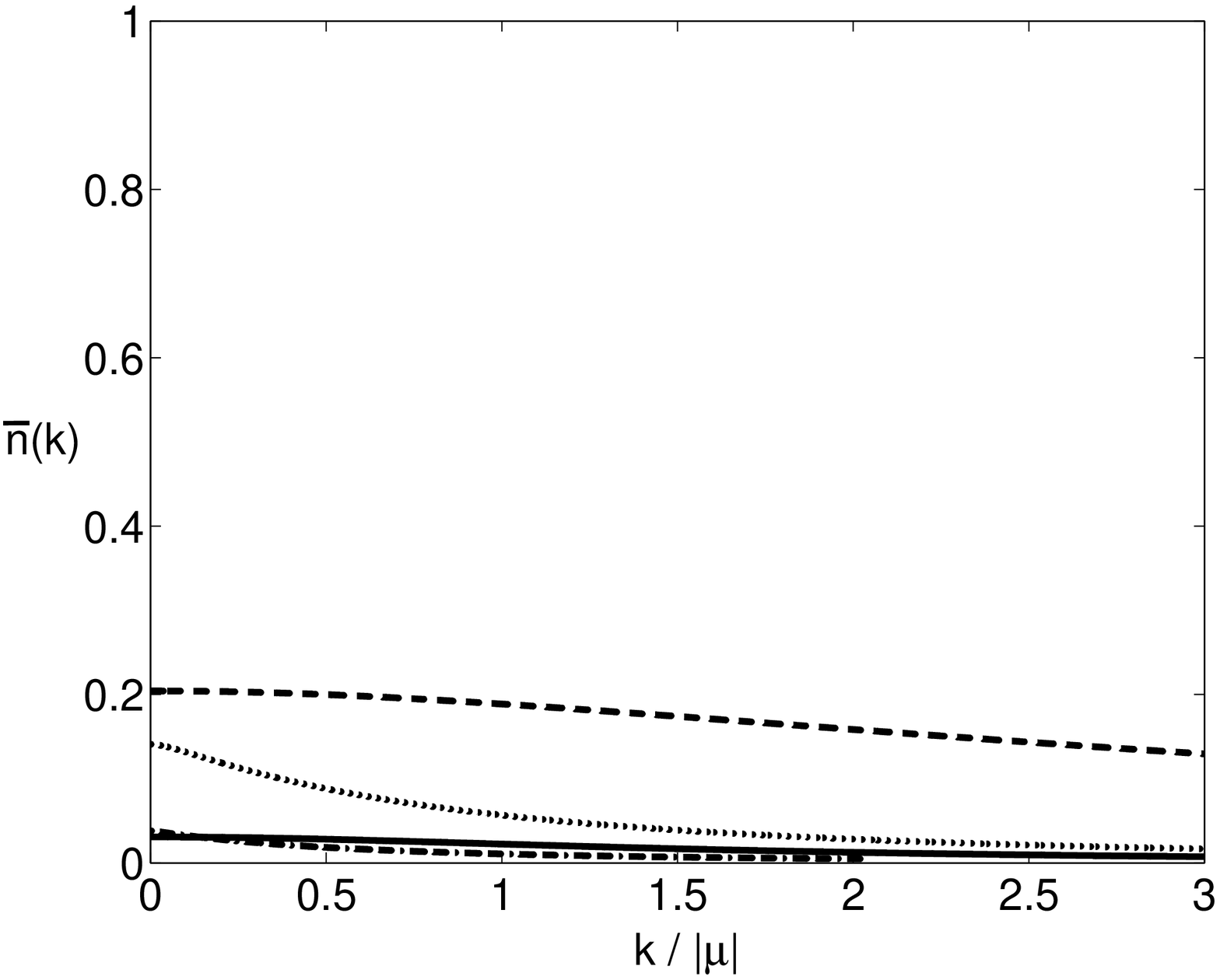}
\caption{The momentum distributions for quarks (upper panel) and
antiquarks (lower panel) for various values of $\mu_{\text B}$. The
momentum is scaled by $|\mu|=|\mu_{\text B}/2-M|$. \label{fig6}}
\end{center}
\end{figure}

\end{widetext}
%

%
\begin{widetext}

\begin{figure}[!htb]
\begin{center}
\includegraphics[width=7cm]{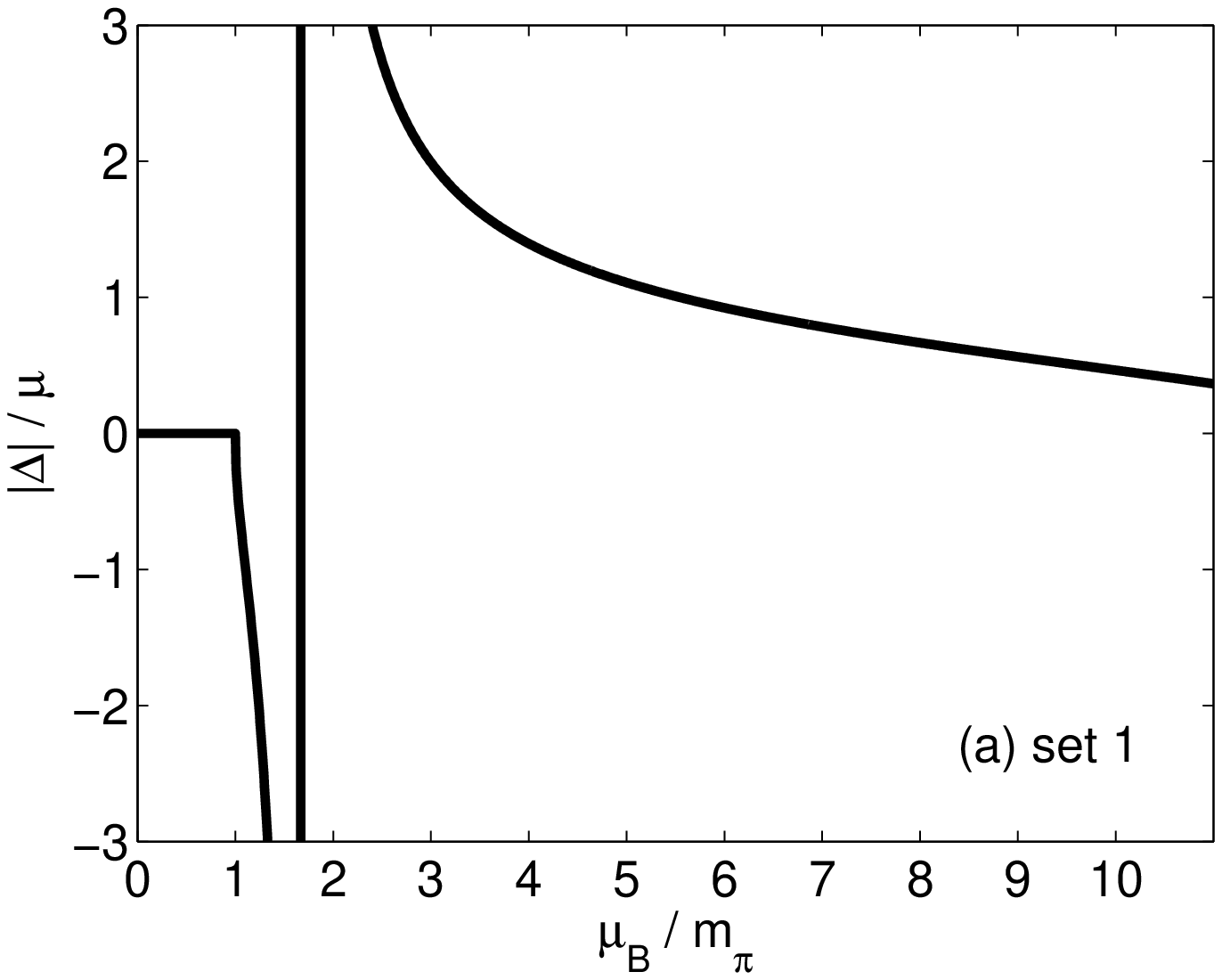}
\includegraphics[width=7cm]{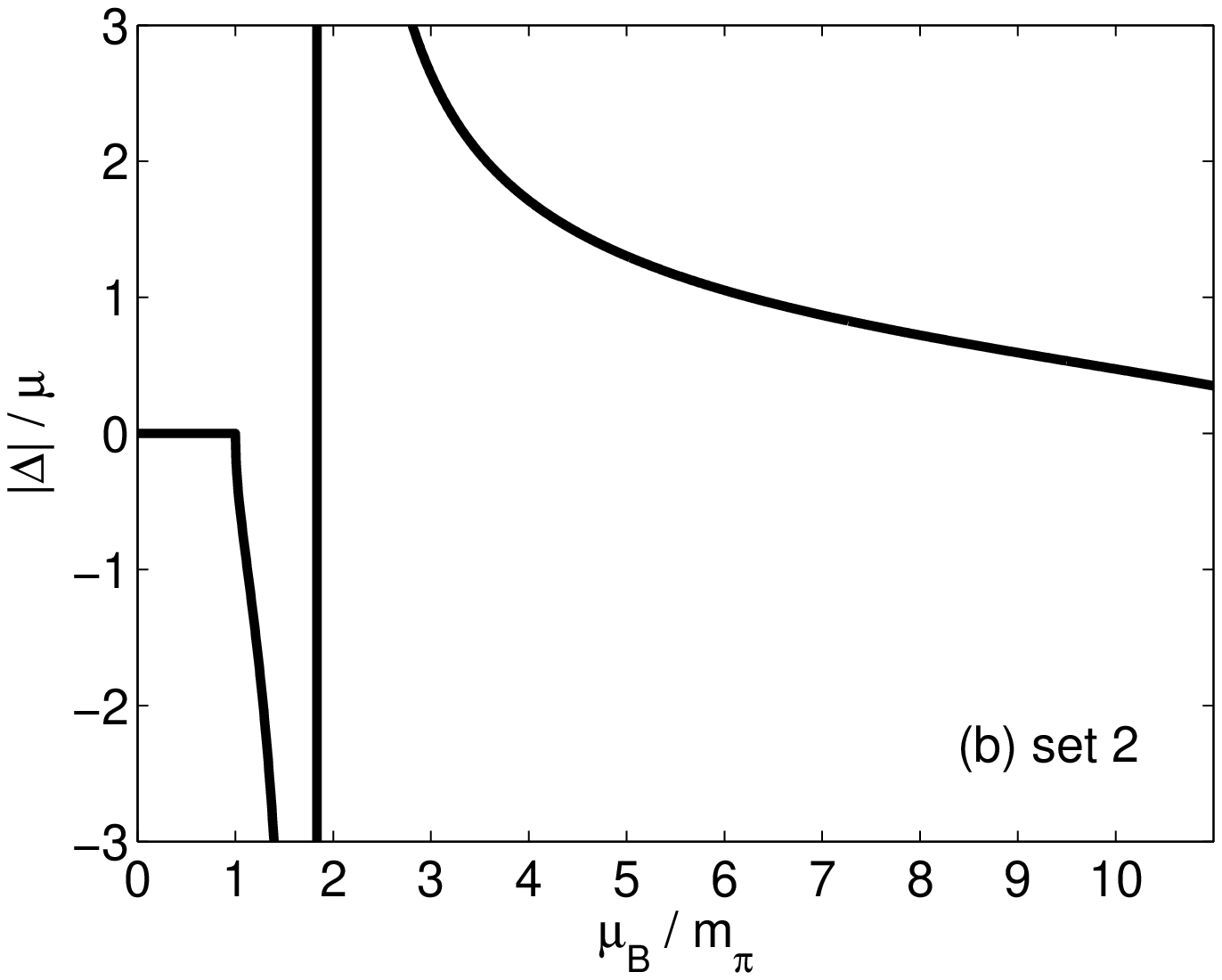}
\includegraphics[width=7cm]{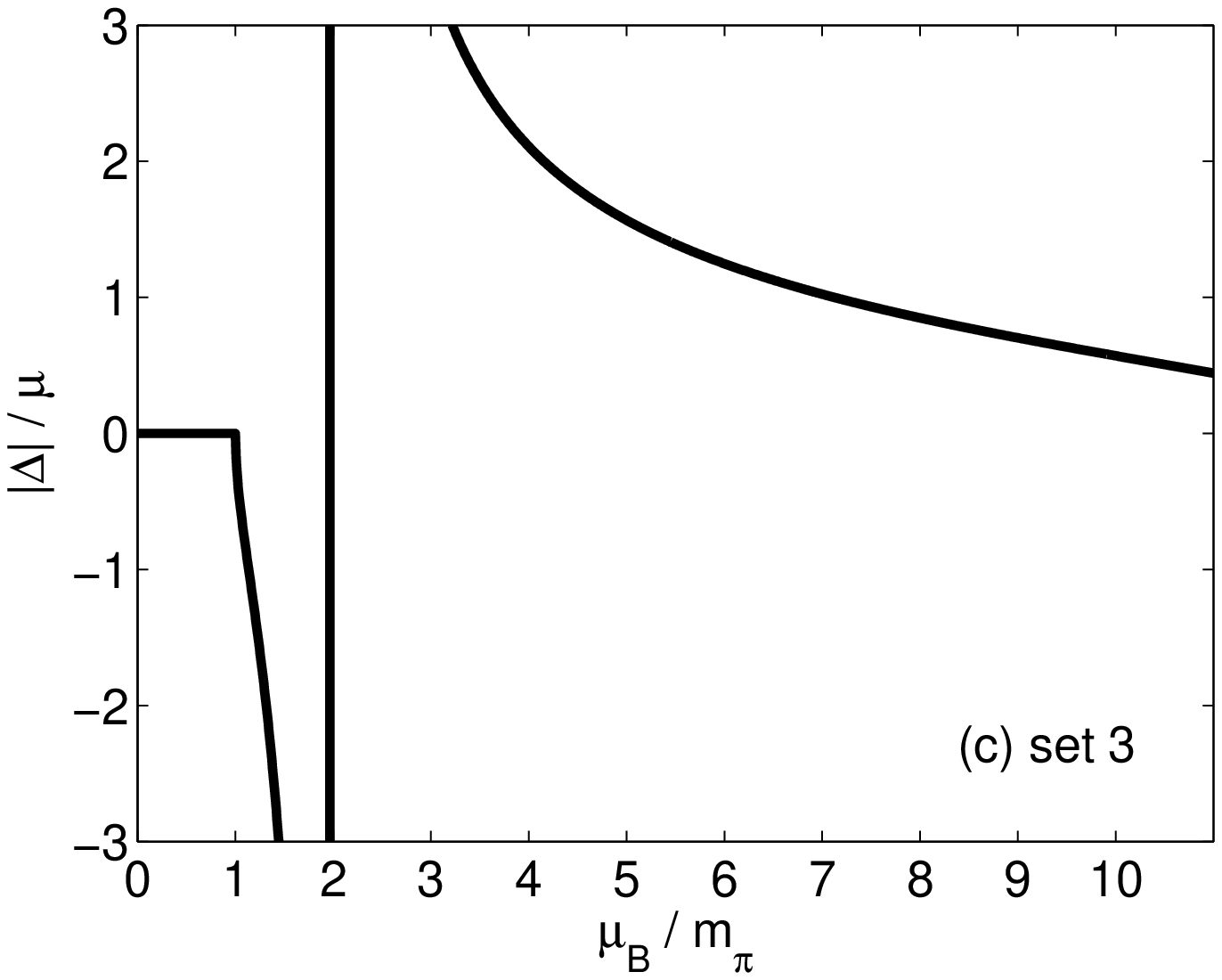}
\includegraphics[width=7cm]{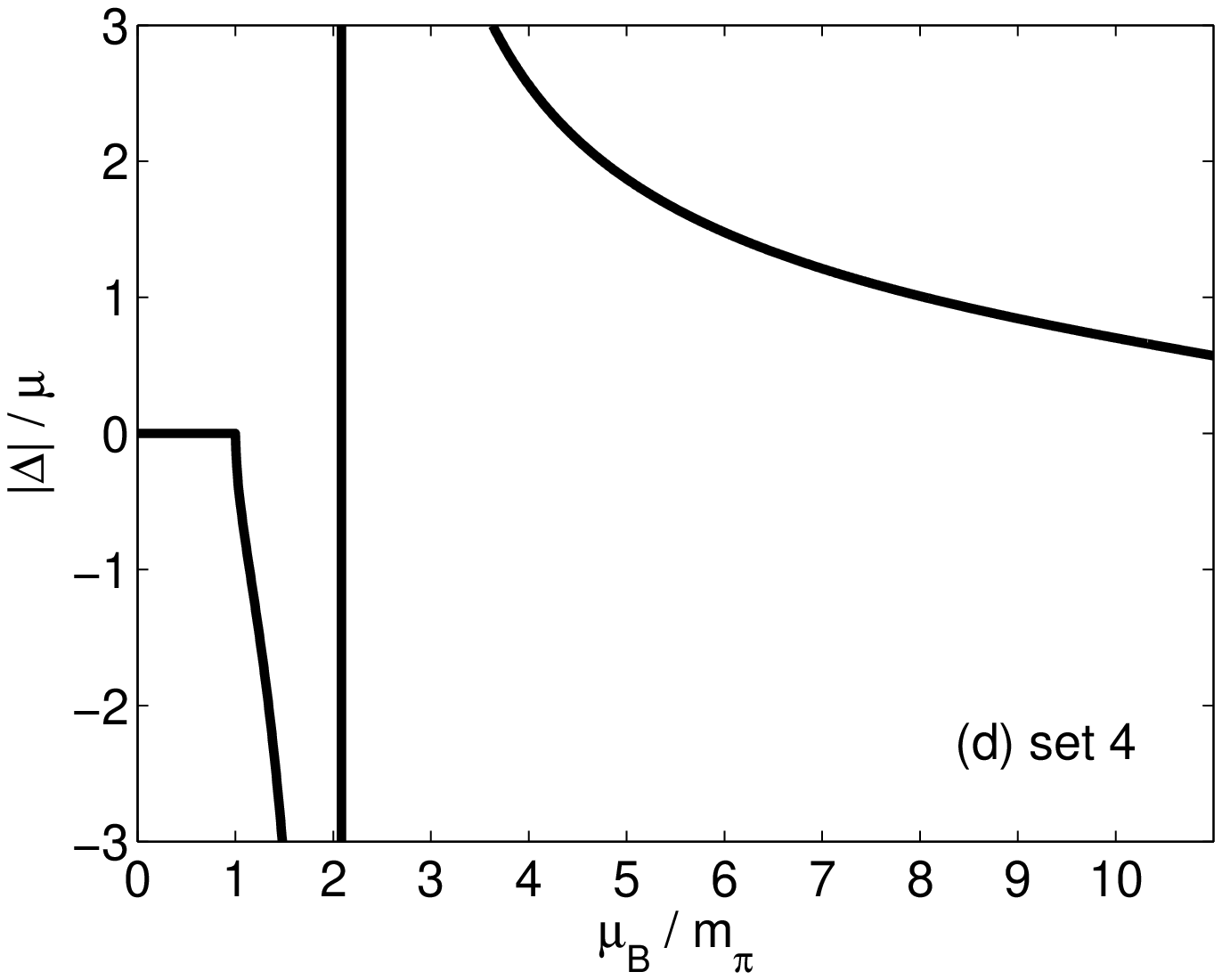}
\caption{The ratio of the pairing gap $|\Delta|$ to the effective
fermionic chemical potential $\mu=\mu_{\text B}/2-M$ as a function
of the baryon chemical potential (in units of $m_\pi$) for different
model parameter sets. The divergent point corresponds to $\mu_{\text
B}=\mu_0$, i.e., the BEC-BCS crossover point. \label{fig7}}
\end{center}
\end{figure}

\end{widetext}
%

\begin{figure}[!htb]
\begin{center}
\includegraphics[width=7cm]{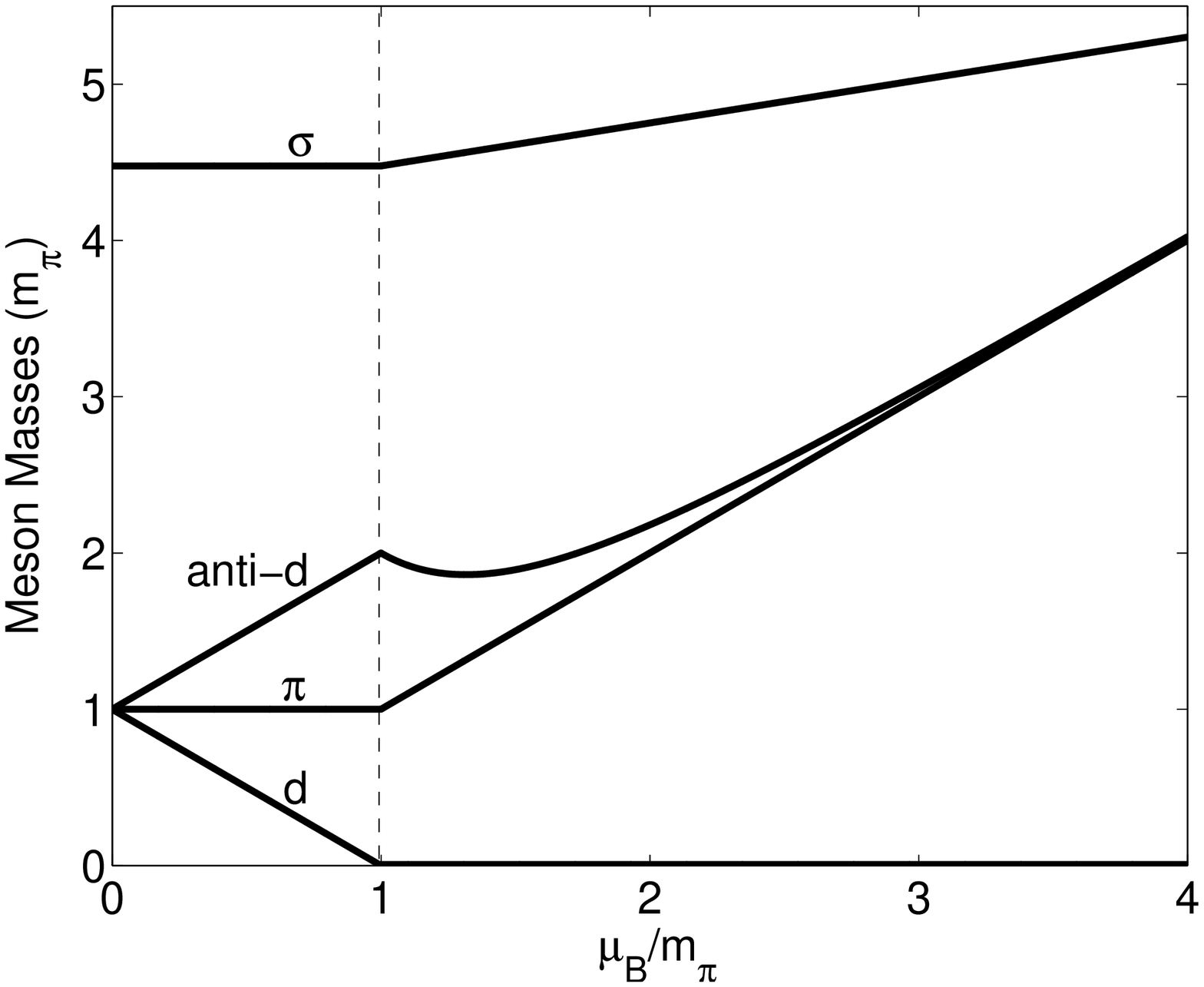}
\caption{The mass spectra of mesons and diquarks (in units of
$m_\pi$) as functions of the baryon chemical potential (in units of
$m_\pi$) for model parameter set 1. For other model parameter sets,
the mass of the heaviest mode is changed but others are almost the
same.\label{fig8}}
\end{center}
\end{figure}

\subsection {Chiral restoration and meson Mott transition}
\label{s4-3}

As in real QCD with two quark flavors, we expect the chiral symmetry
is restored and the spectra of sigma meson and pions become
degenerate at high density \cite{note2}. For the two-flavor case and
with vanishing $m_0$, the residue SU$_{\text L}(2)\otimes$SU$_{\text
R}(2)\otimes$U$_{\text B}(1)$ symmetry group at $\mu_{\text B}\neq0$
is spontaneously broken down to Sp$_{\text L}(2)\otimes$Sp$_{\text
R}(2)$in the superfluid medium with nonzero
$\langle\bar{q}{q}\rangle$ and $\langle qq\rangle$, resulting in one
Goldstone boson. For small nonzero $m_0$, we expect the spectra of
sigma meson and pions become approximately degenerate when the
in-medium chiral condensate $\langle\bar{q}{q}\rangle$ becomes small
enough.

In fact, according to the result
$\langle\bar{q}{q}\rangle_n/\langle\bar{q}{q}\rangle_0\simeq1-n/(2f_\pi^2
m_\pi)$ at low density, we can roughly expect that the chiral
symmetry is approximately restored at $n\sim 2f_\pi^2 m_\pi$. From
the chemical potential dependence of the chiral condensate
$\langle\bar{q}{q}\rangle$ shown in Fig.\ref{fig3}, we find that it
becomes smaller and smaller as the density increases. As a result,
we should have nearly degenerate spectra for the sigma meson and
pions. To show this we need the explicit form of the matrix ${\bf
M}(Q)$ and ${\bf N}(Q)$ given in Appendix \ref{app2}. Since ${\bf
M}_{13},{\bf M}_{32}\propto M\Delta$, at high density where
$\langle\bar{q}{q}\rangle\rightarrow 0$, they can be safely
neglected and the sigma meson decouples from the diquarks. The
propagator of the sigma meson is then given by ${\bf
M}_{33}^{-1}(Q)$. From the explicit form of the polarization
functions $\Pi_\sigma(Q)=\Pi_{33}(Q)$ and $\Pi_\pi(Q)$, we can see
that the inverse propagators of the sigma meson and pions differ
from each other in a term proportional to $M^2$. Thus at high
density their spectra are nearly degenerate, and their masses are
given by the equation
\begin{equation}
1-2G\Pi_\pi(\omega,{\bf 0})=0.
\end{equation}
Using the mean-field gap equation for $\Delta$, we find the solution
is $\omega=\mu_{\text B}$, which means the meson masses are equal to
$\mu_{\text B}$ at large chemical potentials. In Fig.\ref{fig8}, we
show the chemical potential dependence of the meson and diquark mass
spectra determined at zero momentum. We find from the meson spectra
that the chiral symmetry is approximately restored at $\mu_{\text
B}\simeq 3m_\pi$, corresponding to $n\simeq 3.5f_\pi^2m_\pi$. It is
interesting that near the quantum phase transition point $\mu_{\text
B}=m_\pi$ the mixing between the sigma meson and diquarks is very
strong and makes the sigma meson lost its way. Since it is
continuous with the antidiquark mode in the normal phase, it is also
called the ``antidiquark" mode in the superfluid phase
\cite{2CNJL01,tomas}. The ``sigma meson," which is continuous with
the sigma meson in the normal phase, is in fact the Higgs mode
of the BCS superfluid with a mass $2\Delta$ at high density.

%
\begin{widetext}

\begin{figure}[!htb]
\begin{center}
\includegraphics[width=7cm]{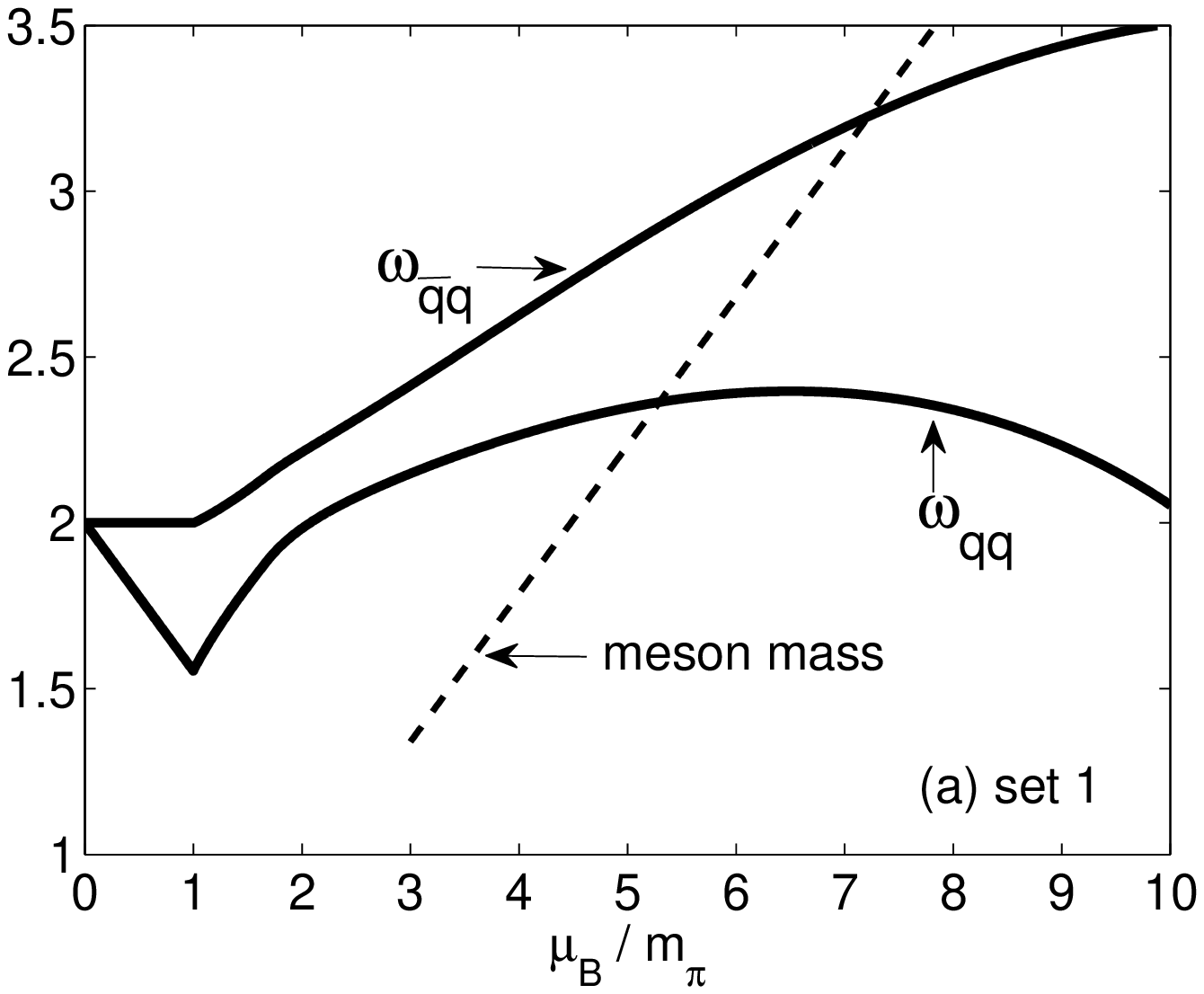}
\includegraphics[width=7cm]{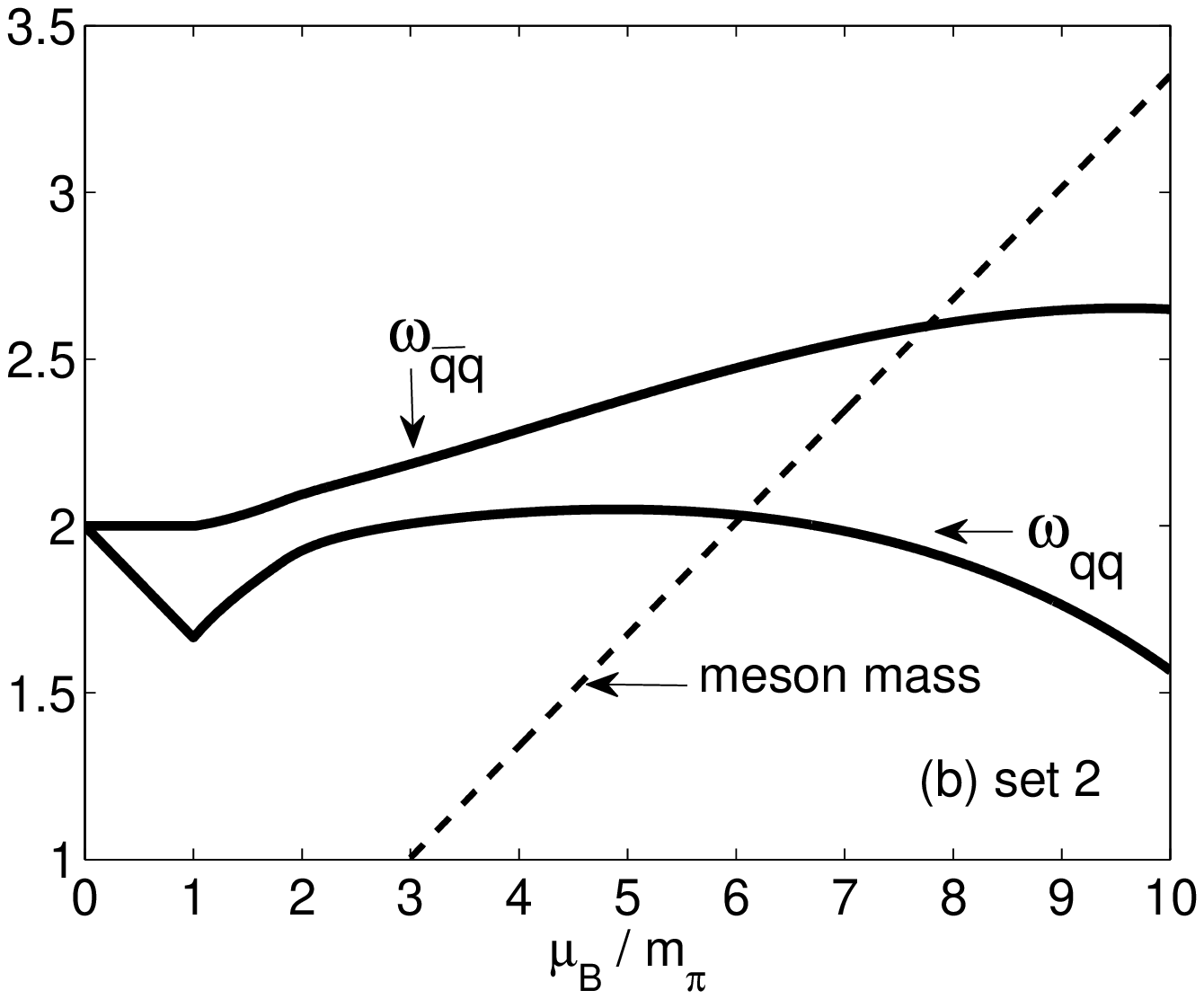}
\includegraphics[width=7cm]{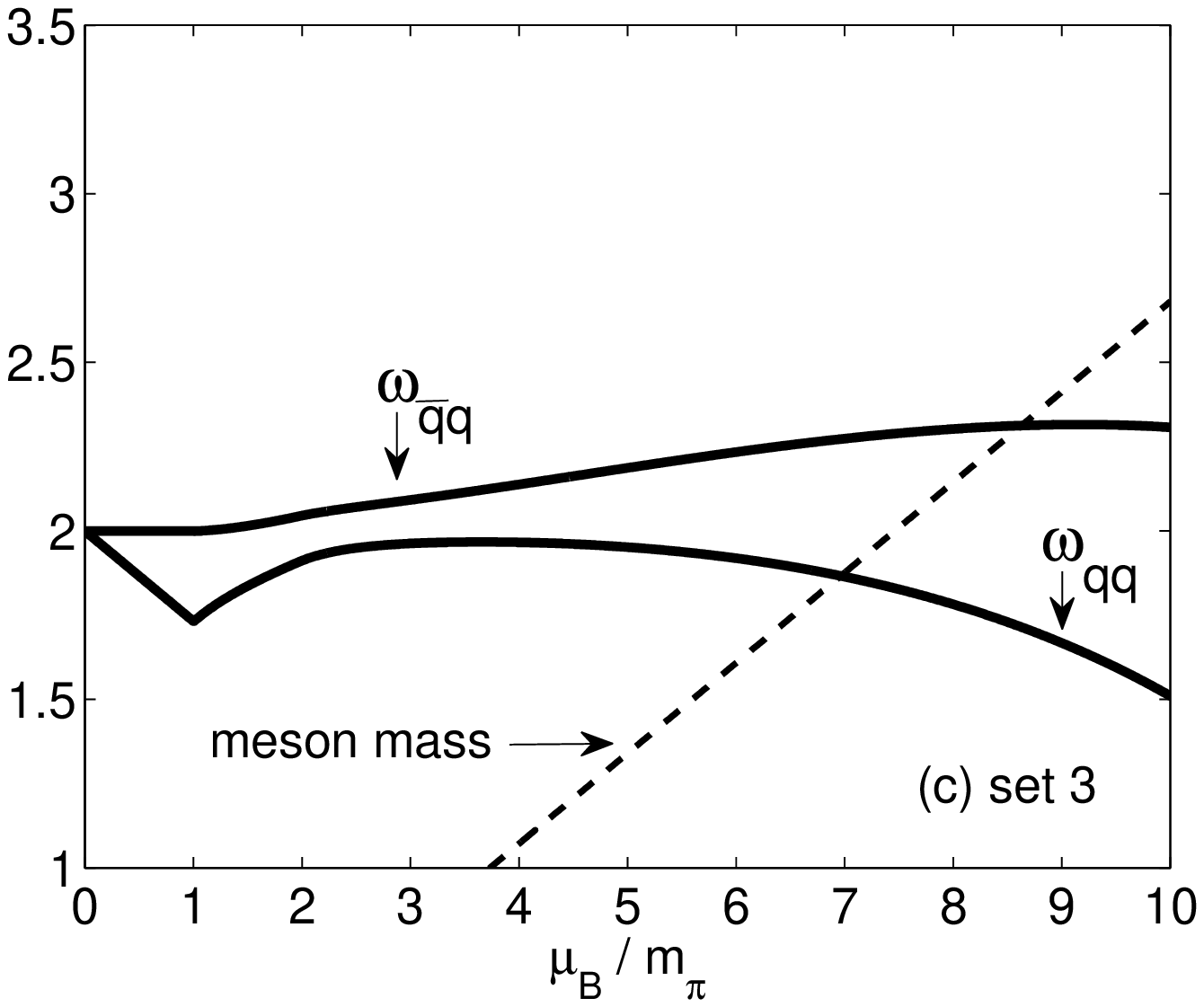}
\includegraphics[width=7cm]{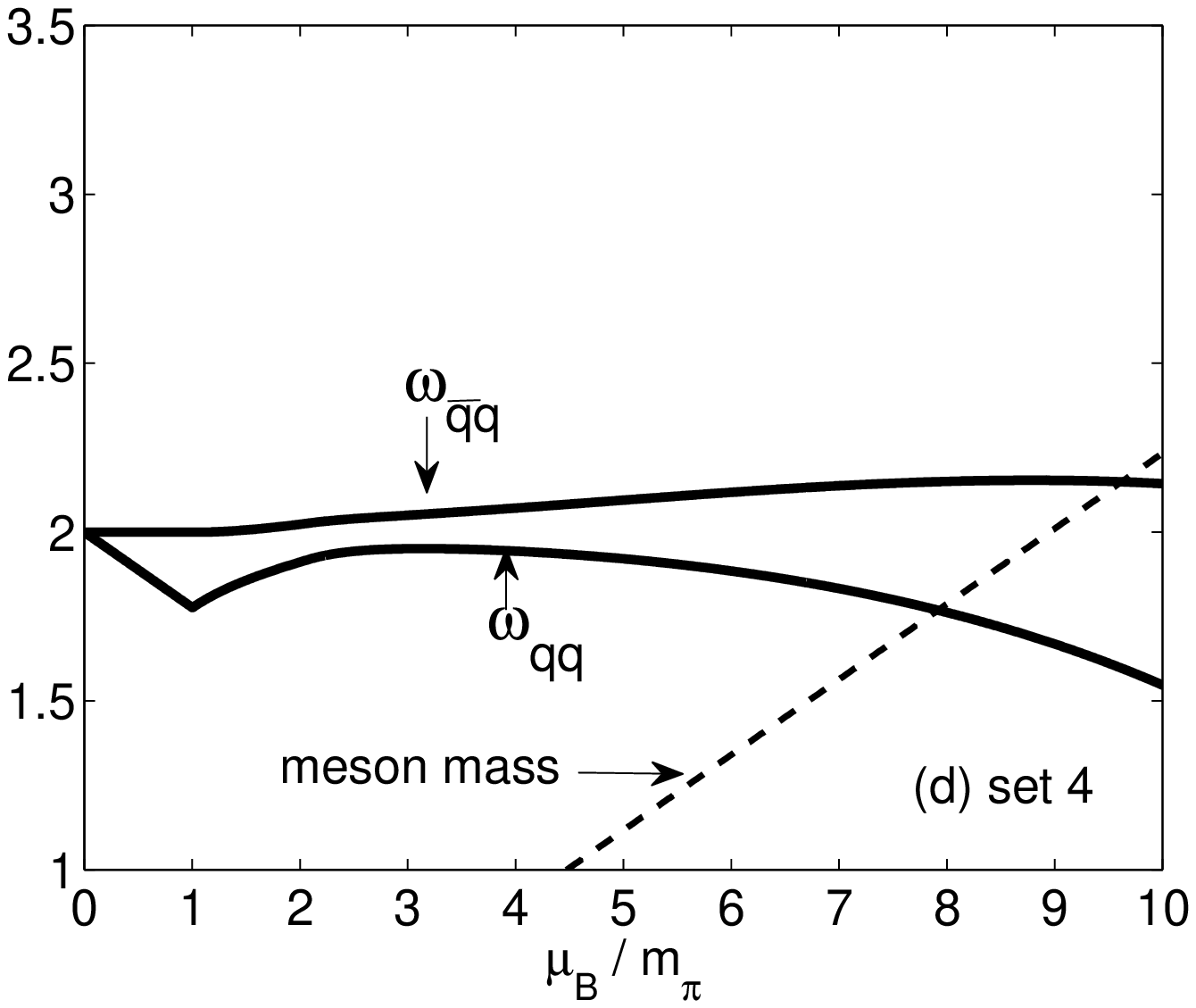}
\caption{The two-particle continua $\omega_{\bar{q}q}$ and
$\omega_{qq}$ (in units of $M_*$) as functions of the baryon
chemical potential (in units of $m_\pi$) for different model
parameter sets. The degenerate mass of pions and sigma meson is
shown by dashed line. \label{fig9}}
\end{center}
\end{figure}

\end{widetext}
%

Even though the deconfinement transition or crossover which
corresponds to the gauge field sector cannot be described in the
NJL model, we can on the other hand study the meson Mott transition
associated with the chiral restoration \cite{mott,precursor,note4}.
The meson Mott transition is defined as the point where the meson
energy becomes larger than the two-particle continuum
$\omega_{\bar{q}q}$ for the decay process $\pi\rightarrow \bar{q}q$
at zero momentum, which means the mesons are no longer bound states.
The two-particle continuum $\omega_{\bar{q}q}$ is different at the
BEC and the BCS sides. From the explicit form of $\Pi_\pi(Q)$, we
find that
\begin{equation}
\omega_{\bar{q}q}=
 \left\{ \begin{array}
{r@{\quad,\quad}l}
 \sqrt{(M-\frac{\mu_{\text
B}}{2})^2+|\Delta|^2}+\sqrt{(M+\frac{\mu_{\text
B}}{2})^2+|\Delta|^2}&
 \mu_{\text B}<\mu_0 \\
 |\Delta|+\sqrt{(M+\frac{\mu_{\text
B}}{2})^2+|\Delta|^2} & \mu_{\text B}>\mu_0.
\end{array}
\right.
\end{equation}
Thus the pions and the sigma meson will undergo a Mott transition
when their masses become larger than the two-particle continuum
$\omega_{\bar{q}q}$, i.e., $\mu_{\text B}>\omega_{\bar{q}q}$. Using
the mean-field results for $\Delta$ and $M$, we can calculate
the two-particle continuum $\omega_{\bar{q}q}$ as a function of
$\mu_{\text B}$, which is shown in Fig.\ref{fig9}. We find that the
Mott transition does occur at a chemical potential $\mu_{\text B}=
\mu_{\text M1}$ which is sensitive to the value of $M_*$. The values
of $\mu_{\text M1}$ for the four model parameter sets are shown in
Table.\ref{mottmu}. For reasonable model parameter sets, the value
of $\mu_{\text M1}$ is in the range $(7-10)m_\pi$. Above this
chemical potential, the mesons are no longer stable bound states and
can decay into quark-antiquark pairs even at zero momentum. We note
that the Mott transition takes place well above the chiral
restoration, in contrast to the pure finite temperature case where
the mesons are dissociated once the chiral symmetry is restored
\cite{mott,precursor}.

\begin{table}[b!]
\begin{center}
\begin{tabular}{|c|c|c|c|c|}
  \hline
  Set & 1 & 2 & 3 & 4 \\
  \hline
  $\mu_{\text M1}$  [$m_\pi$]& 7.22 & 7.76 & 8.63 & 9.62 \\
  \hline
  $\mu_{\text M2}$ [$m_\pi$]& 5.29 & 6.06 & 6.96 & 7.92 \\
  \hline
\end{tabular}
\end{center}
\caption{\small The chemical potentials $\mu_{\text M1}$ and
$\mu_{\text M2}$ (in units of $m_\pi$) for different model parameter
sets.} \label{mottmu}
\end{table}

On the other hand, we find from the explicit forms of the meson
propagators in Appendix \ref{app2} that the decay process
$\pi\rightarrow qq$ is also possible at ${\bf q}\neq0$ (even though
$|{\bf q}|$ is small) due to the presence of superfluidity. Thus, we
have another unusual Mott transition in the superfluid phase. Notice
that this process is not in contradiction to the baryon number
conservation law, since the U$_{\text B}(1)$ baryon number symmetry
is spontaneously broken in the superfluid phase. Quantitatively,
this transition occurs when the meson mass becomes larger than the
two-particle continuum $\omega_{qq}$ for the decay process
$\pi\rightarrow qq$ at ${\bf q}=0^+$. In this case, we have
\begin{equation}
\omega_{qq}=
 \left\{ \begin{array}
{r@{\quad,\quad}l}
 2\sqrt{(M-\frac{\mu_{\text B}}{2})^2+|\Delta|^2}&
 \mu_{\text B}<\mu_0 \\
 2|\Delta| & \mu_{\text B}>\mu_0.
\end{array}
\right.
\end{equation}
The two-particle continuum $\omega_{qq}$ is also shown in
Fig.\ref{fig9}. We find that the unusual Mott transition does occur
at another chemical potential $\mu_{\text B}= \mu_{\text M2}$ which
is also sensitive to the value of $M_*$. The values of $\mu_{\text
M2}$ for the four model parameter sets are also shown in
Table.\ref{mottmu}. For reasonable model parameter sets, this value
is in the range $(5-8)m_\pi$. This process can also occur in the 2SC
phase of quark matter in the $N_c=3$ case \cite{ebert}. In the 2SC
phase, the symmetry breaking pattern is SU$_{\text
c}(3)\otimes$U$_{\text B}(1)\rightarrow$SU$_{\text
c}(2)\otimes\tilde{\text U}_{\text B}(1)$ where the generator of the
residue baryon number symmetry $\tilde{\text U}_{\text B}(1)$ is
$\tilde{\text B}={\text B}-2T_8/\sqrt{3}=\text{diag}(0,0,1)$
corresponding to the unpaired blue quarks. Thus the baryon number
symmetry for the paired red and green quarks are broken and our
results can be applied. To show this explicitly, we write down
the explicit form of the polarization function for pions in the 2SC phase \cite{ebert}
\begin{equation}
\Pi_\pi^{2\text{SC}}(Q)=\Pi_\pi^{\text{2-color}}(Q)+\sum_K\text{Tr}[{\cal G}_0(K)i\gamma_5{\cal G}_0(P)i\gamma_5],
\end{equation}
where ${\cal G}_0(K)$ is the propagator for the unpaired blue quarks.
Here $\Pi_\pi^{\text{2-color}}(Q)$ is given by (\ref{pionpo}) (the effective quarks mass $M$ and the pairing gap $\Delta$ should be
given by the $N_c=3$ case of course) and corresponds to the contribution from the paired red and green sectors. The second
term is the contribution from the unpaired blue quarks. Therefore, the unusual decay process is only available for the paired quarks.

\section { Beyond-Mean-Field Corrections}
\label{s5}
The investigations in Sec. \ref{s3} and \ref{s4} are restricted
in the mean-field approximation, even though the bosonic collective
excitations are studied. In this section, we will include the
Gaussian fluctuations in the thermodynamic potential, and thus
really go beyond the mean field. The scheme of going beyond the mean field
is somewhat like those done in the study of finite temperature
thermodynamics of the NJL model \cite{zhuang01,zhuang02}; however,
in this paper we will focus on the beyond-mean-field corrections at
zero temperature, i.e., the pure quantum fluctuations. We will
first derive the thermodynamic potential beyond the mean field which
is valid at arbitrary chemical potential and temperature, and then
briefly discuss the beyond-mean-field corrections near the quantum
phase transition. The numerical calculations are deferred for future
studies.

\subsection {Thermodynamic potential beyond the mean field}
\label{s5-1}
In the Gaussian approximation, the partition function can be
expressed as
\begin{eqnarray}
Z_{\text{NJL}}\simeq\exp{\left(-{\cal
S}_{\text{eff}}^{(0)}\right)}\int[d\sigma][d\mbox{\boldmath{$\pi$}}][d\phi^\dagger][d\phi]\exp\Big(-{\cal
S}_{\text{eff}}^{(2)}\Big).
\end{eqnarray}
Integrating out the Gaussian fluctuations, we can express the total
thermodynamic potential as
\begin{equation}
\Omega(T,\mu_{\text B})=\Omega_0(T,\mu_{\text
B})+\Omega_{\text{fl}}(T,\mu_{\text B}),
\end{equation}
where the contribution from the Gaussian fluctuations can be written
as
\begin{eqnarray}
\Omega_{\text{fl}}=\frac{1}{2}\sum_Q\big[\ln\det{\bf
M}(Q)+\ln\det{\bf N}(Q)\big].
\end{eqnarray}

However, there is a problem with the above expression, since it is
actually ill-defined: the sum over the boson Matsubara frequency is
divergent and we need appropriate convergent factors to make it
meaningful. In the simpler case without superfluidity, the
convergent factor is simply given by $e^{i\nu_m 0^+}$
\cite{zhuang01,zhuang02}. In our case, the situation is somewhat
different due to the introduction of the Nambu-Gor'kov spinors. Keep
in mind that in the equal time limit, there are additional factors
$e^{i\omega_n0^+}$ for ${\cal G}_{11}(K)$ and $e^{-i\omega_n0^+}$
for ${\cal G}_{22}(K)$. Therefore, to get the proper convergent
factors for $\Omega_{\text{fl}}$, we should keep these factors when
we make the sum over the fermion Matsubara frequency $\omega_n$ in
evaluating the polarization functions $\Pi_{\text{ij}}(Q)$ and
$\Pi_{\pi}(Q)$.

The problem in the expression of $\Omega_{\text{fl}}$ is thus from
the opposite convergent factors for ${\bf M}_{11}$ and ${\bf
M}_{22}$. From the above arguments, we find that there is a factor
$e^{i\nu_m 0^+}$ for ${\bf M}_{11}$ and $e^{-i\nu_m 0^+}$ for ${\bf
M}_{22}$. Keep in mind that the Matsubara sum $\sum_m$ is converted
to a standard contour integral ($i\nu_m\rightarrow z$). The
convergence for $z\rightarrow+\infty$ is automatically guaranteed by
the Bose distribution function $b(z)=1/(e^{\beta z}-1)$, we thus
should treat only the problem for $z\rightarrow-\infty$. To this
end, we write the first term of $\Omega_{\text{fl}}$ as
\begin{eqnarray}
&&\sum_Q\ln\det{\bf M}(Q)=\sum_Q\bigg[\ln{\bf M}_{11}e^{i\nu_m
0^+}+\ln{\bf M}_{22}e^{-i\nu_m
0^+}\nonumber\\
&&\ \ \ \ \ \ \ \ \ \ \ \ \ \ \ \ \ \ \ \ \ \ \ \ \ \
+\ln\left(\frac{\det{\bf M}}{{\bf M}_{11}{\bf
M}_{22}}\right)e^{i\nu_m 0^+}\bigg].
\end{eqnarray}
Using the fact that ${\bf M}_{22}(Q)={\bf M}_{11}(-Q)$, we obtain
\begin{eqnarray}
\sum_Q\ln\det{\bf M}(Q)=\sum_Q\ln\left[\frac{{\bf M}_{11}(Q)}{{\bf
M}_{22}(Q)}\det{\bf M}(Q)\right]e^{i\nu_m 0^+}.
\end{eqnarray}
Therefore, the well-defined form of $\Omega_{\text{fl}}$ is given by
the above formula together with the other term $\sum_Q\ln\det{\bf
N}(Q)$ associated with a factor $e^{i\nu_m 0^+}$.

The Matsubara sum can be written as the contour integral via the
theorem $\sum_m g(i\nu_m)=\oint_{\text C}dz/(2\pi i)b(z)g(z)$, where
${\text C}$ runs on either side of the imaginary $z$ axis, enclosing
it counterclockwise. Distorting the contour to run above and below
the real axis, we obtain
\begin{eqnarray}
&&\Omega_{\text{fl}}=\sum_{\bf
q}\int_{-\infty}^{+\infty}\frac{d\omega}{2\pi}b(\omega)\big[\delta_{\text
M}(\omega,{\bf q})+\delta_{11}(\omega,{\bf
q})\nonumber\\
&&\ \ \ \ \ \ -\delta_{22}(\omega,{\bf q})+3\delta_\pi(\omega,{\bf
q})\big],
\end{eqnarray}
where the scattering phases are defined as
\begin{eqnarray}
&&\delta_{\text M}(\omega,{\bf q})=\text{Im}\ln\det{\bf
M}(\omega+i0^+,{\bf
q}),\nonumber\\
&&\delta_{11}(\omega,{\bf q})=\text{Im}\ln{\bf
M}_{11}(\omega+i0^+,{\bf q}),\nonumber\\
&&\delta_{22}(\omega,{\bf q})=\text{Im}\ln{\bf
M}_{22}(\omega+i0^+,{\bf q}),\nonumber\\
&&\delta_\pi(\omega,{\bf
q})=\text{Im}\ln\left[(2G)^{-1}+\Pi_\pi(\omega+i0^+,{\bf q})\right].
\end{eqnarray}
Keep in mind the pressure of the vacuum should be zero,  the
physical thermodynamic potential at finite temperature and chemical
potential should be defined as
\begin{equation}
\Omega_{\text{phy}}(T,\mu_{\text B})=\Omega(T,\mu_{\text
B})-\Omega(0,0).
\end{equation}

\subsection{Thermodynamic consistency of the vacuum}
\label{s5-2}
As we have shown in the mean-field theory, at $T=0$, the vacuum
state is restricted in the region $|\mu_{\text B}|<m_\pi$. In this
region, all thermodynamic quantities should keep zero, no matter how
large the value of $\mu_{\text B}$ is. While this should be an
obvious physical conclusion, it is important to check whether our
beyond-mean-field theory satisfies this condition.

Notice that the physical thermodynamic potential is defined as
$\Omega_{\text{phy}}(\mu_{\text B})=\Omega(\mu_{\text
B})-\Omega(0)$, we therefore should prove that the thermodynamic
potential $\Omega(\mu_{\text B})$ keeps a constant in the region
$|\mu_{\text B}|<m_\pi$. For the mean-field part $\Omega_0$, the
proof is quite easy. Because of the fact that $M_*>m_\pi/2$, the
solution for $M$ is always given by $M=M_*$. Thus $\Omega_0$ keeps
its value at $\mu_{\text B}=0$ in the region $|\mu_{\text
B}|<m_\pi$.

Now we turn to the complicated part $\Omega_{\text{fl}}$. Since
$\Delta=0$, all the off-diagonal elements of ${\bf M}$ vanishes, and
$\Omega_{\text{fl}}$ is reduced to
\begin{eqnarray}
\Omega_{\text{fl}}&=&\frac{1}{2}\sum_Q\ln\left[\frac{1}{2G}+\Pi_{\sigma}(Q)\right]e^{i\nu_m
0^+}\nonumber\\
&+&\frac{3}{2}\sum_Q\ln\left[\frac{1}{2G}+\Pi_{\pi}(Q)\right]e^{i\nu_m 0^+}\nonumber\\
&+&\sum_Q\ln\left[\frac{1}{4G}+\Pi_{\text d}(Q)\right]e^{i\nu_m
0^+},
\end{eqnarray}
where $\Pi_\sigma(Q)=\Pi_{33}(Q)$ and we should set $\Delta=0$ and
$M=M_*$ in evaluating the polarization functions. First, we can
easily show that the contributions from the sigma meson and pions do
not have explicit $\mu_{\text B}$ dependence and thus keep the same
values as those at $\mu_{\text B}=0$. In fact, since the effective
quark mass $M$ keeps its vacuum value $M_*$ guaranteed by the mean-field
part, all the $\mu_{\text B}$ dependence in
$\Pi_{\sigma,\pi}(Q)$ is included in the Fermi distribution
functions $f(E\pm\mu_{\text B}/2)$. Since $M_*>\mu_{\text B}/2$,
they vanish automatically at $T=0$. In fact, from the explicit
expressions for $\Pi_{\sigma,\pi}(Q)$ in Appendix \ref{app2}, we can
check that there is no $\mu_{\text B}$ independence in
$\Pi_{\sigma,\pi}(Q)$.

The diquark contribution, however, has an explicit $\mu_{\text
B}$ dependence through the combination $i\nu_m+\mu_{\text B}$ in the
polarization function $\Pi_{\text d}(Q)$. The diquark contribution
(at $T=0$) can be written as
\begin{eqnarray}
&&\Omega_{\text{d}}=-\sum_{\bf
q}\int_{-\infty}^{0}\frac{d\omega}{\pi}\delta_{\text
d}(\omega,{\bf q}),\nonumber\\
&&\delta_{\text d}(\omega,{\bf
q})=\text{Im}\ln\left[(4G)^{-1}+\Pi_{\text d}(\omega+i0^+,{\bf
q})\right].
\end{eqnarray}
Making a shift $\omega\rightarrow\omega-\mu_{\text B}$, and noticing
that fact $\Pi_{\text d}(\omega-\mu_{\text B},{\bf
q})=\Pi_\pi(\omega,{\bf q})/2$, we obtain
\begin{eqnarray}
\Omega_{\text{d}}=-\sum_{\bf q}\int_{-\infty}^{-\mu_{\text
B}}\frac{d\omega}{\pi}\delta_{\pi}(\omega,{\bf q}).
\end{eqnarray}
To show the above quantity is in fact $\mu_{\text B}$ independent,
we separate it into a pole part and a continuum part. There is a
well-defined two-particle continuum $E_c({\bf q})$ for pions at
arbitrary momentum ${\bf q}$,
\begin{eqnarray}
E_c({\bf q})=\text{min}_{\bf k}\left(E_{\bf k}^*+E_{\bf
k+q}^*\right).
\end{eqnarray}
The pion propagator has two symmetric poles $\pm\omega_\pi({\bf q})$
when ${\bf q}$ satisfies $\omega_\pi({\bf q})<E_c({\bf q})$. Thus in
the region $|\omega|<E_c({\bf q})$, the scattering phase
$\delta_{\pi}$ can be analytically evaluated as
\begin{eqnarray}
\delta_{\pi}(\omega,{\bf
q})=\pi\left[\Theta\left(-\omega-\omega_\pi({\bf
q})\right)-\Theta\left(\omega-\omega_\pi({\bf q})\right)\right].
\end{eqnarray}
Since $E_c({\bf q})>\omega_\pi({\bf q})>m_\pi>\mu_{\text B}$, the
thermodynamic potential $\Omega_{\text d}$ can be separated as
\begin{eqnarray}
\Omega_{\text{d}}=\sum_{\bf q}\left[\omega_\pi({\bf q})-E_c({\bf
q})\right]-\sum_{\bf q}\int_{-\infty}^{-E_c({\bf
q})}\frac{d\omega}{\pi}\delta_{\pi}(\omega,{\bf q}),
\end{eqnarray}
which is indeed $\mu_{\text B}$ independent. Notice that in the
first term the integral over ${\bf q}$ is restricted in the region
$|{\bf q}|<q_c$ where $q_c$ is defined as
$\omega_\pi(q_c)=E_c(q_c)$.

In conclusion, we have shown that the thermodynamic potential
$\Omega$ in the Gaussian approximation keeps a constant in the
vacuum state, i.e., at $|\mu_{\text B}|<m_\pi$ and at $T=0$. All
other thermodynamic quantities such as the baryon number density
keep zero in the vacuum. The subtraction term $\Omega(0,0)$ in the
Gaussian approximation can be expressed as
\begin{eqnarray}
&&\Omega(0,0)=\Omega_{\text{vac}}(M_*)+\frac{5}{2}\sum_{\bf
q}\left[\omega_\pi({\bf
q})-E_c({\bf q})\right]\nonumber\\
&&\ \ \ -\sum_{\bf q}\int_{-\infty}^{-E_c({\bf
q})}\frac{d\omega}{2\pi}\left[\delta_\sigma(\omega,{\bf
q})+5\delta_{\pi}(\omega,{\bf q})\right].
\end{eqnarray}

\subsection {Quantum corrections near the phase transition}
\label{s5-3}
Now we consider the beyond-mean-field corrections near the quantum
phase transition point $\mu_{\text B}=m_\pi$. Notice that the
effective quark mass $M$ and the diquark condensate $\Delta$ are
determined at the mean-field level, and the beyond-mean-field
corrections are possible only through the equations of state.

Formally, the Gaussian contribution to the thermodynamic
potential $\Omega_{\text{fl}}$ is a function of $\mu_{\text B},M$
and $y=|\Delta|^2$, i.e.,
$\Omega_{\text{fl}}=\Omega_{\text{fl}}(\mu_{\text B},y,M)$. In the
superfluid phase, the total baryon density including the Gaussian
contribution can be evaluated as
\begin{eqnarray}
n(\mu_{\text B})=n_0(\mu_{\text B})+n_{\text{fl}}(\mu_{\text B}),
\end{eqnarray}
where the mean-field part is simply given by $n_0(\mu_{\text
B})=-\partial\Omega_0/\partial\mu_{\text B}$ and the Gaussian
contribution can be expressed as
\begin{eqnarray}
n_{\text{fl}}(\mu_{\text B})=-\frac{\partial
\Omega_{\text{fl}}}{\partial \mu_{\text B}}-\frac{\partial
\Omega_{\text{fl}}}{\partial y}\frac{dy}{d\mu_{\text
B}}-\frac{\partial \Omega_{\text{fl}}}{\partial
M}\frac{dM}{d\mu_{\text B}}.
\end{eqnarray}
The physical values of $M$ and $|\Delta|^2$ should be determined by
their mean-field gap equations. In fact, from the gap equations
$\partial\Omega_0/\partial M=0$ and $\partial\Omega_0/\partial y=0$,
we obtain
\begin{eqnarray}
&&\frac{\partial^2 \Omega_0}{\partial \mu_{\text B}\partial
M}+\frac{\partial^2\Omega_0}{\partial y\partial
M}\frac{dy}{d\mu_{\text B}}+\frac{\partial^2 \Omega_0}{\partial
M^2}\frac{dM}{d\mu_{\text B}}=0,\nonumber\\
&&\frac{\partial^2 \Omega_0}{\partial \mu_{\text B}\partial
y}+\frac{\partial^2\Omega_0}{\partial y^2}\frac{dy}{d\mu_{\text
B}}+\frac{\partial^2 \Omega_0}{\partial M\partial
y}\frac{dM}{d\mu_{\text B}}=0.
\end{eqnarray}
Thus, we can obtain the derivatives $dM/d\mu_{\text B}$ and
$dy/d\mu_{\text B}$ analytically. Finally, $n_{\text{fl}}(\mu_{\text
B})$ is a continuous function of $\mu_{\text B}$ guaranteed by the
properties of second order phase transition, and we have
$n_{\text{fl}}(m_\pi)=0$.

Next we focus on the beyond-mean-field corrections near the quantum
phase transition. Since the diquark condensate $\Delta$ is
vanishingly small, we can expand the Gaussian part
$\Omega_{\text{fl}}$ in powers of $|\Delta|^2$. Notice that
$\mu_{\text B}$ and $M$ can be evaluated as functions of
$|\Delta|^2$ from the Ginzburg-Landau potential and mean-field gap
equations. Thus to order $O(|\Delta|^2)$, the expansion takes the
form
\begin{eqnarray}
\Omega_{\text{fl}}\simeq \eta|\Delta|^2,
\end{eqnarray}
where the expansion coefficient $\eta$ is defined as
\begin{eqnarray}
\eta=\left(\frac{\partial\Omega_{\text{fl}}}{\partial
y}+\frac{\partial\Omega_{\text{fl}}}{\partial\mu_{\text
B}}\frac{d\mu_{\text
B}}{dy}+\frac{\partial\Omega_{\text{fl}}}{\partial
M}\frac{dM}{dy}\right)\Bigg|_{\mu_{\text B}=m_\pi,y=0,M=M_*}.
\end{eqnarray}
Using the definition of $n_{\text{fl}}$, we find that $\eta$ can be
related to $n_{\text{fl}}$ by
\begin{eqnarray}
\eta=n_{\text{fl}}(m_\pi)\frac{d\mu_{\text B}}{dy}\bigg|_{y=0}.
\end{eqnarray}
Thus, the coefficient $\eta$ vanishes, and the leading order of the
expansion should be $O(|\Delta|^4)$.

As shown above, to leading order, the expansion of
$\Omega_{\text{fl}}$ can be formally expressed as
\begin{eqnarray}
\Omega_{\text{fl}}\simeq-\frac{\zeta}{2}\beta|\Delta|^4.
\end{eqnarray}
The method to derive the exact expression of the numerical factor
$\zeta$ is shown in Appendix \ref{app3}. Notice that the factor
$\zeta$ is in fact $\mu_{\text B}$ independent,  thus the total
baryon density to leading order is
\begin{eqnarray}
n=n_0+\zeta\beta|\Delta|^2\frac{d|\Delta|^2}{d\mu_{\text
B}}\bigg|_{\mu_{\text B}=m_\pi}.
\end{eqnarray}
Near the quantum phase transition point, the mean-field contribution
is $n_0=|\psiup_0|^2=2m_\pi {\cal J}|\Delta|^2$ from the
Gross-Pitaevskii free energy. The last term can be evaluated using
the analytical result
\begin{eqnarray}
|\psiup_0|^2=\frac{\mu_{\text d}}{g_0}\Longrightarrow
|\Delta|^2=\frac{2m_\pi{\cal J}}{\beta}\mu_{\text d},
\end{eqnarray}
which is in fact the solution of the mean-field gap equations.
Therefore, to leading order, the total baryon density reads
\begin{eqnarray}
n=(1+\zeta)2m_\pi {\cal J}|\Delta|^2.
\end{eqnarray}
On the other hand, the total pressure $P$ can be expressed as
\begin{eqnarray}
P=(1+\zeta)\frac{\beta}{2}|\Delta|^4.
\end{eqnarray}
Thus we find that the leading order quantum corrections are totally
included in the numerical factor $\zeta$. Setting $\zeta=0$, we
recover the mean-field results obtained in Sec. \ref{s3}.

Including the quantum fluctuations, the equations of state shown in
(\ref{eos}) are modified to be
\begin{eqnarray}
&&P(n)=\frac{1}{1+\zeta}\frac{2\pi
a_{\text{dd}}}{m_\pi}n^2,\nonumber\\
&&\mu_{\text B}(n)=m_\pi+\frac{1}{1+\zeta}\frac{4\pi
a_{\text{dd}}}{m_\pi}n.
\end{eqnarray}
This means, to leading order, the effect of quantum fluctuations is
giving a correction to the diquark-diquark scattering length. The
renormalized scattering length is
\begin{eqnarray}
a_{\text{dd}}^\prime=\frac{a_{\text{dd}}}{1+\zeta}.
\end{eqnarray}
Generally, we have $\zeta>0$ and the renormalized scattering length
is smaller than the mean-field result.

An exact calculation of the numerical factor $\zeta$ can be
performed using the method shown in Appendix \ref{app3}. However,
this needs huge numerical power and we defer it to future work
\cite{he}. In this paper we will give an analytical estimation of
$\zeta$ based on the fact that the quantum fluctuations are
dominated by the gapless Goldstone mode. To this end, we approximate
the Gaussian contribution $\Omega_{\text{fl}}$ as
\begin{eqnarray}
&&\Omega_{\text{fl}}\simeq\frac{1}{2}\sum_Q\ln\bigg[{\cal D}_{\text
d}^{-1}(Q){\cal D}_{\text d}^{-1}(-Q)+3\beta^2|\Delta|^4\nonumber\\
&&\ \ \ \ \ \ +2\beta|\Delta|^2\left({\cal D}_{\text
d}^{-1}(Q)+{\cal D}_{\text d}^{-1}(-Q)\right)\bigg],
\end{eqnarray}
where ${\cal D}_{\text d}^{-1}(Q)$ is given by (\ref{dipro}) and can
be approximated by (\ref{diproa}). Subtracting the value of
$\Omega_{\text{fl}}$ at $\mu_{\text B}=m_\pi$ with $\Delta=0$, and
using the result $\mu_{\text B}=m_\pi+g_0|\psiup_0|^2$ from the
Gross-Pitaevskii equation, we find that $\zeta$ can be evaluated as
\begin{eqnarray}
\zeta=\frac{\beta}{{\cal
J}^2}\left(I_1+I_2\right)\simeq\frac{m_\pi^2}{f_\pi^2}\left(I_1+I_2\right),
\end{eqnarray}
where the numerical factors $I_1$ and $I_2$ are given by
\begin{eqnarray}
I_1&=&\frac{1}{2}\sum_m\sum_{\bf X}\frac{Z_m^2+{\bf
X}^2}{(Z_m^2-{\bf
X}^2)^2-4Z_m^2},\nonumber\\
I_2&=&4\sum_m\sum_{\bf X}\frac{(3Z_m^2-{\bf
X}^2)^2}{\left[(Z_m^2-{\bf X}^2)^2-4Z_m^2\right]^2}.
\end{eqnarray}
Here the dimensionless notations $Z_m$ and ${\bf X}$ are defined as
$Z_m=i\nu_m/m_\pi$ and ${\bf X}={\bf q}/m_\pi$ respectively. Notice
that the integral over ${\bf X}$ is divergent and hence such an
estimation has no prediction power due to the fact that the NJL
model is nonrenormalizable. However, regardless of the numerical
factor $I_1+I_2$, we find that $\zeta\propto m_\pi^2/f_\pi^2$. Thus,
the correction should be small in the nonlinear sigma model limit
$m_\pi\ll 2M_*$.

\subsection {Transition temperature}
\label{s5-4}
While the effect of the Gaussian fluctuations at zero temperature is
to give a small correction to the diquark-diquark scattering length
and the equations of state, it can be significant at finite
temperature. In fact, as the temperature approaches the critical
value of superfluidity, the Gaussian fluctuations should dominate.
In this part, we will show that to get a correct critical
temperature in terms of the baryon density $n$, we must go beyond
the mean field. The situation is analogous to the Nozieres--Schmitt-Rink treatment of
molecular condensation in strongly interacting Fermi gases
\cite{NSR01,NSR02,RNSR01,RNSR02}.

The transition temperature $T_c$ is determined by the Thouless
criterion ${\cal D}_{\text d}^{-1}(0,{\bf 0})=0$ which can be shown
to be consistent with the saddle point condition $\delta {\cal
S}_{\text{eff}}/\delta \phi|_{\phi=0}=0$. Its explicit form is a
BCS-type gap equation
\begin{eqnarray}\label{Tc1}
\frac{1}{4G}=N_cN_f\sum_{e=\pm}\int\frac{d^3{\bf
k}}{(2\pi)^3}\frac{1-2f(\xi_{\bf k}^e)}{2\xi_{\bf k}^e}.
\end{eqnarray}
Meanwhile, the dynamic quark mass $M$ satisfies the mean-field gap
equation
\begin{eqnarray}\label{Tc2}
\frac{M-m_0}{2GM}=N_cN_f\int\frac{d^3{\bf k}}{(2\pi)^3}{1-f(\xi_{\bf
k}^-)-f(\xi_{\bf k}^+)\over E_{\bf k}}.
\end{eqnarray}

To obtain the transition temperature as a function of $n$, we need
the so-called number equation given by $n=-\partial\Omega/\partial
\mu_{\text B}$, which includes both the mean-field contribution
$n_0(\mu_{\text B},T)=2N_f\sum_{\bf k}\left[f(\xi_{\bf
k}^-)-f(\xi_{\bf k}^+)\right]$ and the Gaussian contribution
$n_{\text{fl}}(\mu_{\text B},T)=-\partial\Omega_{\text{fl}}/\partial
\mu_{\text B}$. At the transition temperature where $\Delta=0$,
$\Omega_{\text{fl}}$ can be expressed as
\begin{eqnarray}
&&\Omega_{\text{fl}}=\int\frac{d^3{\bf
q}}{(2\pi)^3}\int_{-\infty}^\infty\frac{d\omega}{2\pi
}b(\omega)\nonumber\\
&&\ \ \ \ \ \ \times\ [2\delta_{\text d}(\omega,{\bf
q})+\delta_{\sigma}(\omega,{\bf q})+3\delta_{\pi}(\omega,{\bf q})],
\end{eqnarray}
where the scattering phases are defined as $\delta_{\text
d}(\omega,{\bf q})=\text{Im}\ln[1/(4G)+\Pi_{\text
d}(\omega+i0^+,{\bf q})]$ for the diquarks,
$\delta_{\sigma}(\omega,{\bf
q})=\text{Im}\ln[1/(2G)+\Pi_\sigma(\omega+i0^+,{\bf q})]$ for the
sigma meson and $\delta_{\pi}(\omega,{\bf
q})=\text{Im}\ln[1/(2G)+\Pi_\pi(\omega+i0^+,{\bf q})]$ for the
pions. Obviously, the polarization functions should take their forms
at finite temperature in the normal phase.

The transition temperature $T_c$ at arbitrary baryon number density
$n$ can be determined numerically via solving simultaneously the gap
and number equations. However, in the dilute limit $n\rightarrow 0$
which we are interested in this section, analytical result can be
achieved. Keep in mind that $T_c\rightarrow 0$ when $n\rightarrow0$,
we find that the Fermi distribution functions $f(\xi_{\bf k}^\pm)$
vanish exponentially (since $M_*-m_\pi/2\gg T_c$) and we obtain
$\mu_{\text B}=m_\pi$ and $M=M_*$ from the gap Eqs. (\ref{Tc1})
and (\ref{Tc2}), respectively. Meanwhile the mean-field contribution
of the density $n_0$ can be neglected and the total density $n$ is
thus dominated by the Gaussian part $n_{\text{fl}}$. When
$T\rightarrow0$ we can show that $\Pi_\sigma(\omega,{\bf q})$ and
$\Pi_\pi(\omega,{\bf q})$ are independent of $\mu_{\text B}$, and
the number equation is reduced to
\begin{eqnarray}
n=-\sum_{\bf q}\int_{-\infty}^\infty\frac{d\omega}{\pi
}b(\omega)\frac{\partial\delta_{\text d}(\omega,{\bf
q})}{\partial\mu_{\text B}}.
\end{eqnarray}
Since $T_c\rightarrow0$, the inverse diquark propagator can be
reduced to ${\cal D}_{\text d}^{-1}(\omega,{\bf q})$ in
(\ref{dipro}). Thus the scattering phase $\delta_{\text d}$ can be
well approximated by $\delta_{\text d}(\omega,{\bf
q})=\pi[\Theta(\mu_{\text B}-\epsilon_{\bf
q}-\omega)-\Theta(\omega-\mu_{\text B}-\epsilon_{\bf q})]$ with
$\epsilon_{\bf q}=\sqrt{{\bf q}^2+m_\pi^2}$. Therefore, the number
equation can be further reduced to the well-known equation for ideal
Bose-Einstein condensation,
\begin{eqnarray}
n=\sum_{\bf q}\left[b(\epsilon_{\bf q}-\mu_{\text
B})-b(\epsilon_{\bf q}+\mu_{\text B})\right]\bigg|_{\mu_{\text
B}=m_\pi}.
\end{eqnarray}
Since the above equation is valid only in the low density limit
$n\rightarrow0$, the critical temperature is thus given by the
nonrelativistic result
\begin{eqnarray}
T_c=\frac{2\pi}{m_\pi}\left[\frac{n}{\xi(3/2)}\right]^{2/3}.
\end{eqnarray}
At finite density but $na_{\text{dd}}^3\ll1$, there exists a
correction to $T_c$ which is proportional to
$n^{1/3}a_{\text{dd}}$\cite{Bose01}. Such a correction is hard to
handle analytically in our model since we should consider
simultaneously the corrections to $M$ and $\mu_{\text B}$, as well
as the contribution from the sigma meson and pions.

\section {Summary}
\label{s6}
In summary, we have examined the NJL model description of weakly
interacting Bose condensate and BEC-BCS crossover in QCD-like
theories at finite baryon density. Our main conclusions are as
follows:
\\
(1)Near the quantum phase transition point $\mu_{\text B}=m_\pi$, we
have performed a Ginzburg-Landau expansion of the effective
potential. At the mean-field level, the Ginzburg-Landau free energy is
essentially the Gross-Pitaevskii free energy describing weakly
repulsive Bose condensates after a proper redefinition of the
condensate wave function. The obtained diquark-diquark scattering
length reads $a_{\text{dd}}=m_\pi/(16\pi f_\pi^2)$, which recovers
the tree-level result predicted by chiral Lagrangian.
\\
(2)We have analytically shown that the Goldstone mode near the
quantum phase transition point takes the same dispersion as the
Bogoliubov excitation in weakly interacting Bose condensates, which
gives a diquark-diquark scattering length identical to that in the
Gross-Pitaevskii free energy. The mixing between the sigma meson and
the diquarks plays an important role in recovering the Bogoliubov
dispersion.
\\
(3)The results of baryon number density and in-medium chiral and
diquark condensates predicted by chiral perturbation theory are
analytically recovered near the quantum phase transition point in
the NJL model.
\\
(4)At high density, the superfluid matter undergoes a BEC-BCS
crossover at $\mu_{\text B}\simeq (m_\sigma/m_\pi)^{1/3}m_\pi\simeq
(1.6-2)m_\pi$. At $\mu_{\text B}\simeq 3m_\pi$, the chiral symmetry
is approximated restored and the spectra of pions and sigma meson
become nearly degenerate. Well above the chemical potential of
chiral symmetry restoration, the degenerate pions and sigma meson
undergo a Mott transition, where they become unstable resonances.
Because of the spontaneous breaking of baryon number symmetry, mesons
can decay into quark pairs in the superfluid medium at nonzero
momentum.
\\
(5)The general theoretical framework of the thermodynamics beyond the mean
field is established. It is shown that the vacuum state in the
region $|\mu_{\text B}|<m_\pi$ is thermodynamically consistent in
the Gaussian approximation, i.e., all thermodynamic quantities keep
vanishing for $|\mu_{\text B}|<m_\pi$ even though the Gaussian
fluctuations are included.
\\
(6)Near the quantum phase transition point, we find that the effect
of the leading order beyond-mean-field correction is to renormalize
the diquark-diquark scattering length. The correction to the mean-field
result is estimated to be proportional to $m_\pi^2/f_\pi^2$.
Our theoretical approach provides a new way to calculate the
diquark-diquark or meson-meson scattering lengths in the NJL model
beyond the mean-field approximation. We also find that we can obtain a
correct transition temperature of Bose condensation in the dilute
limit once the beyond-mean-field corrections are included.

Our studies can be generalized to describe pion condensation at
finite isospin chemical potential $\mu_{\text I}$ \cite{ISO01,ISO02}
and kaon condensation at finite strangeness chemical potential
$\mu_{\text S}$ \cite{Kaon}. In the NJL model, pion condensation is
shown to occur at $|\mu_{\text I}|=m_\pi$ when $|\mu_{\text
S}|<m_{\text K}-m_\pi/2$, and kaon is shown to condense at
$|\mu_{\text S}|=m_{\text K}$ when $\mu_{\text I}\rightarrow 0$
\cite{isoNJL04,isoNJL06}. The generalization to pion condensation is
straightforward. The obtained Ginzburg-Landau and Gross-Pitaevskii
free energies are the same as those derived in this paper, if we
replace $\mu_{\text B}\rightarrow \mu_{\text I}$. The results are
valid both for $N_c=2$ and $N_c=3$ cases. At the mean-field level, the
results for diquark condensation at $N_c=2$ and pion condensation at
$N_c=3$ are formally identical in the NJL model. Significant
difference may appear if we consider the beyond-mean-field
corrections, since for the $N_c=3$ case the scalar diquarks are not
pseudo-Goldstone bosons. A calculation of the pion-pion scattering
length in the $I=2$ channel can be performed within our theoretical
framework. The calculations of kaon condensation and kaon-kaon
scattering length are also possible, but somewhat complicated due to
the large mass difference between the light and strange quarks.

We are also interested in how the beyond-mean-field corrections
modify the superfluid equations of state. As we learn from the
knowledge of BEC-BCS crossover in cold Fermi gases, the superfluid
equations of state can be strongly modified in the crossover regime
\cite{HU,Diener}, corresponding to the moderate baryon density in
our case. This issue is also important to the color-superconducting
quark matter \cite{CSC01,CSC02,CSCreview} at moderate density, i.e.,
for quark chemical potential around $400$MeV where the pairing gap
can be of order $O(100$MeV$)$. The numerical works are in progress.

{\bf Acknowledgement:} The work is supported by the Alexander von
Humboldt Foundation. L. He would like to thank Dr. Tomas Brauner for
valuable discussions and critical reading of the manuscript, and
Prof. Dr. Dirk Rischke for encouragement during the work.

\appendix
\section {Fermionic Model Description of Dilute Bose Condensate}
\label{app1}
In this appendix, we briefly review the theory of molecular Bose
condensation in two-component Fermi gases in the strong coupling
limit. While there exist many theoretical approaches
\cite{Tmatrix01,Tmatrix02,Tmatrix03} to deal with this problem, we
employ the field theoretical approach \cite{HU,Diener} parallel to
that used in this paper.

The Lagrangian density of the system can be written as
\begin{equation}
{\cal
L}=\sum_{\sigma=\uparrow,\downarrow}\psi_{\sigma}^\dagger\left(i\partial_
t+\frac{\nabla^2}{2m}+\mu\right)\psi_{\sigma}+g\psi_{\uparrow}^\dagger\psi_{\downarrow}^\dagger\psi_{\downarrow}^{\phantom{\dag}}
\psi_{\uparrow}^{\phantom{\dag}},
\end{equation}
where $\psi_{\uparrow,\downarrow}$ denote the two-component
(nonrelativistic) fermion fields with equal masses $m$ and chemical
potentials $\mu$. The gas is assumed to be dilute, and the coupling
constant $g$ can be related to the s-wave fermion-fermion scattering
length $a_s$ as
\begin{equation}
\frac{m}{4\pi a_s}=-\frac{1}{g(\Lambda)}+\sum_{|{\bf
k}|<\Lambda}\frac{1}{2\epsilon_{\bf k}},
\end{equation}
where $\epsilon_{\bf k}={\bf k}^2/(2m)$. In the dilute limit, we can
take the limit $\Lambda\rightarrow\infty$ in the final result.

Performing the Hubbard-Stratonovich transformation with the
auxiliary boson field
$\phi(x)=g\psi_{\downarrow}^{\phantom{\dag}}(x)
\psi_{\uparrow}^{\phantom{\dag}}(x)$, and defining the Nambu-Gor'kov
representation $ \Psi^\dagger=(\psi_\uparrow^\dagger,
 \psi_{\downarrow}^{\phantom{\dag}})$, we can evaluate the partition function of the system as
$Z=\int[d\Psi^\dagger][d\Psi][d\phi^\dagger][d\phi]\exp{\left(-{\cal
A}_{\text{eff}}\right)}$, where
\begin{eqnarray}
{\cal A}_{\text{eff}}=\int dx{|\phi(x)|^2\over g}-\int dx\int
dx^\prime\Psi^\dagger(x) {\bf G}^{-1}(x,x^\prime)\Psi(x)
\end{eqnarray}
and the inverse fermion propagator ${\bf G}^{-1}$ is given by
\begin{eqnarray}
\left(\begin{array}{cc}-\partial_{\tau}+\frac{\nabla^2}{2m}+\mu & \phi(x)\\
\phi^\dagger(x) &
-\partial_{\tau}-\frac{\nabla^2}{2m}-\mu\end{array}\right)\delta(x-x^\prime).
\end{eqnarray}
Then integrating out the fermionic degree of freedom, we get $Z=\int
[d\phi^\dagger][d\phi] \exp{\left(-{\cal S}_{\text{eff}}\right)}$
where the bosonized effective action reads
\begin{eqnarray}
{\cal S}_{\text{eff}}[\phi^\dagger,\phi]=\int
dx\frac{|\phi(x)|^2}{g}-\text{Tr}\ln {\bf G}^{-1}(x,x^\prime).
\end{eqnarray}

\subsection {Mean-field theory}
First, we consider the mean-field theory where the auxiliary boson
field $\phi(x)$ is replaced by its expectation value
$\langle\phi(x)\rangle=\Delta(x)$. In the strong coupling limit
$a_s\rightarrow 0^+$, the fermion chemical potential $\mu$
approaches $-E_b/2$ with $E_b=1/(ma_s^2)$ being the molecular
binding energy. Since the pairing gap $|\Delta|\ll|\mu|$, we can
expand the effective action in powers of $|\Delta|$, which resulting
in a Ginzburg-Landau free energy functional
\begin{eqnarray}
&&V_{\text{GL}}[\Delta(x)]=\int dx\Bigg[\Delta^\dagger(x)\left(
\kappa\frac{\partial}{\partial\tau}-\gamma\mbox{\boldmath{$\nabla$}}^2\right)\Delta(x)\nonumber\\
&&\ \ \ \ \ \ \ \ \ \ \ \ \ \ \ \ \ +\ \
\alpha|\Delta(x)|^2+\frac{1}{2}\beta|\Delta(x)|^4\Bigg].
\end{eqnarray}

The coefficients $\alpha,\beta$ of the potential terms can be
obtained from the mean-field thermodynamic potential
$\Omega_0=(T/V){\cal S}_{\text{eff}}[\Delta^\dagger,\Delta]$ which
can be evaluated as
\begin{eqnarray}
\Omega_0=-\frac{m}{4\pi a_s}|\Delta|^2-\sum_{\bf k}\left(E_{\bf
k}-\xi_{\bf k}-\frac{|\Delta|^2}{2\epsilon_{\bf k}}\right),
\end{eqnarray}
where $\xi_{\bf k}=\epsilon_{\bf k}-\mu$ and $E_{\bf
k}=\sqrt{\xi_{\bf k}^2+|\Delta|^2}$. After a simple algebra, the
coefficients $\alpha$ and $\beta$ can be evaluated as
\begin{eqnarray}
&&\alpha=\frac{m}{4\pi}\left(\sqrt{-2m\mu}-\frac{1}{a_s}\right),\nonumber\\
&&\beta=\frac{m^3}{8\pi}\frac{1}{(-2m\mu)^{3/2}}.
\end{eqnarray}
From the expression of $\alpha$, we find that a quantum phase
transition from vacuum to Bose condensation takes place at
$\mu=-1/(2ma_s^2)=-E_b/2$. Thus, near the phase transition, $\alpha$
can be simplified as
\begin{eqnarray}
\alpha\simeq-\frac{m^2a_s}{8\pi}\mu_{\text b},
\end{eqnarray}
where $\mu_{\text b}=2\mu+E_b$ is the boson chemical potential.
Further, setting $\mu= -E_b/2$, $\beta$ can be simplified as
\begin{eqnarray}
\beta\simeq-\frac{m^3a_s^3}{16\pi}.
\end{eqnarray}

The coefficients $\gamma,\kappa$ of the kinetic terms can be
obtained from the inverse boson propagator ${\cal D}^{-1}(Q)$ with
$\Delta=0$. It can be evaluated as
\begin{eqnarray}
{\cal D}^{-1}(Q)=-\frac{m}{4\pi a_s} +\sum_{\bf
k}\left(\frac{1}{i\nu_m-\xi_{{\bf k}}-\xi_{{\bf
k}+{\bf q}}}+\frac{1}{2\epsilon_{\bf k}}\right).\nonumber\\
\end{eqnarray}
In the strong coupling limit, it can be well approximated
as\cite{Tmatrix01}
\begin{eqnarray}
{\cal D}^{-1}(Q)\simeq-\frac{m^2a_s}{8\pi}\left(i\nu_m -\frac{{\bf
q}^2}{4m}\right).
\end{eqnarray}

In summary, if we define the new condensate wave function
$\psiup(x)$ by
\begin{eqnarray}
\psiup(x)=\sqrt{\frac{m^2a_s}{8\pi}}\Delta(x),
\end{eqnarray}
the Ginzburg-Landau free energy can be reduced to the
Gross-Pitaevskii free energy of dilute Bose gases,
\begin{eqnarray}
&&V_{\text{GP}}[\psiup(x)]=\int dx\Bigg[\psiup^\dagger(x)\left(
\frac{\partial}{\partial\tau}-\frac{\mbox{\boldmath{$\nabla$}}^2}{2m_{\text b}}\right)\psiup(x)\nonumber\\
&&\ \ \ \ \ \ \ \ \ \ \ \ \ \ \ \ \ -\ \ \mu_{\text
b}|\psiup(x)|^2+\frac{1}{2}\frac{4\pi a_{\text{bb}}}{m_{\text
b}}|\psiup(x)|^4\Bigg],
\end{eqnarray}
where $m_{\text b}=2m$ is the boson mass and $a_{\text{bb}}=2a_s$ is
the boson-boson scattering length. Since $a_s\rightarrow 0^+$, the
interactions among the composite bosons are repulsive and weak.

\subsection {Beyond-mean-field corrections}
To study the beyond-mean-field corrections, we consider the
fluctuations around the mean field. Making the field shift
$\phi(x)\rightarrow\Delta+\phi(x)$, we can expand the effective
action ${\cal S}_{\text{eff}}$ in powers of the fluctuations. The
zeroth order term ${\cal S}_{\text{eff}}^{(0)}$ is just the mean-field
result, and the linear terms vanish automatically guaranteed
by the saddle point condition for $\Delta$. The quadratic terms,
corresponding to Gaussian fluctuations, can be evaluated as
\begin{eqnarray}
{\cal S}_{\text{eff}}^{(2)}=\frac{1}{2}\sum_Q\left(\begin{array}{cc}
\phi^\dagger(Q) & \phi(-Q)\end{array}\right){\bf M}(Q)\left(\begin{array}{cc} \phi(Q)\\
\phi^\dagger(-Q)
\end{array}\right),
\end{eqnarray}
where the inverse boson propagator ${\bf M}$ is given by
\begin{eqnarray}
{\bf M}_{11}(Q)={\bf M}_{22}(-Q)=\frac{1}{g}+\sum_K{\cal
G}_{22}(K){\cal G}_{11}(K+Q)\nonumber\\
=\frac{1}{g}+\sum_{\bf k}\left(\frac{u_{\bf k}^2u_{{\bf k}+{\bf
q}}^2}{i\nu_m-E_{{\bf k}}-E_{{\bf k}+{\bf q}}}-\frac{v_{\bf
k}^2v_{{\bf k}+{\bf q}}^2}{i\nu_m+E_{{\bf
k}}+E_{{\bf k}+{\bf q}}} \right)\nonumber\\
\end{eqnarray}
and
\begin{eqnarray}
{\bf M}_{12}(Q)={\bf M}_{21}(Q)=\sum_K{\cal
G}_{12}(K){\cal G}_{21}(K+Q)\nonumber\\
=\sum_{\bf k}\left(\frac{u_{\bf k}v_{\bf k}u_{{\bf k}+{\bf
q}}v_{{\bf k}+{\bf q}}}{i\nu_m+E_{{\bf k}}+E_{{\bf k}+{\bf
q}}}-\frac{u_{\bf k}v_{\bf k}u_{{\bf k}+{\bf q}}v_{{\bf k}+{\bf
q}}}{i\nu_m-E_{{\bf
k}}-E_{{\bf k}+{\bf q}}} \right).\nonumber\\
\end{eqnarray}
Here the fermion Green function ${\cal G}$ is defined as ${\cal
G}^{-1}={\bf G}^{-1}[\Delta]$ and the BCS distribution functions
$v_{\bf k}^2=(1-\xi_{\bf k}/E_{\bf k})/2$ and $u_{\bf k}^2=1-v_{\bf
k}^2$ are used.

In the strong coupling limit where $|\Delta|/|\mu|\ll 1$, the matrix
elements of ${\bf M}$ can be analytically evaluated. We have
\cite{HU}
\begin{eqnarray}
&&{\bf M}(Q)\simeq \frac{m^2a_s}{8\pi}\nonumber\\
&&\times\left(\begin{array}{cc} -\omega+\frac{{\bf q}^2}{2m_{\text b}}-\mu_{\text b}+2g_0|\psiup_0|^2&g_0|\psiup_0|^2\\
g_0|\psiup_0|^2&\omega+\frac{{\bf q}^2}{2m_{\text b}}-\mu_{\text
b}+2g_0|\psiup_0|^2\end{array}\right),\nonumber\\
\end{eqnarray}
where $g_0=4\pi a_{\text{bb}}/m_{\text b}$ and $|\psiup_0|^2$ is the
minimum of the Gross-Pitaevskii free energy. Together with the 
mean-field result for the boson density $n_{\text b}=|\psiup_0|^2$, we
can show that the Goldstone mode takes a dispersion relation given
by
\begin{eqnarray}
\omega({\bf q})=\sqrt{\frac{{\bf q}^2}{2m_{\text b}}\left(\frac{{\bf
q}^2}{2m_{\text b}}+\frac{8\pi a_{\text{bb}}n_{\text b}}{m_{\text
b}}\right)},
\end{eqnarray}
which is just the Bogoliubov excitation in a dilute Bose condensate.

To evaluate the thermodynamic potential beyond the mean field, we
express the partition function in the Gaussian approximation as
\begin{eqnarray}
Z\simeq\exp{\left(-{\cal
S}_{\text{eff}}^{(0)}\right)}\int[d\phi^\dagger][d\phi]\exp\Big(-{\cal
S}_{\text{eff}}^{(2)}\Big).
\end{eqnarray}
Integrating out the Gaussian fluctuations, the total thermodynamic
potential can be expressed as
\begin{equation}
\Omega(\mu)=\Omega_0(\mu)+\Omega_{\text{fl}}(\mu),
\end{equation}
where the contribution from the Gaussian fluctuations can be
evaluated as \cite{Diener}
\begin{eqnarray}
\Omega_{\text{fl}}=\frac{1}{2}\sum_Q\ln\left[\frac{{\bf
M}_{11}(Q)}{{\bf M}_{22}(Q)}\det{\bf M}(Q)\right]e^{i\nu_m0^+}.
\end{eqnarray}

Near the quantum phase transition point $\mu=-E_b/2$, we can expand
$\Omega_{\text{fl}}$ in powers of $|\Delta|^2$. Because of the
properties of second order phase transition, the terms of order
$O(|\Delta|^2)$ vanish. To leading order, the result is
\cite{Diener}
\begin{eqnarray}
\Omega_{\text{fl}}&\simeq&-\frac{\zeta}{256\pi}\left(\frac{2m}{-\mu}\right)^{3/2}|\Delta|^4\nonumber\\
&\simeq&-\zeta\frac{m^3a_s^3}{32\pi}|\Delta|^4,\nonumber\\
\end{eqnarray}
where the numerical factor $\zeta=2.61$. From the Gross-Pitaevskii
free energy, we find that the pressure $P$ in the mean-field
approximation can be expressed as
\begin{eqnarray}
P=\frac{m^3a_s^3}{32\pi}|\Delta|^4.
\end{eqnarray}
Thus, to leading order, the beyond-mean-field corrections renormalize
the boson-boson scattering length $a_{\text{bb}}$. The new
renormalized scattering length reads
\begin{eqnarray}
a_{\text{bb}}=\frac{2a_s}{1+\zeta}\simeq0.55 a_s.
\end{eqnarray}
Notice that this result is quite close to the exact result for the
four body problem of $0.6a_s$ \cite{fourbody}. This means the
quantum fluctuations are almost correctly included in the present
theoretical approach.

Further, going beyond the leading order we find that we can fit
$\Omega_{\text{fl}}$ to the functional form \cite{Diener}
\begin{eqnarray}
\Omega_{\text{fl}}=\frac{E_b}{2a_s^3}\left(c_1\tilde{\mu}_{\text
b}^2+c_2\tilde{\mu}_{\text b}^{5/2}+\cdots\right),
\end{eqnarray}
where $\tilde{\mu}_{\text b}=\mu_{\text b}/E_b$ and the
dimensionless factors $c_1,c_2$ can be numerically determined.
Solving for the molecular chemical potential $\mu_{\text b}$ one
obtains
\begin{eqnarray}
\mu_{\text b}=\frac{4\pi a_{\text{bb}}n_{\text b}}{m_{\text
b}}\left[1+\xi\frac{32}{3\sqrt{\pi}}\left(n_{\text
b}a_{\text{bb}}^3\right)^{1/2}+\cdots\right],
\end{eqnarray}
with the coefficient $\xi=0.94$\cite{Diener} which is $6\%$ smaller
than the Lee-Huang-Yang result $\xi=1$ \cite{Bose03}.

\section {The One-Loop Susceptibilities}
\label{app2}

In this appendix, we evaluate the explicit forms of the one-loop
susceptibilities $\Pi_{\text{ij}}(Q)$ (${\text i},{\text j}=1,2,3$)
and $\Pi_\pi(Q)$. At arbitrary temperature, their expressions are
rather huge. However, at $T=0$, they can be written in rather
compact forms. For convenience, we define
$\Delta=|\Delta|e^{i\theta}$ in this appendix.

\emph{(I) Diquark sector.} First, the polarization functions
$\Pi_{11}(Q)$ and $\Pi_{12}(Q)$ can be evaluated as
\begin{widetext}
\begin{eqnarray}
\Pi_{11}(Q)&=&N_cN_f\sum_{\bf k}\Bigg[\left(\frac{(u_{\bf
k}^-)^2(u_{\bf p}^-)^2}{i\nu_m-E_{\bf k}^- -E_{\bf
p}^-}-\frac{(v_{\bf k}^-)^2(v_{\bf p}^-)^2}{i\nu_m+E_{\bf
k}^-+E_{{\bf p}}^-}-\frac{(u_{\bf k}^+)^2(u_{\bf
p}^+)^2}{i\nu_m+E_{\bf k}^++E_{\bf p}^+}
+\frac{(v_{\bf k}^+)^2(v_{\bf p}^+)^2}{i\nu_m-E_{\bf k}^+-E_{\bf p}^+}\right){\cal T}_+\nonumber\\
&&\ \ \ \ \ \ \ \ \ \ \ \ \ \ +\left(\frac{(u_{\bf k}^-)^2(v_{\bf
p}^+)^2}{i\nu_m-E_{\bf k}^--E_{\bf p}^+}-\frac{(v_{\bf
k}^-)^2(u_{\bf p}^+)^2}{i\nu_m+E_{\bf k}^-+E_{\bf
p}^+}-\frac{(u_{\bf k}^+)^2(v_{\bf p}^-)^2}{i\nu_m+E_{\bf
k}^++E_{\bf p}^-}+\frac{(v_{\bf k}^+)^2(u_{\bf p}^-)^2}{i\nu_m-E_{\bf k}^+-E_{\bf p}^-}\right){\cal T}_-\Bigg],\nonumber\\
\Pi_{12}(Q)&=&N_cN_f\sum_{\bf k}\Bigg[\left(\frac{u_{\bf k}^-v_{\bf
k}^-u_{\bf p}^-v_{\bf p}^-}{i\nu_m+E_{\bf k}^-+E_{\bf
p}^-}-\frac{u_{\bf k}^-v_{\bf k}^-u_{\bf p}^-v_{\bf
p}^-}{i\nu_m-E_{\bf k}^--E_{\bf p}^-}+\frac{u_{\bf k}^+v_{\bf
k}^+u_{\bf p}^+v_{\bf p}^+}{i\nu_m+E_{\bf k}^++E_{\bf
p}^+}-\frac{u_{\bf k}^+v_{\bf k}^+u_{\bf p}^+v_{\bf
p}^+}{i\nu_m-E_{\bf k}^+-E_{\bf p}^+}\right){\cal T}_+
\nonumber\\
&&\ \ \ \ \ \ \ \ \ \ \ \ \ \ +\left(\frac{u_{\bf k}^-v_{\bf
k}^-u_{\bf p}^+v_{\bf p}^+}{i\nu_m+E_{\bf k}^-+E_{\bf
p}^+}-\frac{u_{\bf k}^-v_{\bf k}^-u_{\bf p}^+v_{\bf
p}^+}{i\nu_m-E_{\bf k}^--E_{\bf p}^+}+\frac{u_{\bf k}^+v_{\bf
k}^+u_{\bf p}^-v_{\bf p}^-}{i\nu_m+E_{\bf k}^++E_{\bf
p}^-}-\frac{u_{\bf k}^+v_{\bf k}^+u_{\bf p}^-v_{\bf
p}^-}{i\nu_m-E_{\bf k}^+-E_{\bf p}^-}\right){\cal
T}_-\Bigg]e^{2i\theta},
\end{eqnarray}
\end{widetext}
where ${\bf p}={\bf k}+{\bf q}$. Here ${\cal T}_\pm$ are factors
arising from the trace in spin space,
\begin{equation}
{\cal T}_\pm=\frac{1}{2}\pm\frac{{\bf k}\cdot{\bf p}+M^2}{2E_{\bf
k}E_{\bf p}},
\end{equation}
and $u_{\bf k}^{\pm}, v_{\bf k}^{\pm}$ are the BCS distribution
functions defined as
\begin{equation}
(u_{\bf k}^{\pm})^2=\frac{1}{2}\left(1+\frac{\xi_{\bf k}^\pm}{E_{\bf
k}^\pm}\right),\ \ \ \ \ (v_{\bf
k}^{\pm})^2=\frac{1}{2}\left(1-\frac{\xi_{\bf k}^\pm}{E_{\bf
k}^\pm}\right).
\end{equation}
At $Q=0$, we find that
\begin{equation}
\Pi_{12}(0)=\Delta^2\frac{1}{4}N_cN_f\sum_{\bf
k}\left[\frac{1}{(E_{\bf k}^-)^3}+\frac{1}{(E_{\bf k}^+)^3}\right].
\end{equation}
Thus, near the quantum phase transition point, we have
$\Pi_{12}(0)=\Delta^2\beta_1+O(|\Delta|^4)$. On the other hand, a
simple algebra shows that
\begin{equation}
\frac{1}{4G}+\Pi_{11}(0)-|\Pi_{12}(0)|=\frac{\partial\Omega_0}{\partial|\Delta|^2}.
\end{equation}
Therefore, the mean-field gap equation for $\Delta$ ensures the
Goldstone's theorem in the superfluid phase.

\emph{(II) Diquark-sigma mixing terms.} The term $\Pi_{13}$ standing
for the mixing between the sigma meson and the diquarks reads
\begin{widetext}
\begin{eqnarray}
\Pi_{13}(Q)&=&N_cN_f\sum_{\bf k}\Bigg[\bigg(\frac{u_{\bf k}^+v_{\bf
k}^+(v_{\bf p}^+)^2+u_{\bf p}^+v_{\bf p}^+(v_{\bf
k}^+)^2}{i\nu_m-E_{\bf k}^+-E_{\bf p}^+} +\frac{u_{\bf k}^+v_{\bf
k}^+(u_{\bf p}^+)^2+u_{\bf p}^+v_{\bf p}^+(u_{\bf
k}^+)^2}{i\nu_m+E_{\bf k}^++E_{\bf p}^+}
\nonumber\\
&&\ \ \ \ \ \ \ \ \ \ \ \ \ \ \ \ -\frac{u_{\bf k}^-v_{\bf
k}^-(u_{\bf p}^-)^2+u_{\bf p}^-v_{\bf p}^-(u_{\bf
k}^-)^2}{i\nu_m-E_{\bf k}^--E_{{\bf p}}^-} -\frac{u_{\bf k}^-v_{\bf
k}^-(v_{\bf p}^-)^2+u_{\bf p}^-v_{\bf p}^-(v_{\bf
k}^-)^2}{i\nu_m+E_{\bf k}^-
+E_{\bf p}^-}\bigg){\cal I}_+\nonumber\\
&&\ \ \ \ \ \ \ \ \ \ \ \ \ \ \ \ +\bigg(\frac{u_{\bf k}^+v_{\bf
k}^+(u_{\bf p}^-)^2+u_{\bf p}^-v_{\bf p}^-(v_{\bf
k}^+)^2}{i\nu_m-E_{\bf k}^+-E_{{\bf p}}^-} +\frac{u_{\bf k}^+v_{\bf
k}^+(v_{\bf p}^-)^2+u_{\bf p}^-v_{\bf p}^-(u_{\bf
k}^+)^2}{i\nu_m+E_{\bf k}^+ +E_{\bf p}^-}
\nonumber\\
&&\ \ \ \ \ \ \ \ \ \ \ \ \ \ \ \ -\frac{u_{\bf k}^-v_{\bf
k}^-(v_{\bf p}^+)^2+u_{\bf p}^+v_{\bf p}^+(u_{\bf
k}^-)^2}{i\nu_m-E_{\bf k}^--E_{\bf p}^+} -\frac{u_{\bf k}^-v_{\bf
k}^-(u_{\bf p}^+)^2+u_{\bf p}^+v_{\bf p}^+(v_{\bf
k}^-)^2}{i\nu_m+E_{\bf k}^-+E_{\bf p}^+}\bigg){\cal
I}_-\Bigg]e^{i\theta},
\end{eqnarray}
\end{widetext}
where the factors ${\cal I}_\pm$ are defined as
\begin{equation}
{\cal I}_\pm=\frac{M}{2}\left(\frac{1}{E_{\bf k}}\pm\frac{1}{E_{\bf
p}}\right).
\end{equation}
One can easily find that $\Pi_{13}\sim M\Delta$, thus it vanishes
when $\Delta$ or $M$ approaches zero. At $Q=0$, we have
\begin{equation}
\Pi_{13}(0)=\Delta\frac{1}{2}N_cN_f\sum_{\bf k}\frac{M}{E_{\bf
k}}\left[\frac{\xi_{\bf k}^-}{(E_{\bf k}^-)^3}+\frac{\xi_{\bf
k}^+}{(E_{\bf k}^+)^3}\right].
\end{equation}
Thus the quantity $H_0$ defined in (\ref{bexp}) can be evaluated as
\begin{eqnarray}
H_0&=&\frac{1}{2}N_cN_f\sum_{e=\pm}\sum_{\bf k}\frac{M_*}{E_{\bf
k}^*}\frac{1}{(E_{\bf k}^*-em_\pi/2)^2}\nonumber\\
&=&\frac{\partial^2\Omega_0(y,M)}{\partial M\partial y}\Bigg|_{y=0}.
\end{eqnarray}

\emph{(III) Sigma meson and pions.} The polarization function
$\Pi_{33}$ which stands for the sigma meson can be evaluated as
\begin{widetext}
\begin{eqnarray}
\Pi_{33}(Q)&=&
 N_cN_f\sum_{\bf k}
\Bigg[(v_{\bf k}^-u_{\bf p}^-+u_{\bf k}^-v_{\bf p}^-)^2\left(
\frac{1}{i\nu_m-E_{\bf k}^--E_{\bf p}^-}
-\frac{1}{i\nu_m+E_{\bf k}^-+E_{\bf p}^-}\right){\cal T}^\prime_-\nonumber\\
&&\ \ \ \ \ \ \ \ \ \ \ \ \ \ \  +(v_{\bf k}^+u_{\bf p}^++u_{\bf
k}^+v_{\bf p}^+)^2\left( \frac{1}{i\nu_m-E_{\bf k}^+-E_{\bf p}^+}
-\frac{1}{i\nu_m+E_{\bf k}^++E_{\bf p}^+}\right){\cal T}^\prime_-\nonumber\\
&&\ \ \ \ \ \ \ \ \ \ \ \ \ \ \  +(v_{\bf k}^+v_{\bf p}^-+u_{\bf
k}^+u_{\bf p}^-)^2\left( \frac{1}{i\nu_m-E_{\bf k}^+-E_{\bf p}^-}
-\frac{1}{i\nu_m+E_{\bf k}^++E_{\bf p}^-}\right){\cal T}^\prime_+\nonumber\\
&&\ \ \ \ \ \ \ \ \ \ \ \ \ \ \ +(v_{\bf k}^-v_{\bf p}^++u_{\bf
k}^-u_{\bf p}^+)^2\left( \frac{1}{i\nu_m-E_{\bf k}^--E_{\bf p}^+}
-\frac{1}{i\nu_m+E_{\bf k}^-+E_{\bf p}^+}\right){\cal
T}^\prime_+\Bigg],
\end{eqnarray}
\end{widetext}
where the factors ${\cal T}^\prime_\pm$ are defined as
\begin{equation}
{\cal T}^\prime_\pm=\frac{1}{2}\pm\frac{{\bf k}\cdot{\bf
p}-M^2}{2E_{\bf k}E_{\bf p}}.
\end{equation}
At $Q=0$ and for $\Delta=0$, we find that
\begin{eqnarray}
{\bf M}_{33}(0)&=&\frac{1}{2G}-2N_cN_f\sum_{\bf k}\frac{1}{E_{\bf
k}^*}+2N_cN_f\sum_{\bf k}\frac{M_*^2}{E_{\bf k}^{*3}}\nonumber\\
&=&\frac{\partial^2\Omega_0(y,M)}{\partial M^2}\Bigg|_{y=0}.
\end{eqnarray}
Finally, the polarization function $\Pi_\pi(Q)$ for pions can be
obtained by replacing ${\cal T}^\prime_\pm\rightarrow{\cal T}_\pm$.
Thus, when $M\rightarrow 0$, the sigma meson and pions become
degenerate and chiral symmetry is restored.

\section {Expansion of $\Omega_{\text{fl}}$ in Terms of $|\Delta|^2$}
\label{app3}
In this appendix, we derive the expression of the Taylor expansion
of $\Omega_{\text{fl}}$ in terms of $|\Delta|^2\equiv y$. As we have
shown in Sec. \ref{s5}, the leading-order term should be
$O(|\Delta|^4)$. Thus, we need to evaluate the numerical factor
$\zeta$. A key problem here is that the effective quark mass $M$ and
the chemical potential $\mu_{\text B}$ are both functions of
$|\Delta|^2$ determined at the mean-field level.

First, we expand the matrix elements of ${\bf M}$ and ${\bf N}$ in
terms of $y$. Any of these elements denoted by $F$ is a function of
$\mu_{\text B}$, $M$ and $y$. Our method of expansion is as follows.
We firstly expand $F(\mu_{\text B},M,y)$ in terms of $y$ formally
with $\mu_{\text B}$ and $M$ being fixed parameters, i.e.,
\begin{equation}
F(\mu_{\text B},M,y)=F_0(\mu_{\text B},M)+F_1(\mu_{\text
B},M)y+F_2(\mu_{\text B},M)y^2+O(y^3),
\end{equation}
where $F_0(\mu_{\text B},M)\equiv F(\mu_{\text B},M,0)$ and the
expansion coefficients are defined as
\begin{eqnarray}
&&F_1(\mu_{\text B},M)=\frac{\partial F(\mu_{\text B},M,y)}{\partial y}\bigg|_{y=0},\nonumber\\
&&F_2(\mu_{\text B},M)=\frac{1}{2}\frac{\partial^2 F(\mu_{\text
B},M,y)}{\partial y^2}\bigg|_{y=0}.
\end{eqnarray}
We then expand the coefficients $F_i(\mu_{\text B},M)$ ($i=0,1,2$)
at $(\mu_{\text B},M)=(m_\pi,M_*)$, using the fact that
\begin{eqnarray}
&&M\simeq M_*-\frac{1}{2M_*}y,\nonumber\\
&&\mu_{\text B}\simeq m_\pi+\frac{\beta}{2m_\pi{\cal J}}y.
\end{eqnarray}
Doing this we formally obtain
\begin{eqnarray}
F_i(\mu_{\text B},M)&=&F_i(m_\pi,M_*)+F_i^1(m_\pi,M_*)y\nonumber\\
&+&F_i^2(m_\pi,M_*)y^2+O(y^3).
\end{eqnarray}
Finally, up to order $O(y^2)$, we have
\begin{widetext}
\begin{eqnarray}
F(\mu_{\text
B},M,y)=F_0(m_\pi,M_*)+\left[F_0^1(m_\pi,M_*)+F_1(m_\pi,M_*)\right]y+\left[F_0^2(m_\pi,M_*)+F_1^1(m_\pi,M_*)+F_2(m_\pi,M_*)\right]y^2.
\end{eqnarray}
Using this method, we can expand the matrix elements of ${\bf M}$
and ${\bf N}$ formally as follows:
\begin{eqnarray}\label{C6}
&&{\bf M}_{11}(Q)={\cal D}_{\text d}^{*-1}(Q)+X_1(Q)|\Delta|^2+Y_1(Q)|\Delta|^4+O(|\Delta|^6),\nonumber\\
&&{\bf M}_{12}(Q)=\Delta^2Z(Q)+O(|\Delta|^4),\nonumber\\
&&{\bf M}_{13}(Q)=\Delta W(Q)+O(|\Delta|^3),\nonumber\\
&&{\bf M}_{33}(Q)={\cal D}_\sigma^{*-1}(Q)+X_2(Q)|\Delta|^2+Y_2(Q)|\Delta|^4+O(|\Delta|^6),\nonumber\\
&&\ {\bf N}_{11}(Q)={\cal
D}_\pi^{*-1}(Q)+X_3(Q)|\Delta|^2+Y_3(Q)|\Delta|^4+O(|\Delta|^6).
\end{eqnarray}
Here ${\cal D}_{\text d}^{*-1}(Q)$ is defined as ${\cal D}_{\text
d}^{-1}(Q;\mu_{\text B}=m_\pi)$.

Meanwhile, the thermodynamic potential $\Omega_{\text{fl}}$ can be
expressed as
\begin{eqnarray}
\Omega_{\text{fl}}=\frac{1}{2}\sum_Q\left\{\ln\left[\frac{{\bf
M}_{11}(Q)}{{\bf M}_{22}(Q)}\det{\bf M}(Q)\right]+\ln\det{\bf
N}(Q)\right\}e^{i\nu_m 0^+}-\frac{1}{2}\sum_Q\left[2\ln{\cal
D}_{\text d}^{*-1}(Q)+\ln{\cal D}_\sigma^{*-1}(Q)+3{\cal
D}_\pi^{*-1}(Q)\right]e^{i\nu_m 0^+}.
\end{eqnarray}
Using the expansion (\ref{C6}), we find that the factor $\zeta$  is
given by
\begin{eqnarray}
\beta\zeta&=&\sum_Q\left\{\frac{X_2(Q)}{{\cal D}_{\sigma}^{*-1}(Q)}
+\frac{X_1(Q)}{{\cal D}_{\text d}^{*-1}(Q)}+\frac{X_1(-Q)}{{\cal
D}_{\text d}^{*-1}(-Q)}
+\frac{W^2(Q)}{{\cal D}_{\text d}^{*-1}(Q){\cal D}_{\sigma}^{*-1}(Q)}+\frac{W^2(-Q)}{{\cal D}_{\text d}^{*-1}(-Q){\cal D}_{\sigma}^{*-1}(Q)}\right\}^2\nonumber\\
&-&\sum_Q\left\{\frac{Y_2(Q)}{{\cal
D}_{\sigma}^{*-1}(Q)}+\frac{Y_1(Q)}{{\cal D}_{\text d}^{*-1}(Q)}+\frac{Y_1(-Q)}{{\cal D}_{\text d}^{*-1}(-Q)}
+\frac{X_1(Q)X_1(-Q)-Z^2(Q)}{{\cal D}_{\text d}^{*-1}(Q){\cal D}_{\text d}^{*-1}(-Q)}\right\}\nonumber\\
&+&\sum_Q\frac{W^2(-Q)X_1(Q)+W^2(Q)X_1(-Q)-2W(Q)W(-Q)Z(Q)}{{\cal
D}_{\text d}^{*-1}(Q){\cal D}_{\text d}^{*-1}(-Q){\cal
D}_{\sigma}^{*-1}(Q)}\nonumber\\
&+&\sum_Q\left\{\left[\frac{X_3(Q)}{{\cal
D}_{\pi}^{*-1}(Q)}\right]^2-\frac{Y_3(Q)}{{\cal
D}_{\pi}^{*-1}(Q)}\right\}.
\end{eqnarray}

\end{widetext}
It is obvious that $Z(Q)=B(Q)$ and $W(Q)=H(Q)$ where $B(Q)$ and
$H(Q)$ are defined in (\ref{bexp}). On the other hand, since
\begin{eqnarray}
\frac{\beta}{2m_\pi{\cal J}}\simeq\frac{m_\pi}{2M_*}\frac{1}{2M_*},
\end{eqnarray}
to leading order of $O(m_\pi/2M_*)$, we can set $\mu_{\text
B}=m_\pi$ in all equations and identify $X_1(Q)=A(Q)$ defined in
(\ref{bexp}).


\begin{thebibliography}{99}
\small
\bibitem{itoh}          {N. Itoh, Prog. Theor. Phys. {\bf 44}, 291(1970); E. Witten, Phys. Rev. {\bf D30}, 272(1984).}
\bibitem{Lr01}          {F. Karsch, Lect. Notes Phys. {\bf 583}, 209(2002).}
\bibitem{Lr02}          {S. Muroya, A. Nakamura, C. Nonaka and T. Takaishi, Prog. Theor. Phys. {\bf 110}, 615(2003).}
\bibitem{ISO01}         {D. T. Son and M. A. Stephanov, Phys. Rev. Lett. {\bf 86}, 592(2001).}
\bibitem{ISO02}         {D. T. Son and M. A. Stephanov, Phys. Atom. Nucl. {\bf 64}, 834(2001).}
\bibitem{QL01}          {J. B. Kogut, M. A. Stephanov and D. Toublan, Phys. Lett. {\bf B464}, 183(1999).}
\bibitem{QL02}          {J. B. Kogut, M. A. Stephanov, D. Toublan, J. J. M. Verbaarschot and A. Zhitnitsky, Nucl. Phys. {\bf B582}, 477(2000).}
\bibitem{QL03}          {K. Splittorff, D. T. Son and M. A. Stephanov, Phys. Rev. {\bf D64}, 016003(2001).}
\bibitem{QL04}          {J. T. Lenaghan, F. Sannino and K. Splittorff, Phys. Rev. {\bf D65}, 054002(2002).}
\bibitem{QL05}          {K. Splittorff, D. Toublan and J. J. M. Verbaarschot, Nucl. Phys. {\bf B620}, 290(2002).}
\bibitem{QL06}          {K. Splittorff, D. Toublan and J. J. M. Verbaarschot, Nucl. Phys. {\bf B639}, 524(2002).}
\bibitem{QL07}          {T. Zhang, T. Brauner and D. H. Rischke, JHEP {\bf 1006}, 064(2010).}
\bibitem{Bose01}        {For review, see J. O. Andersen, Rev. Mod. Phys. {\bf 76}, 599(2004).}
\bibitem{Bose02}        {N. N. Bogoliubov, J. Phys. USSR {\bf 11}, 23(1947).}
\bibitem{Bose03}        {T. D. Lee, K. Huang and C. N. Yang, Phys. Rev. {\bf 106}, 1135(1957).}
\bibitem{PiC01}         {R. F. Sawyer, Phys. Rev. Lett. {\bf 29}, 382(1972).}
\bibitem{PiC02}         {D. J. Scalapino, Phys. Rev. Lett. {\bf 29}, 386(1972).}
\bibitem{PiC03}         {G. Baym, Phys. Rev. Lett. {\bf 30}, 1340(1973).}
\bibitem{PiC04}         {D. K. Campbell, R. F. Dashen and J. T. Manassah, Phys. Rev. {\bf D12}, 979(1975); \emph{ibid}{\bf D12}, 1010(1975).}
\bibitem{pQ01}          {D. T. Son, Phys. Rev. D {\bf 59}, 094019(1999).}
\bibitem{pQ02}          {T. Sch{\"a}fer and F. Wilczek, Phys. Rev. {\bf D60}, 114033(1999). }
\bibitem{pQ03}          {D. Pisarski and D. H. Rischke, Phys. Rev. D {\bf 61}, 074017(2000);
                         \emph{ibid}{\bf 61}, 051501(2000).}
\bibitem{Yama}          {T. Kanazawa, T. Wettig and N. Yamamoto, JHEP {\bf 0908}, 003(2009).}
\bibitem{Eagles}        {D. M. Eagles, Phys. Rev. {\bf 186}, 456(1969).}
\bibitem{Leggett}       {A. J. Leggett, in {\it Modern Trends in the Theory of Condensed Matter},
                         edited by A. Pekalski and R. Przystawa, Springer-Verlag, Berlin(1980).}
\bibitem{BCSBECexp}     {M. Greiner, et.al, Nature 426, 537(2003);
                         S. Jochim, et.al., Science 302, 2101(2003);
                         M. W. Zwierlein, et.al., Nature 435, 1047(2003).}
\bibitem{L2C01}         {S. Hands, I. Montvay, S. Morrison, M. Oevers, L. Scorzato and J. Skullerud, Eur. Phys. J. {\bf C17}, 285(2000).}
\bibitem{L2C02}         {S. Hands, I. Montvay, L. Scorzato and J. Skullerud, Eur. Phys. J. {\bf C22}, 451(2001).}
\bibitem{L2C03}         {J. B. Kogut, D. K.Sinclair, S. J. Hands and S. E. Morrison, Phys. Rev. {\bf D64}, 094505(2001).}
\bibitem{L2C04}         {J. B. Kogut, D. Toublan and D. K. Sinclair, Phys. Lett. {\bf B514}, 77(2001).}
\bibitem{L2C05}         {S. Hands, S. Kim and J. Skullerud, Eur. Phys. J. {\bf C48}, 193(2006).}
\bibitem{L2C06}         {S. Hands, S. Kim and J. Skullerud, Phys. Rev. {\bf D81}, 091502(R)(2010).}
\bibitem{Liso01}        {J. B. Kogut, D. K. Sinclair, Phys. Rev. {\bf D66}, 034505(2002).}
\bibitem{Liso02}        {J. B. Kogut, D. K. Sinclair, Phys. Rev. {\bf D66}, 014508(2002).}
\bibitem{Liso03}        {J. B. Kogut, D. K. Sinclair, Phys. Rev. {\bf D70}, 094501(2004).}
\bibitem{Liso04}        {P. Forcrand, M. A. Stephanov and U. Wenger, PoSLAT2007, 237(2007).}
\bibitem{CHiso}         {M. Loewe and C. Villavicencio, Phys. Rev. {\bf D67}, 074034(2003).}
\bibitem{Sigma}         {J. O. Andersen, Phys. Rev. {\bf D75}, 065011(2007).}
\bibitem{NJL}           {Y. Nambu and G. Jona-Lasinio, Phys. Rev. {\bf 122}, 345(1961).}
\bibitem{NJLreview01}   {U. Vogl and W. Weise, Prog. Part. and Nucl. Phys. {\bf 27}, 195(1991).}
\bibitem{NJLreview02}   {S. P. Klevansky, Rev. Mod. Phys. {\bf 64(3)}, 649(1992)}
\bibitem{NJLreview03}   {T. Hatsuda and T. Kunihiro, Phys. Rep. {\bf 247}, 221(1994).}
\bibitem{isoNJL01}      {D. Toublan and J. B. Kogut, Phys. Lett. {\bf B564}, 212(2003).}
\bibitem{isoNJL02}      {M. Frank, M. Buballa and M. Oertel, Phys. Lett. {\bf B562}, 221(2003).}
\bibitem{isoNJL03}      {A. Barducci, R. Casalbuoni, G. Pettini and L. Ravagli, Phys. Rev. {\bf D69}, 096004(2004).}
\bibitem{isoNJL04}      {A. Barducci, R. Casalbuoni, G. Pettini and L. Ravagli, Phys. Rev. {\bf D71}, 016011(2005).}
\bibitem{isoNJL05}      {L. He and P. Zhuang, Phys. Lett. {\bf B615}, 93(2005).}
\bibitem{isoNJL06}      {L. He, M. Jin and P. Zhuang, Phys. Rev. {\bf D71}, 116001(2005).}
\bibitem{isoNJL07}      {H. J. Warringa, D. Boer and J. O. Andersen, Phys. Rev. {\bf D72}, 014015(2005).}
\bibitem{isoNJL08}      {L. He, M. Jin and P. Zhuang, Phys. Rev. {\bf D74}, 036005(2006).}
\bibitem{isoNJL09}      {Z. Zhang and Y. -X. Liu, Phys. Rev. {\bf C75}, 064910(2007)}
\bibitem{isoNJL10}      {J. Xiong, M. Jin and J. Li, J. Phys. {\bf G36}, 125005(2009).}
\bibitem{2CNJL01}       {C. Ratti and W. Weise, Phys. Rev. {\bf D70},  054013(2004).}
\bibitem{2CNJL02}       {G. Sun, L. He and P. Zhuang, Phys. Rev. {\bf D75}, 096004(2007).}
\bibitem{2CNJL03}       {T. Brauner, K. Fukushima and Y. Hidaka, Phys. Rev. {\bf D80}, 074035(2009).}
\bibitem{2CNJL04}       {J. O. Andersen and T. Brauner, Phys. Rev. {\bf D81}, 096004(2010).}
\bibitem{HU}            {H. Hu, X.-J. Liu and P. D. Drummond, Europhys. Lett. {\bf 74}, 574(2006);
                         Nature Physics {\bf 3}, 469(2007).}
\bibitem{Diener}        {R. B. Diener, R. Sensarma and M. Randeria, Phys. Rev. {\bf A77}, 023626(2008).}
\bibitem{Leyron}        {X. Leyronas, R. Combescot, Phys. Rev. Lett. {\bf 99}, 170402(2007).}
\bibitem{massive}       {M. Huang, P. Zhuang and W. Chao, Phys. Rev. {\bf D65}, 076012(2002).}
\bibitem{note3}         {To be consistent with the convention used in \cite{BCSBEC0,HU,Diener}, here $\phi^\dagger(Q)$ is defined as $[\phi(Q)]^\dagger$.}
\bibitem{nao}           {N. Nagaosa, {\it Quantum Field Theory in Condensed Matter Physics}, Springer, Heidelberg, Germany(1999).}
\bibitem{GP01}          {L. Pitaevskii and S. Stringari, {\it Bose-Einstein condensation}, Oxford University Press(2003).}
\bibitem{GP02}          {C. J. Pethick and H. Smith, {\it Bose-Einstein condensation in dilute gases}, Cambridge University Press(2002).}
\bibitem{schulze}       {H. J. Schulze, J. Phys. G. {\bf 21}, 185(1995).}
\bibitem{pipi}          {S. Weinberg, Phys. Rev. Lett. {\bf 17}, 616(1966).}
\bibitem{BCSBEC0}       {J. R. Engelbrecht, M. Randeria and C. A. R. S\'{a} de Melo, Phys. Rev. {\bf B55}, 15153(1997).}
\bibitem{RBCSBEC01}     {L. He and P. Zhuang, Phys. Rev. {\bf D75}, 096003(2007).}
\bibitem{RBCSBEC02}     {T. Brauner, Phys. Rev. {\bf D77}, 096006(2008). }
\bibitem{cohen}         {T. D. Cohen, R. J. Furnstahl and D. K. Griegel, Phys. Rev. {\bf C45}, 1881(1992).}
\bibitem{cohen2}        {T. D. Cohen, R. J. Furnstahl, D. K. Griegel and Xuemin Jin, Prog. Part. Nucl. Phys. {\bf 35}, 221(1995).}
\bibitem{HFiso}         {L. He, Y. Jiang and P. Zhuang, Phys. Rev. {\bf C79}, 045205(2009).}
\bibitem{dson}          {D. T. Son and M. A. Stephanov, Phys. Rev. {\bf A74}, 013614(2006).}
\bibitem{kita}          {M. Kitazawa, D. H. Rischke and I. A. Shovkovy, Phys. Lett. {\bf B663}, 228(2008).}
\bibitem{NSR01}         {P. Nozi\`{e}res and S. Schmitt-Rink, J. Low Temp. Phys. {\bf 59}, 195(1985).}
\bibitem{NSR02}         {C. A. R. S\'{a} de Melo, M. Randeria and J. R. Engelbrecht, Phys. Rev. Lett. {\bf 71}, 3202(1993).}
\bibitem{RNSR01}        {Y. Nishida and H. Abuki, Phys. Rev. {\bf D72}, 096004(2005)}
\bibitem{RNSR02}        {H. Abuki, Nucl. Phys. {\bf A791}, 117(2007).}
\bibitem{G0G}           {L. He and P. Zhuang, Phys. Rev. {\bf D76}, 056003(2007).}
\bibitem{note1}         {In nonrelativistic systems, the BCS and BEC states are distinguished by $\mu>0$ and $\mu<0$,
                         respectively, where $\mu$ is the fermion chemical potential.}
\bibitem{decon}         {D. Toublan and A. R. Zhitnitsky, Phys. Rev. {\bf D73}, 034009(2006).}
\bibitem{note2}         {This is only possible for our two-flavor case.
                         For $N_f>2$, the chiral symmetry is spontaneously broken at high density\cite{Yama}.}
\bibitem{tomas}         {For a careful explanation of the meson spectra in chiral perturbation theory, see
                         T. Brauner, Mod. Phys. Lett. {\bf A21}, 559(2006).}
\bibitem{mott}          {D. Blaschke, F. Reinholz, G. Ropke and D. Kremp, Phys. Lett. {\bf 151B}, 439(1985).}
\bibitem{precursor}     {T. Hatsuda, T. Kunihiro, Phys. Rev. Lett. {\bf 55}, 158(1985); Phys. Lett. {\bf B185}, 304(1987).}
\bibitem{note4}         {The mesons we called here are the collective excitations which have the same quantum numbers as the pions and the sigma meson
                         in the vacuum. They are also called ``pions" and ``sigma meson" in the chiral restored regime.}
\bibitem{ebert}         {For the studies of meson properties in color-superconducting quark matter, see
                         D. Ebert, K. G. Klimenko and V. L. Yudichev, Phys. Rev. {\bf C72}, 015201(2005) and
                         D. Zablocki, D. Blaschke and R. Anglani, AIPConf. Proc.{\bf 1038}, 159(2008).
                         However, in these papers the meson properties are investigated only at zero momentum.}
\bibitem{zhuang01}      {J. H{\"u}fner, S. P. Klevansky, P. Zhuang and H. Voss, Annals Phys. {\bf 234}, 225(1994). }
\bibitem{zhuang02}      {P. Zhuang, J. H{\"u}fner and S.P. Klevansky, Nucl. Phys. {\bf A576}, 525(1994).}
\bibitem{he}            {L. He, et.al., in preparation.}
\bibitem{Kaon}          {J. B. Kogut and D. Toublan, Phys. Rev. {\bf D64}, 034007(2001).}
\bibitem{CSC01}          {R. Rapp, T. Schaefer, E.V. Shuryak and M. Velkovsky, Phys. Rev. Lett. {\bf 81}, 53(1998).}
\bibitem{CSC02}          {M. Alford, K. Rajagopal and F. Wilczek, Phys. Lett. {\bf B422}, 247(1998). }
\bibitem{CSCreview}      {M. Alford, K. Rajagopal, T. Sch{\"a}fer and A. Schmitt, Rev. Mod. Phys. {\bf 80}, 1455(2008);
                          K. Rajagopal and F. Wilczek, hep-ph/0011333;
                          D. K. Hong, Acta Phys. Pol. {\bf B32}, 1253(2001);
                          M. Alford, Ann. Rev. Nucl. Part. Sci. {\bf 51}, 131(2001);
                          T. Sch{\"a}fer, hep-ph/0304281;
                          D. H. Rischke, Prog. Part. Nucl. Phys. {\bf 52}, 197(2004);
                          M. Buballa, Phys. Rep. {\bf 407}, 205(2005);
                          H. -C. Ren, hep-ph/0404074;
                          M. Huang, Int. J. Mod. Phys. {\bf E14}, 675(2005);
                          I. A. Shovkovy, Found. Phys.{\bf 35}, 1309(2005);
                          Q. Wang, arXiv:0912.2485.}
\bibitem{Tmatrix01}     {A. Perali, P. Pieri, G. C. Strinati and C. Castellani, Phys. Rev. {\bf B66}, 024510(2002);
                         P. Pieri, L. Pisani and G. C. Strinati, Phys. Rev. {\bf B 70}, 094508(2004).}
\bibitem{Tmatrix02}     {R. Haussmann, W. Rantner, S. Cerrito and W. Zwerger, Phys. Rev. {\bf A75}, 023610(2007).}
\bibitem{Tmatrix03}     {Q. Chen, J. Stajic, S. Tan and K. Levin, Phys. Rep. {\bf 412}, 1(2005).  }
\bibitem{fourbody}      {D. S. Petrov, C. Salomon and G. V. Shlyapnikov, Phys. Rev. Lett. {\bf 93}, 090404(2004). }
\end{thebibliography}
\end{document}